\documentclass[prd,twocolumn,showpacs,nofootinbib]{revtex4}

\usepackage{amsfonts}
\usepackage{amsmath}
\usepackage{amssymb}
\usepackage{upgreek}
\usepackage{bm}
\usepackage{dcolumn}
\usepackage{epsfig}
\usepackage{graphicx}
\usepackage{graphics}
\usepackage[latin1]{inputenc}
\usepackage{latexsym}
\usepackage{rotating}
\usepackage{hyperref}
\usepackage{xspace} 
\usepackage[usenames,dvipsnames]{color}

\usepackage{ulem}
\definecolor {darkgreen}{rgb}{0.2,0.7,0.2} 

\newcommand\be{\begin{equation}}
\newcommand\ba{\begin{eqnarray}}
\newcommand\ee{\end{equation}}
\newcommand\ea{\end{eqnarray}}
\newcommand{\bes}{\begin{subequations}}
\newcommand{\ees}{\end{subequations}}
\newcommand{\beqn}{\begin{eqnarray*}}
\newcommand{\eeqn}{\end{eqnarray*}}

\newcommand{\mb}[1]{\mbox{\boldmath $#1$}}

\newcommand{\nn}{\nonumber}

\newcommand{\SP}{{\mbox{\tiny SP}}}
\newcommand{\nonprec}{{\mbox{\tiny nonprec}}}
\newcommand{\precc}{{\mbox{\tiny prec}}}
\newcommand{\FSP}{{\mbox{\tiny FSP}}}
\newcommand{\RSP}{{\mbox{\tiny RSP}}}

\newcommand{\GW}{{\mbox{\tiny GW}}}

\newcommand{\orb}{{\mbox{\tiny orb}}}
\newcommand{\n}{{\mbox{\tiny (n)}}}
\newcommand{\zero}{{\mbox{\tiny (0)}}}
\newcommand{\one}{{\mbox{\tiny (1)}}}
\newcommand{\two}{{\mbox{\tiny (2)}}}
\newcommand{\onezero}{{\mbox{\tiny (1,0)}}}
\newcommand{\zeroone}{{\mbox{\tiny (0,1)}}}
\newcommand{\zerozero}{{\mbox{\tiny (0,0)}}}
\newcommand{\oneone}{{\mbox{\tiny (1,1)}}}
\newcommand{\dd}{{\mbox{\tiny d}}}
\newcommand{\rr}{{\mbox{\tiny rr}}}
\newcommand{\pr}{{\mbox{\tiny pr}}}

\newcommand{\p}{\partial}

\begin{document}

\title{Gravitational waveforms for precessing, quasicircular compact binaries \\with multiple scale analysis: Small spin expansion}

\author{Katerina Chatziioannou}
\affiliation{Department of Physics, Montana State University, Bozeman, Montana 59718, USA.}
\author{Antoine Klein}
\affiliation{Department of Physics, Montana State University, Bozeman, Montana 59718, USA.}
\author{Nicol\'as Yunes}
\affiliation{Department of Physics, Montana State University, Bozeman, Montana 59718, USA.}
\author{Neil Cornish}
\affiliation{Department of Physics, Montana State University, Bozeman, Montana 59718, USA.}

\date{\today}

\begin{abstract}

We obtain analytical gravitational waveforms in the frequency domain for precessing, quasicircular compact binaries with small spins, applicable, for example, to binary neutron star inspirals. We begin by calculating an analytic solution to the precession equations, obtained by expanding in the dimensionless spin parameters and using multiple-scale analysis to separate time scales. We proceed by analytically computing the Fourier transform of time-domain waveform through the stationary phase approximation. We show that the latter is valid for systems with small spins. Finally, we show that these waveforms have a high overlap with numerical waveforms obtained through direct integration of the precession equations and discrete Fourier transformations. The resulting, analytic waveform family is ideal for detection and parameter estimation of gravitational waves emitted by inspiraling binary neutron stars with ground-based detectors.  

\end{abstract}

\maketitle

\section{Introduction}
\label{intro}

Gravitational waves (GWs) are expected to be detected soon  by advanced ground-based interferometric GW detectors, such as Advanced~LIGO~\cite{Abbott:2007kva,ligo} and Advanced~Virgo~\cite{Acernese:2007zze,virgo}. One of the prime candidates for detection are GWs emitted during the inspiral of neutron star (NS) binaries. Ground-based detectors will be sensitive to GWs in the frequency range $(10,10^{3})$ Hz, which corresponds to the last $10^{4}$ orbits prior to merger. Merger itself is expected to occur at kHz frequencies, where second generation detectors will not be very sensitive. Therefore, the inspiral phase is by far the most important one for detection and parameter estimation of NS binary signals.

The inspiral phase of binary NS coalescences can be well modeled with the post-Newtonian (PN) approximation, i.e.~an expansion in powers of $v/c$, where $v$ is the orbital velocity and $c$ is the speed of light. NSs will take a long time to evolve into the sensitive band of ground-based detectors; by then, they are expected to be old, cold, slowly spinning, and with orbits that have circularized. Accurate predictions of the range of eccentricities and spins expected is currently lacking, but estimates suggest values smaller than $0.1$ in dimensionless units~\cite{Hannam:2013uu}. This is why the modeling of NS inspirals has so far been restricted to mostly quasicircular systems that are nonspinning. Limited studies of spinning systems have been performed for systems with spins that are aligned/antialigned with the orbital angular, where analytic Fourier domain waveforms are available

The first GW detections are expected to be buried in detector noise, and thus, detection and parameter estimation will require accurate waveform templates. In particular, if NSs are spinning with their spin angular momenta misaligned with the orbital angular momentum, the precession of the spins and the orbital plane will induce strong deviations from what one would expect in the spin aligned/antialigned case. This complexity can be leveraged to break parameter degeneracies present in spin aligned/antialigned waveforms to better estimate parameters~\cite{Hannam:2013uu,Lang:1900bz,PhysRevD.80.064027,Stavridis:2009ys,Klein:2010ti,Baird:2012cu}. 

What waveform templates should we then use for detection and parameter estimation to describe NS inspirals? For detection, recent studies have shown that nonspinning~\cite{Brown:2012gs} or spin aligned/antialigned templates~\cite{Poisson:1995ef,Arun:2008kb,Lang:2011je,Ajith:2011ec} are sufficient, provided the NS spin magnitudes are small enough. These same template families, however, would be inherently unable to estimate spin parameters, introducing systematic errors~\cite{Grandclement:2002dv}. Numerical waveforms, either computed from a direct integration of the PN equations~\cite{Fischetti:2010hx,Csizmadia:2012wy} or from an integration of Hamilton's equations~\cite{Buonanno:1998gg,Buonanno00,Damour:2001tu,Buonanno06}, would be ideal both for detection and parameter estimation, but they are very computationally expensive. 

Recently, there has been a concerted effort to construct accurate and computationally inexpensive template families for generically precessing, spinning, compact binary systems. These efforts are hindered by two facts: (i) no one has yet managed to analytically solve the precession equations for generic spin configurations, and (ii) the number of parameters needed to describe a generically precessing spinning system is very large. In particular, (i) has prevented a well-defined perturbative waveform solution valid for all spins orientations and magnitudes simultaneously. Keeping these limitations in mind, one can organize all template families for binary NS inspirals into the following categories:
\begin{enumerate} 
\item
{\emph{Phenomenological}}~\cite{Buonanno:2002fy,PhysRevD.69.104017,PhysRevD.70.104003,Buonanno:2005pt,VanDenBroeck:2009gd,PhysRevLett.106.241101,Ajith:2011ec, Ajith:2012mn,PhysRevD.52.605,PhysRevD.54.2421,Grandclement:2002dv,Santamaria:2010yb}: One introduces parameters to describe the waveforms emitted from spinning systems, driven by the requirements that the waveforms be as computationally efficient as possible.
\item
{\emph{Geometrical}}~\cite{PhysRevD.84.124011,Pekowsky:2013ska,Lundgren:2013jla}: One computes the waveforms in a specific frame, chosen such that the precession modulations are minimized. One then numerically calculates how this frame precesses and coordinate transforms the waveforms to this frame.
\end{enumerate}

Although an excellent first step toward the construction of accurate and computationally efficient templates, both phenomenological and geometrical templates have serious drawbacks for modeling NS inspirals. Phenomenological templates, like the BCV ones~\cite{PhysRevD.70.104003,Buonanno:2005pt}, are adept at matching some of the precessional modulations, but due to their intrinsic phenomenological nature, they are difficult to systematically extend to higher order in perturbation theory. Moreover, due to their phenomenological nature, they are not necessarily superior to using simple nonspinning SPA waveforms for detection~\cite{Grandclement:2002dv,VanDenBroeck:2009gd}. Geometrical templates are very promising, but still require a numerical solution\footnote{An exception here is Ref.~\cite{Lundgren:2013jla}, where the restriction to one spinning body allows for a fully analytical waveform.} for the precession of the frame in which waveforms are calculated. The use of a numerical solution then forces one to use discrete Fourier transforms, which become computationally prohibitive in a Bayesian analysis when the dimensionality of the template space is high.  

A third category of waveform families has been proposed: 
\begin{enumerate}
\item [3.]
{\emph{Analytical}}~\cite{Apostolatos:1994mx,PhysRevD.54.2438,Racine:2008qv,Tessmer:2009yx,Cornish:2010cd,Klein:2013qda}: One makes approximations that correspond to certain physical systems and solves the precession equations perturbatively.
\end{enumerate}
These waveforms are constructed so that they are purely analytical, and thus computationally inexpensive, but also so that they can be systematically improved by carrying out calculations at higher order in perturbation theory. Reference~\cite{Klein:2013qda} implemented such a perturbative framework for compact binaries with spins almost aligned with the orbital angular momentum, but of arbitrary magnitude.

This paper proposes a new template family that fits in the analytical class for quasicircular inspiraling compact binaries, whose spin angular momentum is arbitrarily oriented, but small relative to their mass. This family is ideal to efficiently model precessing NS/NS inspirals for detection and parameter estimation purposes. The waveforms are obtained through {\emph{multiple-scale expansions}}, which allow us to accurately solve the precession equations in the time domain, and the well-known {\emph{stationary phase approximation}} (SPA), which allows us to model the Fourier transform of the waveform. Such a procedure is a direct application of the general framework of~\cite{Klein:2013qda} but for a different physical system.

The time-domain waveforms are obtained by expanding the differential equations describing precession in the dimensionless spin parameters, which then leads to a separation of time scales that is amenable to multiple scale analysis: the differential system is reexpanded in the ratio of the precession to the radiation-reaction time scales and solved order by order, while ensuring that resonances are not introduced. One thus obtains an analytic solution to the precession equations for the orbital and spin angular momenta as a bivariate expansion in the dimensionless spin and the ratio of the precession to the radiation-reaction time scales. We work here to first order in both expansion parameters, but the formalism can easily be extended to higher order. An analytic approximation for the time evolution of the momenta then leads to an analytic time-domain waveform that accurately models the precession of the spin and orbital angular momenta.  

The resulting, analytic, time-domain waveform contains pieces of different PN and spin order (see Table~\ref{table:PNorders}). Clearly, the zeroth order in spin terms are the dominant contribution to the waveform, since we are expanding about small dimensionless spin. One can think of the nonspinning terms in the waveform as a ``background solution'' to which we find spin perturbations; thus, we will try to model this background as accurately as possible. In particular, the nonspinning orbital frequency evolution equation is modeled to $3.5$PN order with mass-ratio corrections and up to $5.5$PN order in the test particle limit\footnote{An $N$PN order term is one that scales with $(v/c)^{2N}$ relative to the leading order term, where $v$ is the characteristic orbital velocity.}. We solve this equation to $8$PN order, which artificially extends the series beyond its formal expansion order, so that any differences between our analytic and numerical frequency evolution are exclusively due to spin. We also include nonspinning PN corrections to the time-domain waveform amplitude up to $2.5$PN order.  

The first order in spin terms in the waveforms are perturbations to the nonspinning waveform background and we model them by expanding in the ratio of the precession to the radiation reaction time scale. We model the spin-orbit precession equations to $2$PN order and include first order in spin terms in the frequency evolution equation up to $4$PN order with mass-ratio corrections. Such precession effects introduce corrections in the waveform phase starting at $3$PN order due to a Thomas precession effect. 

\begin{table}
\begin{centering}
\begin{tabular}{ccccc}
\hline
\hline
\noalign{\smallskip}
 {}   &&  \multicolumn{1}{c}{Spin Precession } &&  \multicolumn{1}{c}{Radiation Reaction}  \\
 {}   &&  \multicolumn{1}{c}{Eqs.\eqref{Ldot}-\eqref{S2dot}} &&  \multicolumn{1}{c}{Eqs.\eqref{k-def},\eqref{xidot}, App.\ref{app-coeffomegadot}}  \\
\hline
\noalign{\smallskip}
NonS-full && N/A && 3.5 \\
NonS-pp && N/A && 5.5  \\
SO-full && 2 && 2.5 \\
SO-pp && 2 && 2.5 \\
SS-full && N/A && 0 \\
SS-pp && N/A && 0 \\
\noalign{\smallskip}
\hline
\hline
\end{tabular}
\end{centering}
\caption{PN order (relative to the leading-order Newtonian term) of the various ingredients used in this paper for the spin precession evolution equations and the radiation reaction evolution equations. NonS stands for nonspinning, SO for spin-orbit, SS for spin-spin and pp for point-particle. The spin precession equations are solved to linear order in spin, which is why SS-full and SS-pp are labeled as N/A.}
\label{table:PNorders}
\end{table}
\begin{table*}[htb]
\setlength{\tabcolsep}{4pt}
\begin{centering}
\begin{tabular}{ccc||cccc|cccc|c}
\hline
\hline
\noalign{\smallskip}
 {} & {}  &&  \multicolumn{3}{c}{Full GW} &&  \multicolumn{3}{c}{Restricted GW}& & \\
Quantity & Symbol  &&  Eq.~or App. &&  PN order && Eq.~or App. &&  PN order && Spin order \\
\noalign{\smallskip}
\hline
\noalign{\smallskip}
Gravitational wave & $\tilde{h}(f)$ && \eqref{full-SPA-family},\eqref{Fourier-Amplitude},\eqref{Fourier-phase} && X && \eqref{restricted-SPA-family},\eqref{non-prec-waveform},\eqref{prec-waveform} && X && X \\
\noalign{\smallskip}
\hline
\noalign{\smallskip}
Nonprecessing phase &${\Psi}^{\nonprec}_n$&& \eqref{sum-SPA-Phase},\ref{app-PhiofF} && 8 && \eqref{sum-SPA-Phase},\ref{app-PhiofF} && 8 && 8\\
Phase log correction &$\Phi^{{\mbox{\tiny log}}}_n$ && \eqref{Philog-def} && 1 && \eqref{Philog-def} && 1 && 0\\
SPA phase correction &$\delta \Psi_n$ && \eqref{SPA-Cor} && 3 && \eqref{SPA-Cor} && 3 && 2\\
Thomas phase & $\delta \phi_n$ && \eqref{deltaphi-atan},\eqref{sec-thomas} && 1 && \eqref{deltaphi-atan},\eqref{sec-thomas} && 1 && 2\\
Inclination angle &$\iota_n$ && \eqref{iota-def} && X && \eqref{iota-def} && X && 1\\
Polarization angle & $\psi_n$ && \eqref{pol-def} && X && \eqref{pol-def} && X && 1\\
\noalign{\smallskip}
\hline
\noalign{\smallskip}
PN amplitudes & ${\cal{A}}_{n,k,m}$ && \ref{app-gwamp} && 2.5 && \ref{app-gwamp} && 0 && 0\\
$2\text{nd}$ derivative of $\Phi^{\orb}_n$ &$\ddot{\Phi}^{\orb}_n$ && \eqref{ddot-Phi} && 5.5 && \eqref{ddot-Phi} && 0 && 2\\
$2\text{nd}$ derivative of $\delta \phi_n$ &$\delta\ddot{\phi}_n$ && \eqref{phiddot-lin-S} && 0 && \eqref{phiddot-lin-S} && 0 && 1\\
$2\text{nd}$ derivative of $\iota_n$ &$\ddot{\iota}_n$ && \eqref{iddot-lin-S} && 0 && \eqref{iddot-lin-S} && 0 && 1\\
$2\text{nd}$ derivative of $\psi_n$ &$\ddot{\psi}_n$ && \eqref{psiddot-lin-S} && 0 && \eqref{psiddot-lin-S} && 0 && 1\\
\noalign{\smallskip}
\hline
\noalign{\smallskip}
Angular momentum & $\bm{L}$ && \eqref{Lxfinal},\eqref{Lyfinal},\eqref{Lzfinal} && X && \eqref{Lxfinal},\eqref{Lyfinal},\eqref{Lzfinal} && X && 1\\
Precession phase &$\phi_{1,2}$ && \eqref{phii-eq} && 4 && \eqref{phii-eq} && 4 && 4\\
Stationary point & $\xi^{\SP}_n$ && \eqref{SP-xi} && 0 &&\eqref{SP-xi} && 0 && 0\\
\noalign{\smallskip}
\hline
\hline
\end{tabular}
\end{centering}
\caption{Ingredients of the full and restricted PN, frequency-domain gravitational wave response function, together with the PN and spin order. All PN orders are given relative to the leading order of each quantity. The symbol X means that the particular term is composed of ingredients of different PN order. The PN orders listed here are not necessarily complete, in the sense that terms at certain high-PN orders may be missing because they have not yet been calculated. The high powers of spin that enter ${\Psi}^{\nonprec}_n$ and $\phi_{1,2}$ come from cross terms of the spin-orbit and spin-spin couplings that appear at lower PN order, when one artificially extends the series to high PN order.}
\label{table:GWorders}
\end{table*}

The second order in spin terms are even smaller perturbations to the nonspinning background, which we will not systematically include in this paper. For example, the spin precession equations contain spin-spin interactions that we will ignore when analytically solving for the angular momenta. However, we will keep the second-order in spins terms in the frequency evolution, since they lead to better agreement between our analytical results and numerical solutions (see Sec.~\ref{firstorder}). This results in the inclusion of a 2PN second-order in the spin term in the waveform phase that is evaluated assuming the spins are time independent. 

Table~\ref{table:GWorders} summarizes the various equations that are combined in the full and the restricted PN \footnote{The restricted PN approximation keeps all PN terms in the waveform phase, but truncates the amplitude to leading PN order.}, frequency-domain waveform together with their PN and spin order. All PN orders are counted relative to the leading-order (Newtonian) term of each quantity. The frequency-domain waveform is obtained by Fourier transforming the time-domain waveform in the SPA~\cite{PhysRevD.49.2658}. Previous work has shown that the SPA might break for precessing systems~\cite{Klein:2013qda}. However, we find that for systems with small spin, the SPA remains valid with probability higher than $99.8 \%$. 

\begin{figure}[htb]
\begin{center}
\includegraphics[width=\columnwidth,clip=true]{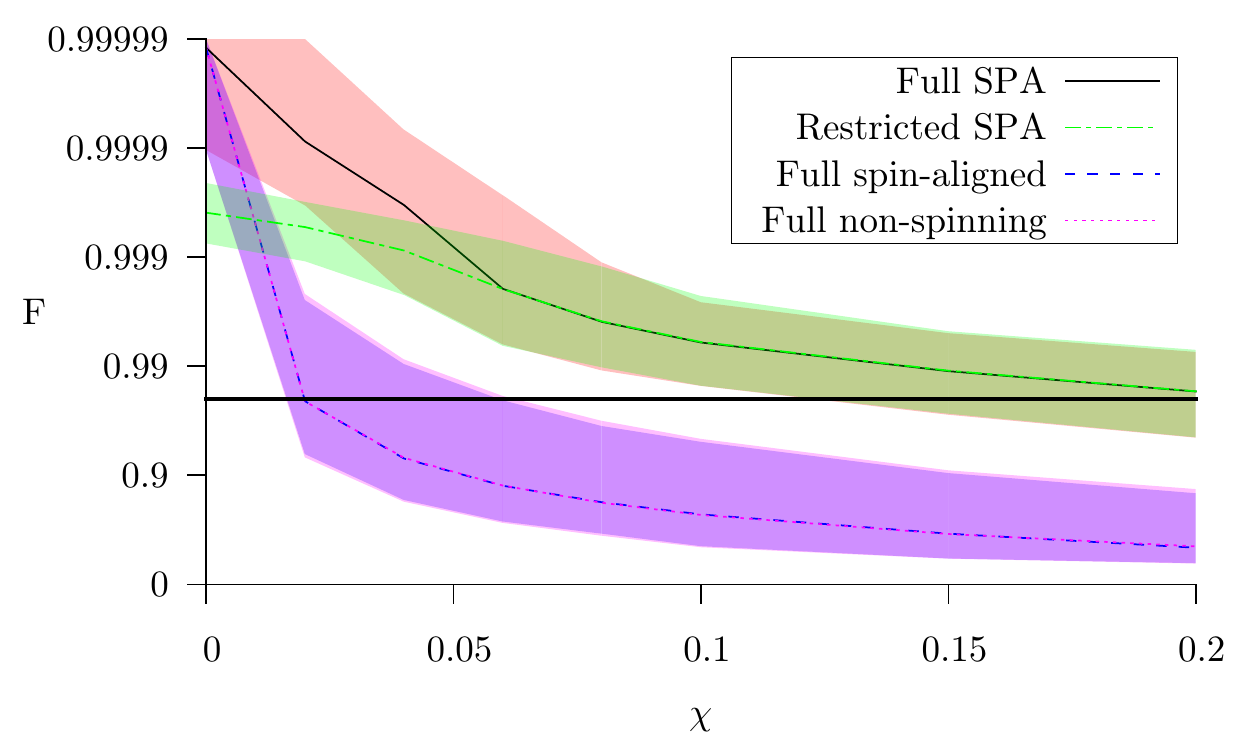}
\caption{\label{fig:F} Mean faithfulness of the full SPA family (red, solid line), the restricted PN SPA family (green dot-dashed line), the full spin-aligned SPA family (blue dashed line) and the full nonspinning SPA family (red dotted line) against numerical waveforms for 1000 different systems with randomized parameters. The shaded areas show the faithfulness regions for $68\%$ of the systems considered. For reference, the black solid line corresponds to a faithfulness of $97\%$. The overlaps presented here were not maximized over parameters, except for the time and phase of coalescence.}
\end{center}
\end{figure}

The main product of this paper is an analytic, frequency-domain template family for neutron star binary inspirals that accounts for spin precession. The restricted PN version of this family leads to faithfulnesses\footnote{The faithfulness between two waveforms is the noise-weighted cross-correlation of their Fourier transforms only maximizing over the time and phase of coalescence.} above $99\%$ relative to fully numeric templates for systems with dimensionless spin parameters below $0.2$. Figure~\ref{fig:F} shows the averaged faithfulness of the full SPA family (red, solid line), the restricted PN SPA family (green dot-dashed line), the full spin-aligned SPA family (blue, dashed line) and the full nonspinning SPA family (magenta dotted line) relative to numerical waveforms as a function of dimensionless spin parameter, for $1,000$ systems with random parameters. This figure also shows faithfulness intervals (shaded regions) for $68\%$ of the systems considered (1$\sigma$ quantiles). 

Observe that the performance of the new analytic, restricted PN family is as good as that of the full family. This is because amplitude corrections become important for systems with large dimensionless mass differences, which is never large in NS binaries. Higher harmonic effects become important for $F \gtrsim 0.999$. Observe also that both of these families perform dramatically better than spin-aligned or nonspinning SPA families for dimensionless spin parameters larger than $0.01$. This last two lead to almost indistinguishable faithfulnesses on a logarithmic scale. This figure suggests that the new template families constructed here may be sufficiently accurate for parameter estimation with advanced ground detectors. 

The remainder of this paper deals with the details of our calculation and it is organized as follows.
In Sec.~\ref{L-S} we solve for the time evolution of the spin and orbital angular momenta. 
In Sec.~\ref{gw-t} we obtain time-domain waveforms, while
in Sec.~\ref{gw-f} we obtain frequency-domain waveforms.
In Sec.~\ref{gw-comps} we compare the analytic waveforms to numerical ones by calculating their match.
In Sec.~\ref{finite-size} we discuss the finite-size corrections that arise at high frequencies for neutron star binaries.
In Sec.~\ref{conclusion} we conclude and point to future research.
The appendices present all details necessary to reproduce the results of this paper. 

Throughout this paper we use geometric units with $G=c=1$; we will retain powers of $c$ to denote the PN order of any given term. We also employ the following conventions:
\begin{itemize}
\item
Vectors are written in boldface, while their components are denoted by $\bm{A}=[A_x,A_y,A_z]$. Unit vectors are denoted with a hat, e.g. $\bm{\hat{A}}$.
\item
The masses of the two objects are $M_A$, with $A \in (1,2)$, the total mass is $M\equiv M_1+M_2$, the reduced mass is $\mu\equiv M_1M_2/M$, the symmetric mass ratio is $\eta\equiv\mu/M$ and the dimensionless mass difference is $\delta M = (M_{1} - M_{2})/M$. Capital letters are used here to distinguish masses from the $m$th harmonic number. 
\item
The orbital angular frequency in a frame fixed to the orbital plane is $\omega$, while the PN expansion parameter will here be chosen to be $\xi\equiv(M\omega)^{1/3}$.
\item
The orbital angular momentum of the system is $\bm{L}$, while the spin angular momenta of each object is $\bm{S}_A$, with $S_A$ the (constant) spin magnitude. The dimensionless spin parameter of each object is then $\chi_A\equiv S_A/M_A^2$ with $A \in \{1,2\}$. 
\item
The unit vector pointing form the GW detector to the source is $\bm{\hat{N}}$.
\item
Vectors with a subscript $d$, such as $\bm{A}_d$, are defined in the detector frame, while those with a subscript $s$, such as $\bm{A}_s$, are defined in the source frame.
\end{itemize}

\section{Evolution of the Spin and Orbital Angular Momentum}
\label{L-S}

In this section, we solve for the time evolution of the spin and orbital angular momentum as a series in the dimensionless spin parameters and the ratio of the precession to the radiation-reaction time scales. We begin by presenting the precession equations and expanding them in small spin magnitude. We continue by solving the precession equations order by order in this perturbation parameter. At linear order in spin, we solve the precession equations using multiple scale analysis. We conclude with a summary of results and a comparison between the numerical solution and the analytic approximation to the evolution of the angular momenta.

\subsection{Precession equations and the small spin expansion}

In the PN approximation, the equations that govern the evolution of the orbital angular momentum $\bm{L}$ and the spin angular momenta $\bm{S}_i$ averaged over one orbit are~\cite{springerlink:10.1007/BF00756587,Bohe:2012mr}
\begin{align}
\dot{\bm{L}}&= \frac{\omega^2}{M}\sum_{n = 0}^{2} \eta^n (M \omega)^{2n/3}  \left[C_1^{\n} \bm{S}_1 + C_2^{\n} \bm{S}_2\right] \times \bm{L} \nn 
\\
&- \frac{3}{2}\frac{\omega^2}{M}[( \bm{S}_2\cdot \bm{\hat{L}})\bm{S}_1+(\bm{S}_1 \cdot \bm{\hat{L}})\bm{S}_2]\times \bm{\hat{L}}-k \bm{L} \label{Ldot} \,,
\\
\dot{\bm{S}}_1 &= \frac{\omega^2}{M} \sum_{n = 0}^{2} C_1^{\n} \eta^n (M \omega)^{2n/3}  \bm{L} \times \bm{S}_1\nn
\\
&+ \frac{\omega^2}{M}\left[\frac{1}{2}\bm{S}_2-\frac{3}{2}(\bm{S}_2 \cdot \bm{\hat{L}})\bm{\hat{L}}\right]\times \bm{S}_1 \label{S1dot}\,,
\\
\dot{\bm{S}}_2 &= \frac{\omega^2}{M} \sum_{n = 0}^{2} C_2^{\n} \eta^n (M \omega)^{2n/3} \bm{L}\times \bm{S}_2\nn
\\
&+ \frac{\omega^2}{M}\left[\frac{1}{2}\bm{S}_1-\frac{3}{2}(\bm{S}_1 \cdot \bm{\hat{L}})\bm{\hat{L}}\right] \times\bm{S}_2 \label{S2dot}\,,
\end{align}
where we have defined
\be
k=\frac{1}{3}\frac{a_0}{M}(M\omega)^{8/3}\left\{1+\sum_{i=2}^{11} [a_i+b_i \ln{(M\omega)}] (M\omega)^{i/3}\right\}\,,\label{k-def}
\ee
\begin{align}
C_A^{\zero}&=2+\frac{3}{2}\frac{M_B}{M_A}\label{C0def}\,,
\\
C_A^{\one}&=3\frac{M_A}{M_B}+\frac{35}{6}+4\frac{M_B}{M_A}+\frac{9}{8}\frac{M_B^2}{M_A^2}\label{C1def}\,,
\\
C_A^{\two}&=\frac{27}{4}\frac{M_A^2}{M_B^2}+\frac{31}{2}\frac{M_A}{M_B}+\frac{137}{12}+\frac{19}{4}\frac{M_B}{M_A}
\nn \\
&+\frac{15}{4}\frac{M_B^2}{M_A^2}+\frac{27}{16}\frac{M_B^3}{M_A^3}\label{C2def}\,,
\end{align}
where $(A,B) \in \{1,2\}$, $A \neq B$ and $\omega$ is the orbital frequency of the binary, with coefficients $(a_{i},b_{i})$ given in Appendix~\ref{app-coeffomegadot}.

The evolution equation for $\bm{L}$ contains two types of terms that are valid to different PN orders: {\emph{conservative terms}} and {\emph{dissipative terms}}. The dissipative ones are all contained in the last term of Eq.~\eqref{Ldot}, which changes the magnitude of the orbital angular momentum, and governs the GW frequency evolution. The conservative terms [the $\bm{S}_{i}$-dependent terms in Eq.~\eqref{Ldot} and all terms in Eqs.~\eqref{S1dot}-\eqref{S2dot}] describe spin-spin and spin-orbit interactions to 3.5PN order. These terms do not change the magnitude of $\bm{L}$, but only its direction. We work to the highest PN order known (3.5PN) in the spin-orbit interactions~\cite{Bohe:2012mr}. However, since we are carrying a small spin magnitude analysis we keep here only the leading-order spin-spin interaction terms (2PN). We will later ignore this term when deriving an analytic solution for the angular momenta, but include it in the numerical solutions we will compare against, to show that indeed such a term can be neglected.

The evolution of the magnitude of the angular momentum, and thus, the frequency evolution, is controlled by $k$, which is given in Eq.~\eqref{k-def} by a sum taken to $5.5$PN order. The coefficients $(a_{i},b_{i})$ listed in Appendix~\ref{app-coeffomegadot} are known to $2.5$PN order including all spin terms, to $4$PN to linear-order in spin~\cite{Bohe:2012mr} and to $22$PN order in the test particle limit neglecting spins and black hole (BH) absorption terms~\cite{Tanaka:1997dj,Shibata:1994jx,Mino:1997bx,Fujita:2012cm}. Since the evolution of nonspinning binaries acts as a background upon which we perturbative expand in $\chi_{1,2}$, we will model the former very accurately, keeping nonspinning terms in $(a_{i},b_{i})$ to $3.5$PN order with $\eta$ corrections and to $5.5$PN order without all $\eta$ corrections. To linear-order in spin, we keep terms in $(a_{i},b_{i})$ up $4$PN order, while to quadratic-order we keep terms up to $2$PN order. We express the evolution equation (and its solution later on) in terms of $(a_{i},b_{i})$, so that higher-order PN corrections can be easily incorporated in our analysis by simply modifying these coefficients, when higher PN order terms become available in the future. 
 
For those astrophysically realistic NS binaries that are expected to be detected by advanced ground-based detectors, the dimensionless spin parameter is not expected to exceed $\chi_{A} \sim 0.1$~\cite{Hannam:2013uu}. We can, therefore, define the book-keeping parameter $\epsilon_s$ as a way to count powers of $\chi_A$. When working to leading order in $\epsilon_{s}$, we can neglect the spin-spin interactions (the last terms) in Eqs.~\eqref{S1dot} and \eqref{S2dot}, and the first term of the second line in Eq.~\eqref{Ldot}, which leads to the precession equations
\allowdisplaybreaks
\begin{align}
\dot{\bm{L}} &= \frac{\omega^2}{M}\sum_{n \ge 0}^{2} \eta^n (M \omega)^{2n/3}  \left[C_1^{\n} \bm{S}_1 + C_2^{\n} \bm{S}_2\right] \times \bm{L}-k \bm{L} \label{Ldot-noss} \,,
\\
\dot{\bm{S}}_1 &= \frac{\omega^2}{M}\sum_{n \ge 0}^{2} \eta^n (M \omega)^{2n/3}  C_1^{\n}\bm{L} \times \bm{S}_1 \label{S1dot-noss}\,,
\\
\dot{\bm{S}}_2&=\frac{\omega^2}{M}\sum_{n \ge 0}^{2} \eta^n (M \omega)^{2n/3} C_2^{\n} \bm{L}\times \bm{S}_2\label{S2dot-noss}\,. 
\end{align}
where the coefficients $C_i^{\n}$ are given in Eqs.~\eqref{C0def}-\eqref{C2def}.  In Eq.~\eqref{Ldot-noss}, we can rewrite $\dot{\bm{L}}=L\dot{\bm{\hat{L}}}+\dot{L}\bm{\hat{L}}$ to separate precession effects from radiation-reaction effects. The magnitude of the spin angular momentum is conserved due to the particular choice of variables~\cite{Bohe:2012mr}. 
 
The precession equations will be solved as a function of the independent variable $\xi$, a PN expansion parameter defined by
\be
\xi\equiv(M\omega)^{1/3} = M^{2} \eta \; L^{-1}\,,\label{xidef}
\ee
where $\omega$ is the orbital frequency and $L$ is the Newtonian expression for the magnitude of the orbital angular momentum, $L = M^{2} \eta  (M \omega)^{-1/3}$. The radiation-reaction equation for the magnitude of the orbital angular moment allows us to write an evolution equation for $\xi$ 
\be
\dot{\xi}=\frac{a_0}{3M}\xi^9\left\{1+\sum_{i=2}^{11}\left[a_i+3b_i \ln({\xi})\right]\xi^i\right\}\,,\label{xidot}
\ee
where the coefficients $(a_i,b_i)$ are the same as before. Just like velocity or angular frequency, our PN expansion parameter is time dependent, approaching $\xi \sim v \sim (M/r_{12})^{1/2} \leq [M/(2 R)]^{1/2} \sim 0.3$ by the end of the inspiral, where $R$ is the NS radius. 

Having set up the problem, the remainder of this section solves Eqs.~\eqref{Ldot-noss}-\eqref{S2dot-noss} perturbatively in $\chi_{A}$. To do so, we perturbatively expand all quantities in $\epsilon_{s}$:
\be
\bm{A} = \sum_{n=0}^{N} \epsilon_{s}^{n} \; \bm{A}^{\n}\,,
\label{formal-exp}
\ee
where $\bm{A}$ is any of $\bm{L}$, $\bm{S}_{1}$ or $\bm{S}_{2}$, while $\bm{A}^{\n}$ is a term proportional to $(\chi_{A})^{n}$. Equation~\eqref{formal-exp} is nothing but the mathematical definition of the small-spin expansion, where we will here work to ${\cal{O}}(\epsilon_{s})$, i.e.~to $N=1$. 

\subsection{${\cal{O}}(\epsilon_s^0)$ solution}
\label{zerothorder}

At this order, the NS's spin angular momenta vanish:
\begin{align}
\label{S10sol} \bm{S}^{\zero}_1&=0\,,
\qquad
\bm{S}^{\zero}_2=0\,.
\end{align}
while the orbital angular momentum evolves according to
\begin{align}
\dot{\bm{L}}^{\zero} &= -k^{\zero} \bm{L}^{\zero} \label{L0dot} \,,
\end{align}
This equation implies that the angular momentum does not change in direction, but only shrinks in magnitude due to radiation reaction. 

Let us then work in a coordinate system that is adapted to the problem at hand by choosing $\bm{\hat{z}}=\bm{\hat{J}}(0)$. Since $\bm{J}$ is evolving at higher orders in $\epsilon_s$, the $\bm{\hat{z}}$ axis of the frame will not remain aligned with $\bm{J}$ at later times. However, to ${\cal{O}}(\epsilon_s^{0})$, $\bm{\hat{J}}^{\zero}$ is not evolving, and thus, $\bm{\hat{z}}=\bm{\hat{J}}^{\zero}=\bm{\hat{L}}^{\zero}$. With this choice of coordinate system, $L_x$ and $L_y$ simply vanish to this order.  

The $\bm{\hat{z}}$ component of the orbital angular momentum $L_z^{\zero}=L^{\zero}$ satisfies the evolution equation
\begin{align}
\dot{L}^{\zero}\!\!&=\!\!-\frac{a_0}{3}\frac{M^7\mu^8}{L^{\zero7}}\left\{1\!+\!\sum_{i=2}^{11}\left[a^{\zero}_i\!+\!3b^{\zero}_i \ln\!{\left(\frac{M\mu}{L^{\zero}}\right)}\right]\!\frac{M^i\mu^i}{L^{\zero i}}\right\}\!,\label{Lz0dot}
\end{align}
where recall that $L^{\zero}$ is the magnitude of the orbital angular momentum to order ${\cal{O}}(\epsilon_s^0)$ and the coefficients $a_i^{\zero} = a_{i}(\bm{S}_A=0)$. Since all the coefficients are constant, we can directly integrate the above equation, invert the PN expansion, and obtain $L^{\zero}$ as a function of time. We rewrite Eq.~\eqref{Lz0dot} in terms of the PN parameter $\xi^{\zero}$, related to $L_{z}^{\zero}$ through Eq.~\eqref{xidef}, namely,
\begin{align}
L_z^{\zero}&=\frac{M^2 \eta}{\xi^{\zero}}\,,\label{L0sol}
\end{align}
and solve this equation to obtain
\begin{align}
\xi^{\zero}&=\zeta\left[1+\sum_{i=2}^{11}\xi_i^{\zero}\zeta^i+\sum_{i=6}^{11} \xi^{\ell,\zero}_i \zeta^i\ln{(\zeta)}\right.\nn\\
&\left.+\sum_{i=8}^{11} \xi^{\ell^{2},\zero}_i \zeta^i(\ln{\zeta})^2
+{\cal{O}}(c^{-12})\right]\,,\label{xi0sol}
\end{align}
where we have defined the function of time
\be
\zeta \equiv \left[\frac{3M}{8a_0(t_c-t)}\right]^{1/8}\,,
\ee
with $t_c$ the time of coalescence. The PN coefficients $(\xi_i^{\zero}, \xi^{\ell,\zero}_i, \xi^{\ell^{2},\zero}_i)$ can be obtained from Appendix~\ref{app-coeffxi} by setting $\bm{S}_{A} = 0$. Combining Eq.~\eqref{xi0sol} with Eq.~\eqref{L0sol} completes the solution for the time evolution of the spin and angular momenta to ${\cal{O}}(\epsilon_{s}^{0})$. 
 
As explained before, this solution keeps terms beyond $3.5$PN order, even though the evolution equation is formally only known to that order. We do so to minimize the difference between the numerical solution to the evolution of the angular momentum and the analytic approximation in Eq.~\eqref{xi0sol} in the nonspinning case. In fact, taking this series to $5.5$PN order guarantees that the frequency to time mapping is accurate to roughly $10^{-2}$ Hz during the entire inspiral (from $10$Hz up to $400$Hz, where finite size effects become important~\cite{Read:2009yp,Hinderer:2009ca,Markakis:2010mp}). Doing so will allow us to isolate any spin precession effects cleanly. Ultimately, however, we will be interested in the frequency-domain waveform, which can be constructed without knowledge of Eq.~\eqref{xi0sol}.

\subsection{${\cal{O}}(\epsilon_s^1)$ solution}
\label{firstorder}

At this order, the orbital and spin angular momenta evolve according to 
\begin{align}
\dot{\bm{L}}^{\one} &= \frac{\xi^6}{M^3}\sum_{n = 0}^{2} \eta^n \xi^{2n}  \left[C_1^{\n} \bm{S}^{\one}_1 + C_2^{\n} \bm{S}^{\one}_2\right] \times \bm{L}^{\zero}
\nonumber \\
&-k^{\zero} \bm{L}^{\one}-k^{\one} \bm{L^{\zero}} \label{L1dot} \,,
\\
\dot{\bm{S}}^{\one}_1 &= \frac{\xi^6}{M^3}\sum_{n = 0}^{2} \eta^n \xi^{2n} C_1^{\n} \bm{L}^{\zero} \times \bm{S}^{\one}_1 \label{S11dot}\,,
\\
\dot{\bm{S}}^{\one}_2 &= \frac{\xi^6}{M^3}\sum_{n = 0}^{2} \eta^n \xi^{2n} C_2^{\n} \bm{L}^{\zero} \times \bm{S}^{\one}_2 \label{S21dot}\,,
\end{align}
where $\bm{L}^{\zero}$ is given by Eqs.~\eqref{L0sol} and~\eqref{xi0sol}. The above equations are easy to decouple: we first use Eqs.~\eqref{S11dot} and \eqref{S21dot} to solve for $\bm{S}^{\one}_1$ and $\bm{S}^{\one}_2$, and we then substitute these solutions into Eq.~\eqref{L1dot} to solve for the orbital angular momentum. 

\subsubsection{Solution for \texorpdfstring{$\bm{S}_1^{\one}$}{} and \texorpdfstring{$\bm{S}_2^{\one}$}{}}
\label{Si-sol}

Without loss of generality, we focus on $\bm{S}^{\one}_1$; the solution for $\bm{S}^{\one}_2$ can be obtained by exchange symmetry, i.e.~$1\leftrightarrow2$. In term of its components, Eq.~\eqref{S11dot} can be written as
\begin{align}
\dot{S}^{\one}_{1,x}&= -\frac{\xi^6}{M^3}\sum_{n = 0}^{2} \eta^n \xi^{2n} C_1^{\n} L^{}S^{\one}_{1,y}\,,\\
\dot{S}^{\one}_{1,y}&= \frac{\xi^6}{M^3}\sum_{n = 0}^{2} \eta^n \xi^{2n} C_1^{\n} L^{}S^{\one}_{1,x}\,,\\
\dot{S}^{\one}_{1,z}&=0\,.
\end{align}
which we can rewrite as
\be
\frac{d \lambda_1^{\one}}{d \phi_1^{}}=i\lambda_1^{\one}\,,\label{lambda1dot}
\ee
where we have defined 
\be
\lambda_1^{\one}=S^{\one}_{1,x}+iS^{\one}_{1,y}\,,\label{lambda1def}
\ee
and the new independent variable
\be
\frac{d \phi_1^{}}{dt}=\frac{\xi^5 \mu}{M^2}\sum_{n = 0}^{2} \eta^n \xi^{2n} C_1^{\n}\,.\label{phi1def} 
\ee
Notice that we have not included a superscript index in $\phi_{A}$ because this will act as an independent variable, just like time and orbital frequency. 

The combined spin evolution equation can now be integrated directly. Doing so, decoupling Eq.~\eqref{lambda1def} and using exchange symmetry, we are led to the solution 
\begin{align}
S^{\one}_{1,x}&=S^{\one}_{1,x}(0)\cos{\phi_1^{}}-S^{\one}_{1,y}(0)\sin{\phi_1^{}}\,,\label{S1x1sol}\\
S^{\one}_{1,y}&=S^{\one}_{1,y}(0)\cos{\phi_1^{}}+S^{\one}_{1,x}(0)\sin{\phi_1^{}}\,,\label{S1y1sol}\\
S^{\one}_{1,z}&=S^{\one}_{1,z}(0)\,,\label{S1z1sol}\\
S^{\one}_{2,x}&=S^{\one}_{2,x}(0)\cos{\phi_2^{}}-S^{\one}_{2,y}(0)\sin{\phi_2^{}}\,,\label{S2x1sol}\\
S^{\one}_{2,y}&=S^{\one}_{2,y}(0)\cos{\phi_2^{}}+S^{\one}_{2,x}(0)\sin{\phi_2^{}}\,,\label{S2y1sol}\\
S^{\one}_{2,z}&=S^{\one}_{2,z}(0)\,,\label{S2z1sol}
\end{align}
where $\phi_{2}^{}$ is defined by Eq.~\eqref{phi1def} with $1 \leftrightarrow 2$. Observe that this is a simple harmonic oscillator with precession frequency $\dot{\phi}_{A}^{}$. 

To complete the calculation, one must solve Eq.~\eqref{phi1def} for the phase angle $\phi_{A}^{}$. Doing so as an expansion in $\xi$, we find 
\begin{align}
\frac{\phi_{A}^{}}{C_{A}^{(0)}} &=\phi_{0,A} - \frac{\eta}{a_0 \xi^3}\Big[1+\sum_{i=2}^{8}\phi_{i,A}\xi^i \nonumber\\
&+\sum_{i=3}^{8} \phi^{\ell}_{i,A} \xi^i\ln{\xi}+{\cal{O}}(c^{-9})\Big]\,,
\label{phii-eq}
\end{align}
with $A \in \{1,2\}$, where the coefficients $\phi_{i,A}$ are given in Appendix~\ref{app-coeffphi} and $\phi_{0,A}$ is a constant of integration picked such that $\phi_{A}(t=0)=0$ to satisfy Eqs.~(\ref{S1x1sol}-\ref{S2z1sol}).  One could include higher-order PN terms in this expansion. However, we find that truncating it at 4PN order is sufficient to obtain an accurate time-domain waveform phase, i.e.~higher-order PN terms induce phase corrections that are smaller than those induced by neglected terms of ${\cal{O}}(\epsilon_{s}^{2})$. 

In principle, the coefficients $(a_{i},b_{i})$ that appear in Eq.~\eqref{phii-eq} are given in Appendix~\ref{app-coeffomegadot} with all spins set to zero, since $\phi_{i,A}^{}$ should be kept to  ${\cal{O}}(\epsilon_{s}^{0})$. However, we will here use the full expressions in Appendix~\ref{app-coeffomegadot} for the coefficients $(a_{i},b_{i})$, with the spin couplings $\beta_{i}$ and $\sigma_{i}$ evaluated at the initial spin and orbital angular momenta $\bm{L}(t=0)$ and $\bm{S}_{i}(t=0)$. The solution obtained with this substitution differs with the initial ${\cal{O}}(\epsilon_{s}^{0})$ solution by terms of ${\cal{O}}(\epsilon_{s}^{2})$, and is thus equally valid. In practice, we find that using these coefficients leads to better agreement between the analytical approximation and the numerical solution of the orbital phase, as quantitatively presented in Sec.~\ref{GW-time-numerical-comp}.

\subsubsection{Solution for \texorpdfstring{$\bm{L}^{\one}$}{}}

The evolution equation [Eq.~\eqref{L1dot}] contains terms that change on two different time scales: a radiation-reaction time scale $t_{\rr}$ and a precession time scale $t_{\pr}$. The former is associated with the last two terms in Eq.~\eqref{L1dot} and it is defined by
\be
t_{\rr} \equiv \frac{\xi}{\dot{\xi}} \sim \frac{M}{\eta} \xi^{-8}\,,
\ee
while the latter is associated with the first two terms in Eq.~\eqref{L1dot} and it is defined by
\be
t_{\pr} \equiv \frac{M}{\omega^{2} L} \sim \frac{M}{\eta} \xi^{-5}\,.
\ee
The ratio of these time scales is $t_{\pr}/t_{\rr}\sim\xi^{3} = {\cal{O}}(c^{-3})$, which then suggests one should use multiple scale analysis to solve Eq.~\eqref{L1dot} (see e.g.~\cite{Klein:2013qda}). 

Let us then define a new perturbative (bookkeeping) parameter $\epsilon_{p}$ that counts powers of $(t_{\pr}/t_{\rr})$ and expand all quantities in a bivariate series
\be
\bm{A} = \sum_{n=0}^{N} \sum_{m=0}^{M} \epsilon_{s}^{n} \epsilon_{p}^{m} \; \bm{A}^{(n,m)}\,,
\label{formal-bivariate-exp}
\ee
where $\bm{A}$ is any of $\bm{L}$, $\bm{S}_{1}$ or $\bm{S}_{2}$, while $\bm{A}^{(n,m)}$ is a term proportional to $\chi_{1,2}^{n} (t_{\pr}/t_{\rr})^{m}$. Of course, $\bm{A}^{(0,m)}$ $\forall \; m$ has already been obtained in Eq.~\eqref{xi0sol}. 

In multiple scale analysis, all quantities must be assumed to depend on all independent time scales, and thus, 
\be
\bm{A}^{(n,m)} = \bm{A}^{(n,m)}(t,\tau)\,, 
\ee
where we have defined the long time scale 
\be
d \tau=\epsilon_p \left(\frac{t_{\pr}}{t_{\rr}}\right) dt\,.
\ee
The differential operator of Eq.~\eqref{L1dot} is then
\be
\frac{d}{dt}=\frac{\p}{\p t}+\epsilon_p \left(\frac{t_{\pr}}{t_{\rr}}\right)\frac{\p}{\p \tau}\,.
\ee

The solution for $\bm{L}^{\onezero}$ is more easily obtained if we work with the total angular momentum instead. Recall that the latter is defined by $\bm{J}^{\one}=\bm{L}^{\one}+\bm{S}_1^{\one}+\bm{S}_2^{\one}$ to ${\cal{O}}(\epsilon_{s})$, and satisfies the equation
\be
\dot{\bm{J}}^{\one}=-k^{\zero} \bm{L}^{\one}-k^{\one} \bm{L}^{\zero} \,.\label{J1dot}
\ee
This last equation can in turn be expanded in $\epsilon_{p}$ to obtain a bivariate series. Now we can proceed to solve Eq.~\eqref{J1dot} order by order in $\epsilon_p$.

\vspace{0.5cm}
\begin{center}
\emph{2a. Solution to ${\cal{O}}(\epsilon_{s}, \epsilon_p^0)$}
\end{center}
\vspace{0.5cm}

To zeroth order in radiation reaction we have the simple {\emph{partial}} differential equation
\be
\frac{\p \bm{J}^{\onezero}}{\p t}=0\,,
\ee
the solution to which is  
\be
\bm{J}^{\onezero}=[J^{\onezero}_x(\tau),J^{\onezero}_y(\tau),J^{\onezero}_z(\tau)]\,.\label{J10sol} 
\ee
The quantity $\bm{J}^{\onezero}(\tau)$ are functions of the long time scale, i.e. they change over the radiation-reaction time scale, but are constant on a precession time scale. The functional form of these quantities can only be determined by going to next order in $\epsilon_{p}$. 

\vspace{0.5cm}
\begin{center}
\emph{2b. Solution to ${\cal{O}}(\epsilon_{s},  \epsilon_p)$}
\end{center}
\vspace{0.5cm}

To ${\cal{O}}(\epsilon_p)$, the evolution equation for the total angular momentum is
\be
\frac{\p \bm{J}^{\oneone}}{\p t}+\frac{t_{\pr}}{t_{\rr}}\frac{\bm{J}^{\onezero}}{\p \tau}=-k^{\zeroone} \bm{L}^{\onezero}-k^{\oneone} \bm{L}^{\zerozero}\,,
\label{nasty-eq}
\ee
where $\bm{L}^{\onezero}$ is to be understood as shorthand for $\bm{J}^{\onezero} - \bm{S}_{1}^{\onezero} - \bm{S}_{2}^{\onezero}$. Equation~\eqref{nasty-eq} is a differential equation for $\bm{J}^{\oneone}(t)$, whose solution will grow linearly (a behavior that characterizes a resonance) if sourced by a $t$-independent term. One can eliminate such t-independent terms by requiring that 
\begin{align}
\frac{t_{\pr}}{t_{\rr}}\frac{d J^{\onezero}_x(\tau)} {d \tau}&=-k^{\zeroone} J^{\onezero}_x(\tau)\,,\\
\frac{t_{\pr}}{t_{\rr}}\frac{d J^{\onezero}_y(\tau)} {d \tau}&=-k^{\zeroone} J^{\onezero}_y(\tau)\,,
\end{align}
whose solution is
\begin{align}
J^{\onezero}_x&=J^{\onezero}_x(0) \exp\left[-\int k^{\zeroone} \left(\frac{t_{\rr}}{t_{\pr}}\right) d\tau\right]\,,
\nn \\
&=J^{\onezero}_x(0) \exp\left[\int \frac{\dot{L}^{\zeroone}}{L^{\zeroone}} dt \right]\,,
\nn\\
&=J^{\onezero}_x(0)\frac{L^{\zeroone}}{L^{\zeroone}(0)}
=J^{\onezero}_x(0)\frac{\xi^{\zeroone}(0)}{\xi^{\zeroone}}\,,
\end{align}
and $J^{\onezero}_y$ is obtained by replacing $x \leftrightarrow y$. Collecting all the results, after imposing the initial conditions $\bm{\hat{z}}=\bm{\hat{J}}(0)$, the solutions for $L_x^{\onezero}$ and $L_y^{\onezero}$ are
\begin{align}
L^{\onezero}_x&=-S^{\one}_{1,x}(0)\cos{\phi_1}+S^{\one}_{1,y}(0)\sin{\phi_1}\nn\\
&-S^{\one}_{2,x}(0)\cos{\phi_2}+S^{\one}_{2,y}(0)\sin{\phi_2}\,,\label{L1x0sol}\\
L^{\onezero}_y&=-S^{\one}_{1,y}(0)\cos{\phi_1}-S^{\one}_{1,x}(0)\sin{\phi_1}\nn\\
&-S^{\one}_{2,y}(0)\cos{\phi_2}-S^{\one}_{2,x}(0)\sin{\phi_2}\,.\label{L1y0sol}
\end{align}

The $z$ component of the orbital angular momentum to first order in the spins can be obtained in the following way. First, we notice that
\be
\xi \equiv\frac{M^{2}\eta}{L} = \frac{M^{2}\eta}{L_z}+{\cal{O}}(\epsilon_s^2)\,.
\ee
Therefore, to this order we have
\be
L_z=L^{\zero}_z+\epsilon_sL^{\one}_z+{\cal{O}}(\epsilon_s^2)=\frac{M^{2}\eta}{\xi}+{\cal{O}}(\epsilon_s^2)\,.
\ee

We can further improve on the solution for $\xi$ as a function of time to ${\cal{O}}(\epsilon_{s})$ by revisiting its evolution equation. Equation~\eqref{xidot} depends on the PN coefficients $(a_i, b_i)$ (see Appendix~\ref{app-coeffomegadot}), which in turn depend on the spins only through combinations $\bm{S}_A\cdot\bm{\hat{L}}$ and $\bm{S}_A\cdot\bm{S}_B$. The spin-spin terms are of ${\cal{O}}(\epsilon_s^{2})$ and can thus be neglected, while the spin-orbit terms are constant to ${\cal{O}}(\epsilon_{s})$. We can then substitute $\bm{S}_A\cdot\bm{\hat{L}}\to S_{A,z}$ in Eq.~\eqref{xidot}, treat all the PN coefficients $(a_i, b_i)$ as constant and integrate to obtain 
\begin{align}
\xi &= \xi^{\zero} + \epsilon_{s} \xi^{\one} + {\cal{O}}(\epsilon_{s}^{2}) =\zeta\left[1+\sum_{i=2}^{11} \left(\xi_i^{\zero} + \epsilon_{s} \xi_i^{\one}\right) \zeta^i
\right.\nn\\
&\left.+\sum_{i=6}^{11} \left(\xi^{\ell,\zero}_i + \epsilon_{s} \xi^{\ell,\one}_i\right) \zeta^i\ln{\zeta}
\right.\nn\\
&\left.+\sum_{i=8}^{11} \left(\xi^{\ell^{2},\zero}_i + \epsilon_{s} \xi^{\ell^{2},\one}_i\right)\zeta^i(\ln{\zeta})^2
+{\cal{O}}(\epsilon_{s}^{2},c^{-12})\right]\,.\label{xi1sol}
\end{align}
Recall that $(\xi_i^{\zero}, \xi^{\ell,\zero}_i, \xi^{\ell^{2},\zero}_i)$ are those in Appendix~\ref{app-coeffxi} with $\bm{S}_{A} = 0$, while the new coefficients $(\xi_i^{\one}, \xi^{\ell,\one}_i, \xi^{\ell^{2},\one}_i)$ are those in Appendix~\ref{app-coeffxi} that are linear in $\bm{S}_{A}$. 

However, as in Eq.~\eqref{phii-eq}, we will here include higher-order terms in $\epsilon_{s}$ to improve the mapping between frequency and time. The coefficients  $(\xi_i^{\zero}+\epsilon_{s} \xi_i^{\one}, \xi^{\ell,\zero}_i + \epsilon_{s} \xi^{\ell,\one}_i, \xi^{\ell^{2},\zero}_i + \epsilon_{s} \xi^{\ell^{2},\one}_i)$ will be replaced by those in Appendix~\ref{app-coeffxi} but with $(a_{i},b_{i})$ coefficients evaluated at the initial spin and orbital angular momenta $\bm{L}(t=0)$ and $\bm{S}_{A}(t=0)$. As before, this replacement adds terms of ${\cal{O}}(\epsilon_{s}^{2})$ and higher to the precession phase that improve the accuracy of the analytical solution, as we show in Sec.~\ref{comparison}. 

\subsection{Summary of results}
\label{summary}

The final solutions for the orbital angular momentum to first order in spin are
\begin{align}
L_x&=\epsilon_s\left\{-S_{1,x}(0)\cos{\phi_1^{}}  +S_{1,y}(0)\sin{\phi_1^{}}\right.
\nn \\
&\left.-S_{2,x}(0)\cos{\phi_2^{}}+S_{2,y}(0)\sin{\phi_2^{}}\right\}+{\cal{O}}(\epsilon_s^2,\epsilon_p)\,,\label{Lxfinal}\\
L_y&=\epsilon_s\left\{-S_{1,y}(0)\cos{\phi_1^{}}-S_{1,x}(0)\sin{\phi_1^{}}\right. 
\nn \\
&\left.-S_{2,y}(0)\cos{\phi_2^{}}-S_{2,x}(0)\sin{\phi_2^{}}\right\}+{\cal{O}}(\epsilon_s^2,\epsilon_p)\,,\label{Lyfinal}\\
L_z&=\frac{M\mu}{\xi}+{\cal{O}}(\epsilon_s^2)\,,\label{Lzfinal}
\end{align}
while for the spin angular momenta we find
\allowdisplaybreaks
\begin{align}
S_{1,x}&=\epsilon_s\left[S_{1,x}(0)\cos{\phi_1^{}}-S_{1,y}(0)\sin{\phi_1^{}}\right]+{\cal{O}}(\epsilon_s^2)\,,\label{S1xfinal}\\
S_{1,y}&=\epsilon_s\left[S_{1,y}(0)\cos{\phi_1^{}}+S_{1,x}(0)\sin{\phi_1^{}}\right]+{\cal{O}}(\epsilon_s^2)\,,\label{S1yfinal}\\
S_{1,z}&=\epsilon_s S_{1,z}(0)+{\cal{O}}(\epsilon_s^2)\,,\label{S1zfinal}\\
S_{2,x}&=\epsilon_s\left[S_{2,x}(0)\cos{\phi_2^{}}-S_{2,y}(0)\sin{\phi_2^{}}\right]+{\cal{O}}(\epsilon_s^2)\,,\label{S2xfinal}\\
S_{2,y}&=\epsilon_s\left[S_{2,y}(0)\cos{\phi_2^{}}+S_{2,x}(0)\sin{\phi_2^{}}\right]+{\cal{O}}(\epsilon_s^2)\,,\label{S2yfinal}\\
S_{2,z}&=\epsilon_sS_{2,z}(0)+{\cal{O}}(\epsilon_s^2)\,,\label{S2zfinal}
\end{align}
The phase angles $\phi_{A}^{}$ are given explicitly in Eq.~\eqref{phii-eq} in terms of $\xi$, which is given explicitly as a function of time in Eq.~\eqref{xi1sol}. Both of these equations depend on the coefficients $(a_{i},b_{i})$ which are given in Appendix~\ref{app-coeffomegadot}, with $[\bm{L}(t),\bm{S}_{A}(t)] \to [\bm{L}(t=0),\bm{S}_{A}(t=0)]$.

\subsection{Numerical comparison}
\label{comparison}

We can now show that the bivariate, analytic solution found above is indeed an accurate representation of the full numerical solution. By the latter, we mean the numerical solution to Eqs.~\eqref{Ldot}-\eqref{S2dot}, with $k$ given by Eq.~\eqref{k-def}. Notice that these equations contain spin-spin interactions, i.e.~terms of ${\cal{O}}(\epsilon_{s}^{2})$ that are neglected in the analytic solution of the previous sections. We use an adaptive Cash-Karp, fifth-order Runge-Kutta method to solve these equations~\cite{Cash:1990:VOR:79505.79507}. We have performed convergence tests to guarantee that the numerical error introduced by the integrator is well controlled and not visible in any of the figures we show in this paper. 

For the comparisons to follow, we choose a particular system with the following properties:
\begin{itemize}
\item[] \hspace{-0.75cm}\vspace{-0.1cm} {\bf{Test System:}}
\item $M_{1} = 1.4 M_{\odot}$ and $M_{2} = 1.6 M_{\odot}$, which then implies $\eta = 0.2489$, $\delta M = -0.067$, $M = 3 M_{\odot}$;
\item $\chi_{1} = 0.1$ and $\chi_{2} = 0.1$, which then implies $S_{1} \approx 0.196 M_{\odot}^{2}$, $S_{2} \approx 0.256 M_{\odot}^{2}$;
\item $\bm{\hat{S}}_{A}(0) = \left( \cos{\phi_{S_{A}}}\sin{\theta_{S_{A}}},\sin{\phi_{S_{A}}}\sin{\theta_{S_{A}}},\cos{\theta_{S_{A}}}\right)$, where $(\phi_{S_{1}},\theta_{S_{1}}) = (\pi/4, 17 \pi/24)$ and $(\phi_{S_{2}},\theta_{S_{2}}) = (\pi/3, -\pi/6)$.
\item The source is located at polar angles $(\theta_{\dd},\phi_{\dd}) = (\pi/3,2 \pi/3)$ in the detector frame.
\end{itemize}
Notice also that $L$ ranges from about $29 M_{\odot}^{2}$ at $10$~Hz to $8.5 M_{\odot}^{2}$ at $400$~Hz, which thus implies $S_{A}/L \ll 1$ during the entire inspiral. This system experiences approximately $55$ precession cycles from $10$Hz to $400$~Hz, during which it accumulates $\sim 14,500$ cycles of GW phase. We choose the integration constants in the analytic solution such that the quantities compared agree at $10$~Hz. Although we choose a particular system for the figures to come, the results are representative of all systems we investigated.

\begin{figure}[th]
\begin{center}
\includegraphics[width=\columnwidth,clip=true]{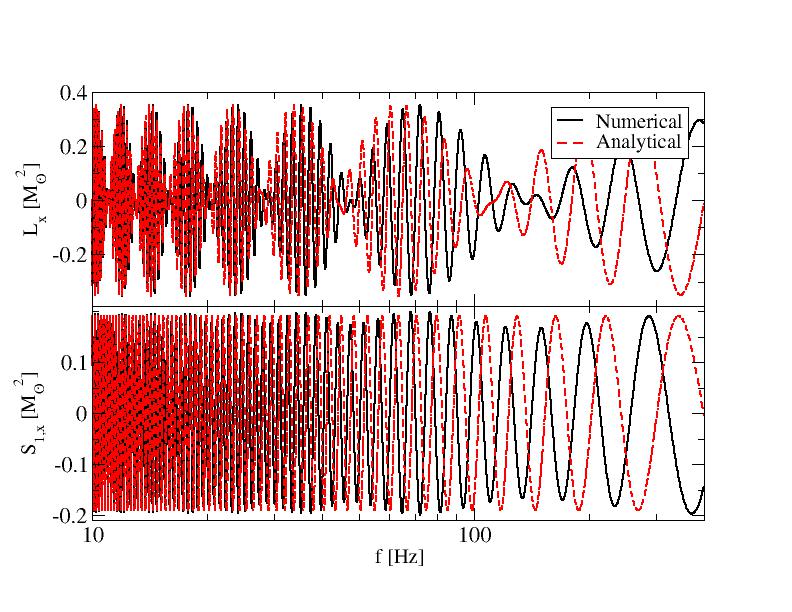}
\caption{\label{fig:Lx}Numerical solution (solid black curve) and analytical approximation (red dashed curve) to $L_{x}$ (top) and $S_{1,x}$ (bottom) as a function of the GW frequency $f$ (twice the orbital frequency). Observe that the analytical solution tracks closely the numerical solution, losing coherence as the frequency increases. This analytical approximation is dramatically better than setting these components at zero, as one would do for spin aligned/antialigned binaries.}
\end{center}
\end{figure}

Figure~\ref{fig:Lx} presents the numerical (black solid curve) and analytical (red dashed curve) approximation to the x component of orbital (top) and spin angular momentum of NS 1 (bottom) as a function of the dominant GW frequency (twice the orbital frequency) in units of Hz.  The $y$ components present similar behavior. Observe that the analytical result tracks the numerical solution closely, becoming out of phase by the end of the inspiral. This analytic solution is dramatically better than that which assumes these components simply vanish, as is done when one neglects precession and uses a spin aligned/antialigned approximation. The analytical approximation can, of course, be improved if taken to next order in $\epsilon_{s}$. 

The dephasing seen here can be significantly reduced by perturbing the system parameters, such as the individual masses and spin magnitudes slightly, as one does when calculating fitting factors. We do not show the minimized dephasing here, because we are interested in calculating faithfulnesses, instead of fitting factors.

\begin{figure}[th]
\begin{center}
\includegraphics[width=\columnwidth,clip=true]{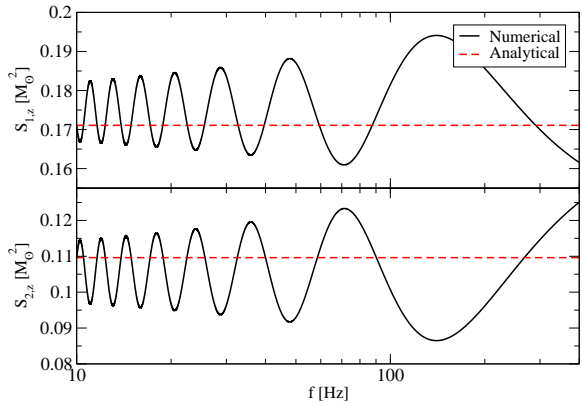}
\caption{\label{fig:Sz} Numerical solution (solid black curve) and analytical approximation (red dashed curve) to $S_{1,z}$ (top) and $S_{2,z}$ (bottom) as a function of the GW frequency. Observe that the numerical and analytical solutions start with the same initial condition, but the former oscillates with an amplitude of ${\cal{O}}(\epsilon_{s}^{2})$, roughly $10^{-2}$ for this system.}
\end{center}
\end{figure}

Figure~\ref{fig:Sz} shows the numerical (black solid curve) and analytical (red dashed curve) approximation to the $z$ component of the spin angular momentum for NS 1 (top) and NS 2 (bottom) as a function of the GW frequency in Hz. Both solutions start at the same initial value, but the analytical approximation remains constant, while the numerical one oscillates. The amplitude of this oscillation is of ${\cal{O}}(\epsilon_{s}^{2})$, which for this system is of ${\cal{O}}(10^{-2})$; the solutions can only be improved by going to next order in spin.

\begin{figure}[th]
\begin{center}
\includegraphics[width=\columnwidth,clip=true]{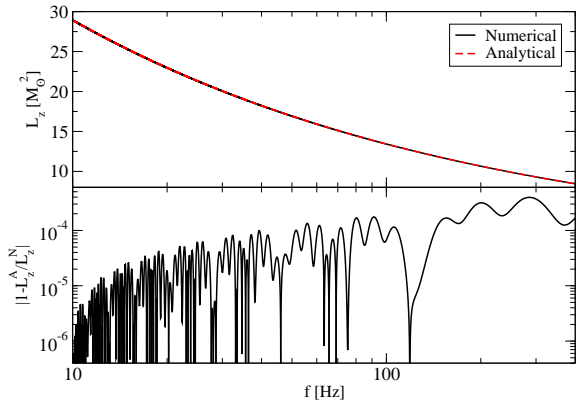}
\caption{\label{fig:Lz} Top: numerical solution (solid black curve) and analytical approximation (red dashed curve) to $L_{z}$ (top) as a function of the GW frequency. Bottom: fractional difference of numerical and analytical $L_{z}$ as a function of GW frequency. Observe that the fractional error is always less than $4 \times 10^{-4} $ during the entire inspiral.}
\end{center}
\end{figure}

The top panel of Fig.~\ref{fig:Lz} presents the numerical (black solid curve) and analytical (red dashed curve) approximation to the $z$ component of the orbital angular momentum as a function of GW frequency. Observe that indeed the analytical approximation is so good that it cannot be distinguished from the numerical result. This is why the bottom panel of Fig.~\ref{fig:Lz} shows the absolute value of the fractional difference between the numerical and the analytical result, again as a function of GW frequency. The fractional error never exceeds one part in $10^{3}$.   

\begin{figure}[th]
\begin{center}
\includegraphics[width=\columnwidth,clip=true]{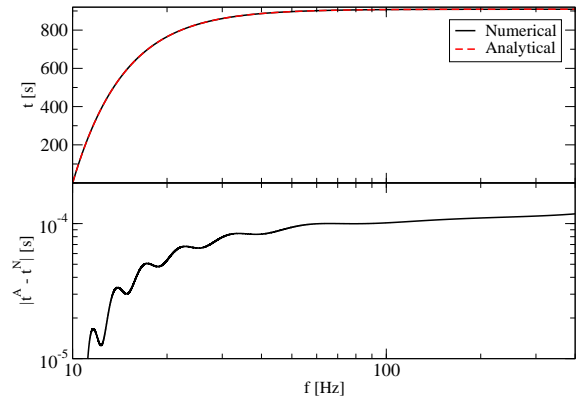}
\caption{\label{fig:t-of-f} Top: numerical solution (solid black curve) and analytical approximation (red dashed curve) to $t(f)$ with time in seconds and frequency in Hz. Bottom: absolute value of the difference between the numerical and analytical times as a function of GW frequency in seconds. Observe that the analytical solution tracks the elapsed time to better than $10^{-4} $ during the entire inspiral.}
\end{center}
\end{figure}

The construction of time-domain waveforms also requires a mapping between time and frequency. The top panel of Fig.~\ref{fig:t-of-f} shows the numerical (solid black curve) and analytical approximation (red dashed curve) to the evolution of time as a function of GW frequency in Hz; the bottom panel of this figure shows their  difference. Observe again that the analytical result tracks the numerical one to a precision better than a few times $10^{-4}$.   

\section{Time-Domain Waveform}
\label{gw-t}

Given the approximate analytical solution to the orbital angular momentum derived in the previous section, we can now construct an analytical approximation to the time-domain GW response function. We begin by defining the basic ingredients that go into the waveform construction. We then derive analytic approximations to the waveform phase, and mode-decompose the response function. We conclude with a comparison between the analytic approximation to the time-domain waveform phase and the phase derived from a numerical solution. 

\subsection{Basics}

Let $\bm{\hat{N}}=[N_x,N_y,N_z]$ be a unit vector that points in the direction of the center of mass of the compact binary relative to the detector. The time-domain response function in the long wavelength approximation can then be written as a sum of harmonics~\cite{Blanchet:2006zz}
\begin{align}
h(t)&=\sum_{n\geq0}(F_+h_{n,+}+F_{\times}h_{n,\times})\,,\label{ht-def}\\
h_{n,+}&={\cal{A}}_{n,+}(\iota)\cos{n\Phi}+{\cal{B}}_{n,+}(\iota)\sin{n \Phi}\,,\\
h_{n,\times}&={\cal{A}}_{n,\times}(\iota)\cos{n \Phi}+{\cal{B}}_{n,\times}(\iota)\sin{n \Phi}\,,
\end{align}
where $n \in \mathbb{N}$ is the harmonic number, the inclination angle is
\be
\iota\equiv \arccos\left[ \bm{\hat{L}}\cdot \bm{\hat{N}}\right]\,,
\label{iota-def}
\ee
the antenna pattern functions are given by
\begin{align}
F_+&=\frac{1}{2}(1+\cos^2{\theta})\cos{2\phi}\cos{2\psi}-\cos{\theta}\sin{2\phi}\sin{2\psi}\,,\label{Fplus-def}\\
F_{\times}&=\frac{1}{2}(1+\cos^2{\theta})\cos{2\phi}\sin{2\psi}+\cos{\theta}\sin{2\phi}\cos{2\psi}\,,\label{Fcross-def}
\end{align}
and $\psi$ is the polarization angle:
\be
\psi=\tan^{-1}{\left(\frac{\bm{\hat{L}}\cdot\bm{\hat{z}}-(\bm{\hat{L}}\cdot\bm{\hat{N}})(\bm{\hat{z}}\cdot\bm{\hat{N}})}{\bm{\hat{N}}\cdot(\bm{\hat{L}}\times\bm{\hat{z}})}\right)}\,.
\label{pol-def}
\ee

The precession of the orbital angular momentum, i.e.~of the orbital plane, has two main effects on the waveform~\cite{Apostolatos:1994mx}: 
\begin{enumerate}
\item[(i)] the inclination and the polarization angles become time dependent and; 
\item[(ii)] the reference frame used to define the orbital frequency in the orbital plane, $\bm{\hat{L}}\times\bm{\hat{N}}$, is no longer constant in time. 
\end{enumerate}
These effects will induce corrections to the waveform phase, as well as amplitude modulations. We investigate these effects below. 

\subsection{Waveform phase}
\label{gw-phase}

The waveform phase can be decomposed as follows:
\be
\Phi=\Phi^\orb+\delta\phi\,,\label{Phidef}
\ee
where $\Phi^\orb$ is the orbital phase and $\delta \phi$ is a precession correction, induced by the changing reference frame. We will refer to the latter as the Thomas precession phase. 

The orbital phase can be computed directly from
\begin{align}
\Phi^\orb &=\int \omega \; dt=\int \frac{\xi^3}{M} \frac{d\xi}{\dot{\xi}}=\phi_c-\frac{3}{5a_0 \xi^5}\left[1+\sum_{i=2}^{16}\Phi^{\orb}_i\xi^i\right.\nn\\
&\left.+\!\sum_{i=5}^{16}\Phi^{{\orb},\ell}_i\xi^n\ln{\xi}+\!\sum_{i=12}^{16}\Phi^{{\orb},\ell^{2}}_i\xi^i(\ln{\xi})^2\!+\!{\cal{O}}(c^{-17})\right]\,,\label{phic}
\end{align}
where $\phi_c$ is a constant of integration (the so-called {\emph{phase of coalescence}}, corresponding to the value of the phase when the frequency diverges) and the PN coefficients $(\Phi^{\orb}_i, \Phi^{{\orb},\ell}_{i}, \Phi^{{\orb},\ell^{2}}_i)$ to 8PN order are given in Appendix~\ref{app-coeffPhic}. As discussed in Sec.~\ref{L-S}, we extend the series here to 8PN order, so that when spins are zero, the error between this analytical phase and the numerical solution is negligibly small. Doing so will allow us to isolate any dephasings induced by spin. 

The Thomas precession phase $\delta \phi$ satisfies the differential equation~\cite{Apostolatos:1994mx}
\be
\delta \dot{\phi}=\frac{1}{L}\frac{\bm{L}\cdot\bm{\hat{N}}}{\bm{L}^2-\left(\bm{L}\cdot\bm{\hat{N}}\right)^2}\left(\bm{L}\times\bm{\hat{N}}\right)\cdot \dot{\bm{L}}\,.
\label{deltaphidot-eq}
\ee
Reference~\cite{Klein:2013qda} found a uniform asymptotic expansion to the solution of this equation to ${\cal{O}}(\epsilon_{s})$, which works both when $\hat{N}_{x}^{2} + \hat{N}_{y}^{2} = {\cal{O}}(\epsilon_{s})$ and when $\hat{N}_{x} = {\cal{O}}(\epsilon_{s}) = \hat{N}_{y}$, namely
\be
\delta \phi^{(1)} = -\hat{N}_{z} \arctan\left[ \frac{\hat{N}_{x} L_{z} - L_{x}}{\hat{N}_{y} L_{z} - L_{y}} \right] + {\cal{O}}(\epsilon_{s}^2)\,,
\label{deltaphi-atan}
\ee
where recall that $\bm{N}$ is constant but $\bm{L}$ varies on a precession time scale as given by Eqs.~\eqref{Lxfinal}-\eqref{Lzfinal}. 

We here improve on this solution by including the ${\cal{O}}(\epsilon_{s}^2)$ secular growth of $\delta \phi$. To do so, we first expand Eq.~\eqref{deltaphidot-eq} to ${\cal{O}}(\epsilon_{s}^{2})$. The ${\cal{O}}(\epsilon_{s}^{2})$ term depends on the $\bm{L}$ and $\bm{S}_{A}$ solutions found in Eqs.~\eqref{Lxfinal}-\eqref{S2zfinal}, and after expanding it in $S_{A,x}/L_{z} \ll 1 \gg S_{A,y}/L_{z}$ and averaging over a precession cycle, we find
\begin{align}
\left< \delta\dot{\phi}^{(2)} \right> &= \frac{1}{4 \pi^{2}} \int _{0}^{2 \pi}\int_{0}^{2 \pi} \delta\dot{\phi}^{(2)} d\phi_{1} d\phi_{2}\,,
\nn \\
&= \frac{1}{2} \frac{\dot{\phi}_{1}}{L_{z}^{2}} S_{1,\perp}^{2}(0) + 1 \leftrightarrow 2\,,
\label{phi2dot}
\end{align}
where we have defined $S_{1,\perp}^{2}(0) = S_{1,x}(0)^{2} + S_{1,y}(0)^{2}$. We can conveniently rewrite this as 
\begin{align}
\left<\frac{d}{d\xi} \delta {\phi}^{(2)} \right> &= \sum_{n=0}^2 \frac{\eta^n \xi^{2n} C_{1}^{\n}}{2 M^{5} \eta} S_{1,\perp}^{2}(0) \frac{\xi^{7}}{\dot{\xi}} + 1 \leftrightarrow 2\,,
\end{align}
Solving this differential equation to 1PN order, we obtain the ${\cal{O}}(\epsilon_{s}^{2})$ secular correction
\begin{align}
\left<\delta \phi^{(2)}\right> &= - \frac{5}{64}  S_{1,\perp}^{2}(0) \frac{1}{M^{4} \eta^{2}} \frac{1}{\xi} 
\nn \\
&\times \left[C_1^{{\mbox{\tiny (0)}}} + \left(a_{2} C_1^{{\mbox{\tiny (0)}}} - \eta C_1^{{\mbox{\tiny (1)}}}  \right)\xi^{2}\right] + 1 \!\leftrightarrow \!2 \!+\! {\cal{O}}(\epsilon_{s}^{3})\,.
\label{sec-thomas}
\end{align}
The constant of integration can be absorbed in the constant $\phi_c$ introduced in Eq.~\eqref{phic}. We empirically find that it is sufficient to keep terms up to 1PN order in this secular approximation, relative to numerical solutions.  

The final expression for the Thomas phase $\delta \phi$ is then the sum of Eq.~\eqref{deltaphi-atan} and Eq.~\eqref{sec-thomas}:
\be
\delta \phi = \delta \phi^{(1)} + \epsilon_{s}^{2} \left<\delta \phi^{(2)}\right> + {\cal{O}}(\epsilon_{s}^{3})\,.
\label{full-Thomas}
\ee
This expression, of course, is missing the nonsecular evolution of $\delta \phi^{(2)}$, but this cannot be computed without knowledge of the evolution of the angular moment to ${\cal{O}}(\epsilon_{s}^{2})$. We will see later that even without these nonsecular terms, Eq.~\eqref{full-Thomas} is an excellent approximation to the numerical Thomas phase. 

In order to calculate the GW signal that will be measured by a ground-based detector on Earth, we need to work in a frame attached to the arms of the detector. We choose the $\bm{\hat{z}}_d$ axis to be perpendicular to the detector plane and the $\bm{\hat{x}}_d$ and $\bm{\hat{y}}_d$ axes to be aligned with the detector's arms. The subscript $d$ denotes the detector frame, while the subscript $s$ denotes the source frame (see e.g.~Fig.~1 in~\cite{Apostolatos:1994mx}). In this frame, the position of the binary in the sky is given by $\bm{\hat{N}}_d=[\sin{\theta_{N}}\cos{\phi_{d}},\sin{\theta_{N}}\sin{\phi_{d}},\cos{\theta_{N}}]$. In order to transform vectors from the source frame to the detector frame, we assume that the binary is oriented in such a way that its total angular momentum at $t=0$ in the detector frame is given by $\bm{\hat{J}}_d(0)\equiv[\sin{\theta_0}\cos{\phi_0},\sin{\theta_0}\sin{\phi_0},\cos{\theta_0}]$. Then, the rotation matrix relating the frames is 
\be
\mathbb{R}_{d\to s} = \left[ {\begin{array}{ccc}
 \cos{\theta_0}\cos{\phi_0} &\cos{\theta_0}\sin{\phi_0} & -\sin{\theta_0} \\
 -\sin{\phi_0} & \cos{\phi_0} & 0 \\
 \sin{\theta_0}\cos{\phi_0} & \sin{\theta_0}\sin{\phi_0} & \cos{\theta_0}  
\end{array} } \right]
\ee  
We apply this matrix to rotate $\bm{N}_{d}$ into $\bm{N}_{s}$ and $\bm{\hat{z}}_{d}$ into $\bm{\hat{z}}_{s}$ when computing the polarization angle and the Thomas precession angle.

\subsection{Mode decomposition}

The analytical approximations to the Fourier transform that we will employ require that we cast the time-domain response function in the following form
\be
h(t) = {\cal{A}}(t)e^{i \Phi_{\GW}(t)}\,,
\ee
where ${\cal{A}}(t)$ is a slowly varying amplitude and $\Phi_{\GW}(t)$ is a rapidly varying phase. Therefore, we must express all the terms that vary in the orbital or the precessional time scales in terms of exponentials. This includes $\Phi^\orb$, $\delta \phi$, $\iota$, and $\psi$ because 
\be
\dot{\Phi}^\orb\sim{\cal{O}}(c^{-3}),\quad \delta\dot{\phi}\sim \dot{\psi} \sim \dot{\iota}\sim{\cal{O}}(c^{-6})\label{dotphases-order}\,,
\ee
and 
\be
\ddot{\Phi}^\orb\sim{\cal{O}}(c^{-{11}}),\quad \delta\ddot{\phi}\sim \ddot{\psi} \sim \ddot{\iota}\sim{\cal{O}}(c^{-{11}})\,.
\ee
We then leave in the amplitude only terms that vary on the radiation-reaction time scale.

Expressing the $\Phi$ dependence of $h_n(t)$ in Eq.~\eqref{ht-def} and the $\psi$ dependence of antenna pattern functions in Eqs.~\eqref{Fplus-def} and \eqref{Fcross-def} as exponentials, we find~\cite{Klein:2013qda}
\begin{align}
h_{n,+}&=\frac{1}{2}\left({\cal{A}}_{n,+}-i{\cal{B}}_{n,+}\right)e^{in\Phi}+ {\rm{c.c.}}\,,\\
h_{n,\times}&=\frac{1}{2}\left({\cal{A}}_{n,\times}-i{\cal{B}}_{n,\times}\right)e^{in\Phi}+ {\rm{c.c.}}\,,\\
F_{+}&=\frac{1}{2}\left({\cal{A}}_F+i{\cal{B}}_F\right)e^{2i\psi}+{\rm{c.c.}}\,,\\
F_{\times}&=\frac{1}{2}\left({\cal{B}}_F-i{\cal{A}}_F\right)e^{2i\psi}+{\rm{c.c.}}\,,
\end{align}
where ${\rm{c.c.}}$ stands for complex conjugate. The amplitudes
\begin{align}
{\cal{A}}_F&\equiv\frac{1}{2}(1+\cos^2{\theta})\cos{2\phi}\,,\\
{\cal{B}}_F&\equiv\cos{\theta}\sin{2\phi}\,,
\end{align}
depend only on the slowly varying sky-location angles $(\theta,\phi)$. The amplitudes $({\cal{A}}_{n,+}, {\cal{A}}_{n,\times}, {\cal{B}}_{n,+},{\cal{B}}_{n,\times})$ depend on the rapidly varying $\iota$ which we also express in terms of complex exponentials.  

Combining all these results and expanding all terms in a Fourier series we obtain~\cite{Klein:2013qda} 
\be
h(t)=\frac{\mu \xi^2}{D_L}\sum_{n\geq0} \sum_{k \in \mathbb{Z}} \sum_{m=\pm2}h_{n,k,m}(t)\,,
\label{full-h-of-t}
\ee
where $D_L$ is the luminosity distance and
\be
h_{n,k,m}(t)={\cal{A}}_{n,k,m}(\theta,\phi)e^{i \Phi^{\GW}_{nkm}(t)}+{\rm{c.c.}} \label{hnkm}\,,
\ee
where we have defined
\be
\Phi^{\GW}_{nkm}(t) \equiv \!n \Phi^{\orb}(t) + \!n \delta \phi(t)+\! n \Phi^{{\mbox{\tiny log}}}(t) +\!k \iota(t) + m\psi(t),
\ee
and
\be
\Phi^{{\mbox{\tiny log}}}(t) \equiv \left(6 - 3 \eta \xi^{2}\right) \xi^{3} \ln{\xi}\label{Philog-def}.
\ee
This last term arises when converting certain log-dependent amplitude terms into phase terms~\cite{Arun:2004ff} and from now on it will be included in $\Phi=\Phi^\orb+\delta\phi+\Phi^{{\mbox{\tiny log}}}$. The slowly varying amplitudes are given in Appendix E of \cite{Klein:2013qda} to 2PN order, while Appendix~\ref{app-gwamp} of this paper presents the 2.5PN contribution.

\subsection{Numerical comparison}
\label{GW-time-numerical-comp}

\begin{figure}[t]
\begin{center}
\includegraphics[width=\columnwidth,clip=true]{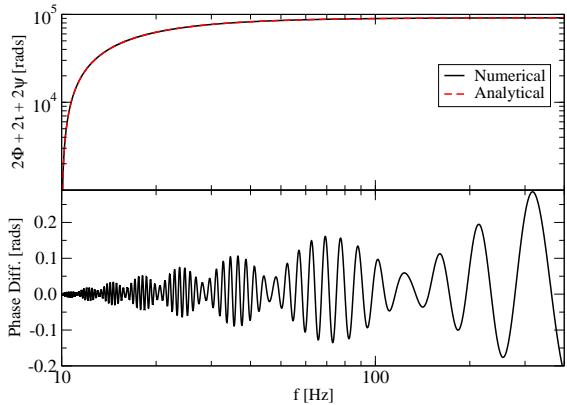}
\caption{\label{fig:TotPhase-Amp} Dominant harmonic of the total waveform phase (top) and phase difference (bottom) in radians as a function of GW frequency in Hz, computed numerically (black solid curve) and analytically (red dashed curve). Observe that the dephasing never exceeds $0.3$ radians in over $9 \times 10^{4}$ radians of inspiral evolution.}
\end{center}
\end{figure}

\begin{figure*}[htb]
\begin{center}
\includegraphics[width=\columnwidth,clip=true]{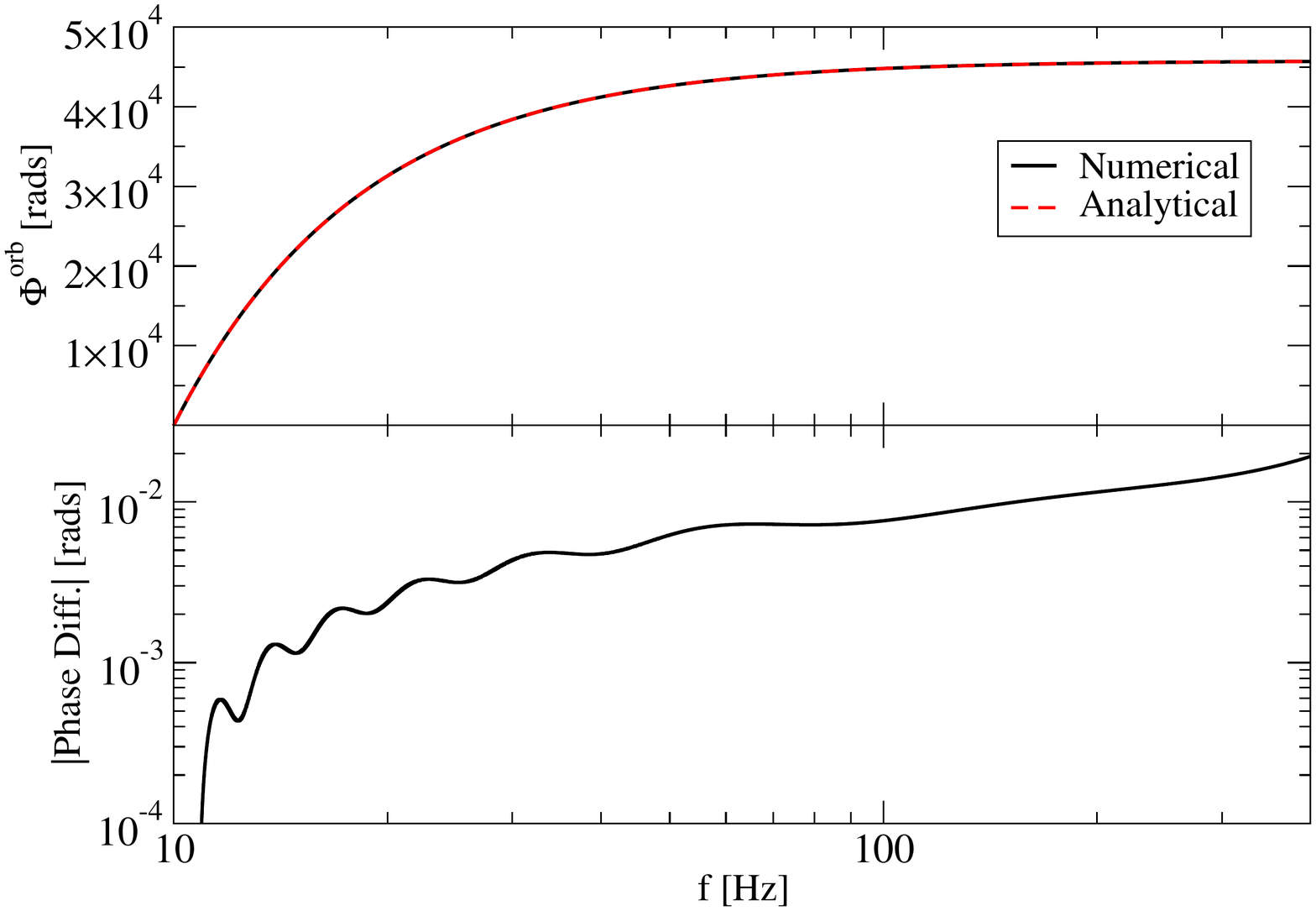} \quad
\includegraphics[width=\columnwidth,clip=true]{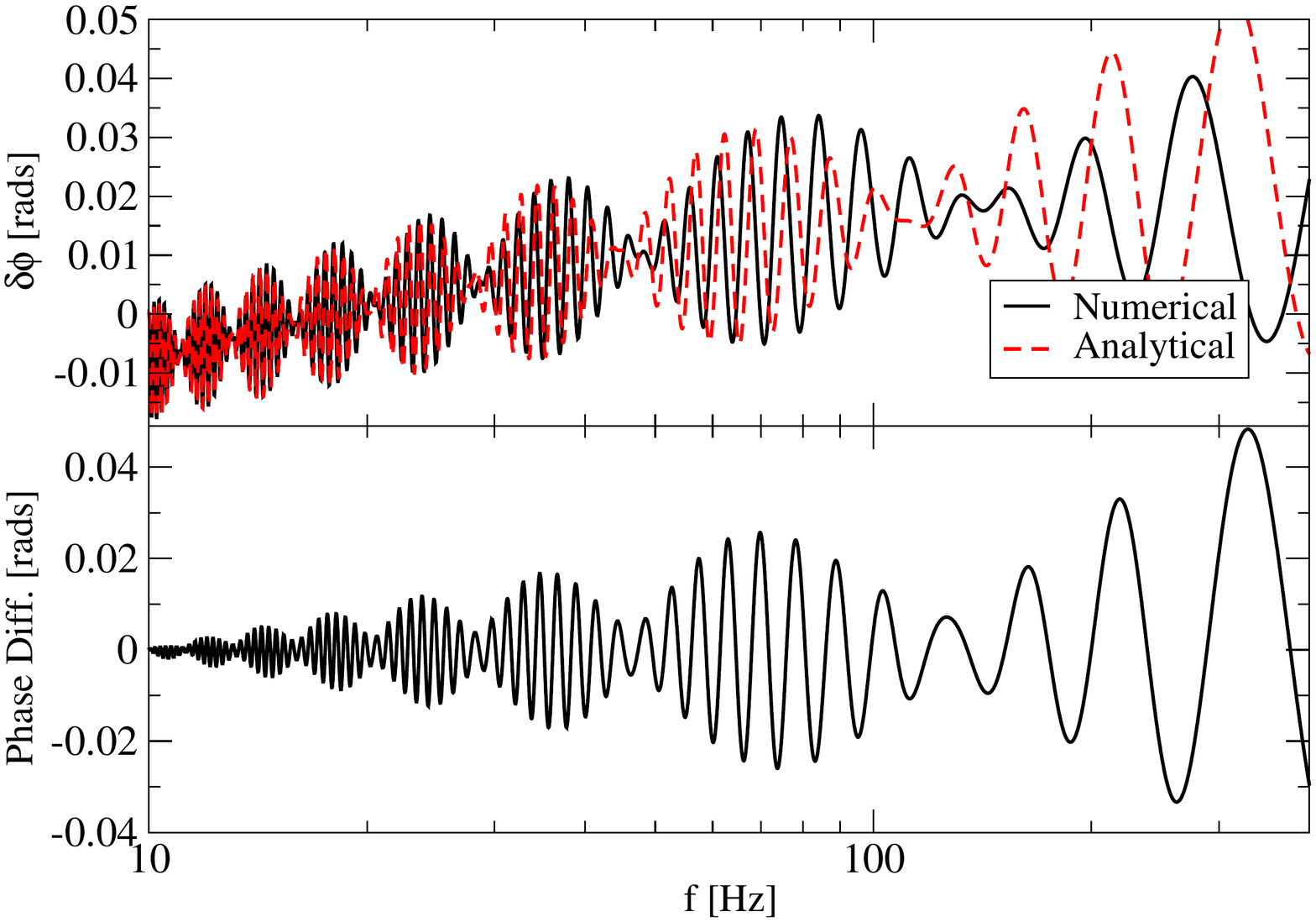} \\
\includegraphics[width=\columnwidth,clip=true]{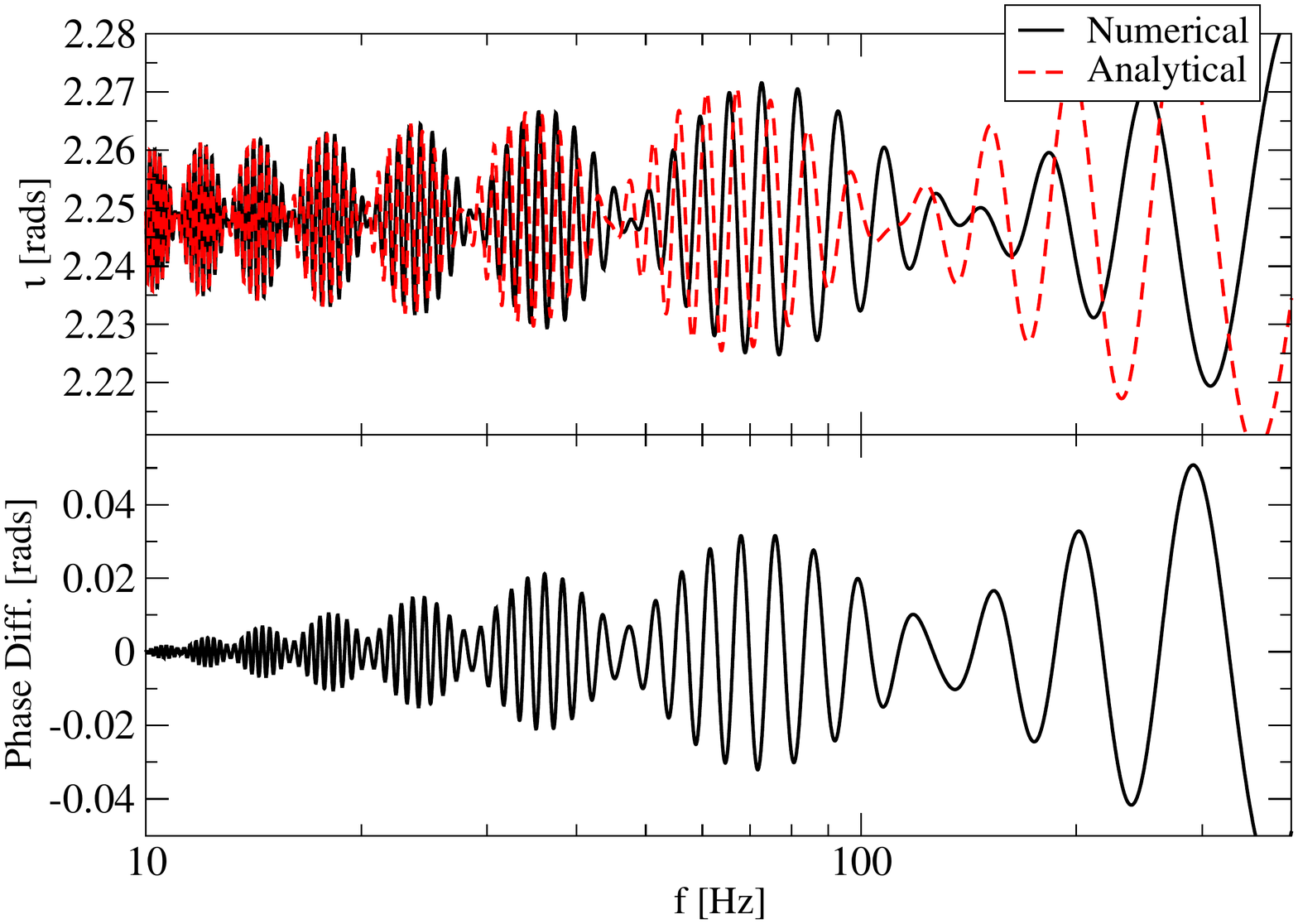} \quad
\includegraphics[width=\columnwidth,clip=true]{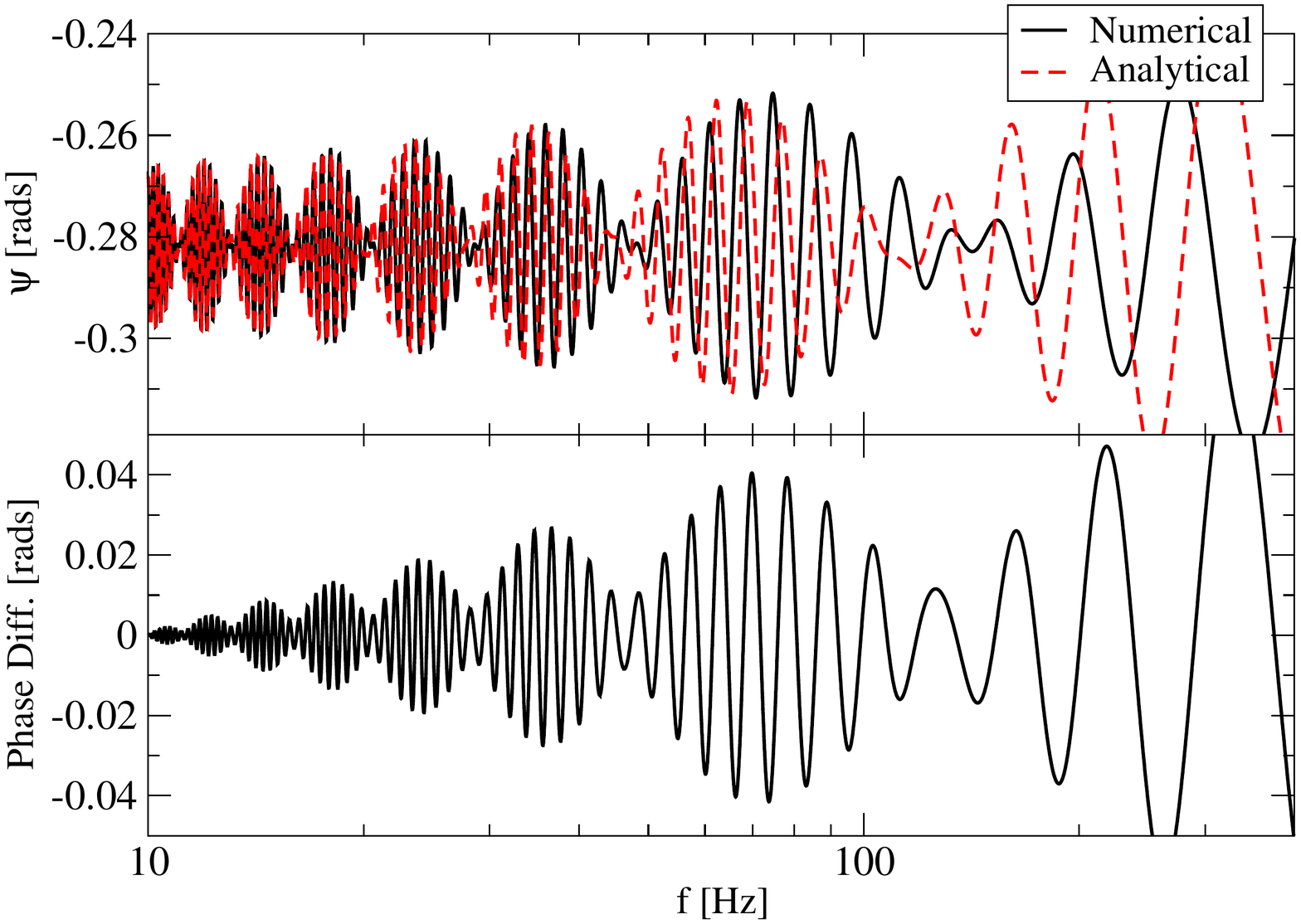} 
\caption{\label{fig:other-phases} Different pieces of the dominant harmonic of the total waveform phase (top panels) in radians and phase difference (bottom panels) in radians and as a function of GW frequency in HZ, computed numerically (black solid curve) and analytically (red dashed curve). These pieces include the orbital phase (top left), the Thomas phase (top right), the inclination angle (bottom left) and the polarization phase (bottom right). Observe that the dephasing is always of the order of $10^{-2}$ radians. }
\end{center}
\end{figure*}

Let us now compare the analytical approximate waveform response of Eq.~\eqref{full-h-of-t} to a numerical one. The latter is computed by numerically solving Eqs.~\eqref{Ldot}-\eqref{S2dot} for the momenta, with $k$ given by Eq.~\eqref{k-def}, and then inserting these into the response function of Eq.~\eqref{ht-def}. The numerical solutions are obtained with the same numerical algorithms discussed in Sec.~\ref{comparison}. Moreover, for the  comparisons to follow, we will choose the same test system as in that section. Additionally, we choose $(\theta_{N},\phi_{N})=(\pi/3,2\pi/3)$ for the polar angles of the line-of-sight unit vector in the detector frame, and $(\theta_{0},\phi_{0})=(2\pi/3,-2\pi/3)$ for the total angular momentum at $t=0$ in the detector frame. Finally, we align the waveform time offset so that the phases agree at $f_{\GW}=10$~Hz. As explained then, this system is representative of a variety of other systems studied; we postpone a more detailed population study to Sec.~\ref{gw-comps}.

Figure~\ref{fig:TotPhase-Amp} shows the dominant ($n = 2$) harmonic of the total waveform phase as a function of the GW frequency in Hz. The bottom panel shows the phase difference in radians as a function of the GW frequency. Observe that the analytical approximation tracks the numerical solution to better than $0.3$ radians over $9 \times 10^{4}$ radians of evolution in the phase. The error is a combination of a secular drift of the mean, superimposed with an oscillation; we shall see below that all of this error is induced by different ${\cal{O}}(\epsilon_{s}^{2})$ effects, which could be recovered if our calculation was carried out to next order.  

Figure~\ref{fig:other-phases} shows different pieces of the dominant ($n=2$) harmonic of the waveform phase as a function of frequency (top panels), together with the dephasing between analytical and numerical expressions. The top-left plot shows the orbital phase in radians, which is by far the dominant contribution to the total phase. Observe that the analytical approximation tracks the numerical result to about $2 \times 10^{-2}$ radians. The oscillatory features of the phase in Fig.~\ref{fig:TotPhase-Amp} are due to the Thomas phase, the inclination angle and the polarization angle, shown in the top-right, bottom-left and bottom-right plots respectively. Observe that in all cases the analytical approximation tracks the numerical solution accurately, with dephasing of order a few times $10^{-2}$. The Thomas phase, in addition, presents a secular drift, which we can see is accurately matched by the analytical approximation due to the correction in Eq.~\eqref{sec-thomas}. We have verified that these errors scale with spin squared, as expected from the fact that the analytic approximation does not consistently account for all ${\cal{O}}(\epsilon_{s}^{2})$ effects.

\section{Frequency-Domain Waveforms}
\label{gw-f}

In this section, we construct an analytical approximation to the Fourier transform of the approximate time-domain response function found in the previous section. We begin by defining the basic tools needed and then we apply them to the approximate time-domain response function of Eq.~\eqref{full-h-of-t}. We conclude with a comparison of this analytical frequency-domain waveform to the discrete Fourier transform (DFT) of the numerical time-domain waveform used in the previous section. 

\subsection{Basics}

Having expressed the time-domain waveform in the desired form, i.e. as a product of a slowly varying amplitude and a rapidly varying phase, we can now Fourier-transform it. The technique that we are going to use is the stationary phase approximation~\cite{Bender}, where one approximates the Fourier transform
\be
\tilde{h}(f)=\int h(t) e^{2\pi i f t} dt\,
\label{FT-def}
\ee
by taking into account only the part of the integrand where the integral accumulates the most. 

Let us rewrite the Fourier transform of Eq.~\eqref{FT-def} as a sum of harmonics
\be
\tilde{h}(f)=\frac{\mu \xi^2}{D_L}\sum_{n\geq0} \sum_{k \in \mathbb{Z}} \sum_{m=\pm2}\tilde{h}_{n,k,m}(f)\,,
\ee
where
\begin{align}
\tilde{h}_{n,k,m}(f)&=\int {\cal{A}}_{n,k,m}e^{i(2\pi f t +n\Phi+k\iota+m\psi)}dt\nn\\
&+\int {\cal{A}}^*_{n,k,m}e^{i(2\pi f t -n\Phi-k\iota-m\psi)}dt\,.\label{hnkmf}
\end{align}
These integrals are dominated by the regions where the phase is stationary, i.e. where the argument of the exponential is nearly constant. Otherwise, the integrand oscillates rapidly and the integral cancels out by the Riemann-Lebesgue Lemma~\cite{Bender}. Given this and the symmetry properties of Fourier transforms of real signals, the first term of Eq.~\eqref{hnkmf} is subdominant for positive frequencies and can be neglected. 

The SPA replaces the argument in the exponential of Eq.~\eqref{hnkmf} by a Taylor expansion about the stationary point $t_{\SP}$ defined by
\be
2\pi f= n\dot{\Phi}(t_{\SP})+k\dot{\iota}(t_{\SP})+m\dot{\psi}(t_{\SP})\,.
\label{SPA-condition-f-t}
\ee
This approximation works provided the amplitude varies much more slowly than the phase:
\begin{align}
\left\lvert\frac{\dot{{\cal{A}}}_{n,k,m}}{{\cal{A}}_{n,k,m}}\right\rvert&\ll\left\lvert n\dot{\Phi}+k\dot{\iota}+m\dot{\psi}\right\rvert\,.
\label{SPA-condition}
\end{align}

In the SPA, one must invert Eq.~\eqref{SPA-condition-f-t} to obtain a relation for the orbital frequency, or equivalently the PN expansion parameter $\xi$, as a function of the Fourier frequency $f$. When precession is present, an exact inversion is not possible because Eq.~\eqref{SPA-condition-f-t} is transcendental. One can, however, take Eq.~\eqref{dotphases-order} into account and approximate the inversion by setting 
\begin{align}
\xi_{n}^{\SP} = \left(\frac{2 \pi M f}{n} \right)^{1/3} \equiv v_{n}\,, 
\label{SP-xi}
\end{align}
which is an excellent approximation to the location of the stationary point. We have investigated perturbative corrections to Eq.~\eqref{SP-xi} due to precession effects and found that these have a very small effect on the Fourier domain waveform. In fact, this effect is much smaller than other errors already contained in the time-domain waveform. 

\subsection{Waveform families}
\subsubsection{Full SPA}

With all of this at hand, the full SPA to the Fourier transform of Eq.~\eqref{FT-def} is
\be
\tilde{h}_{\FSP}(f) = \frac{\eta M \xi^2}{D_L}\sum_{n\geq0} \sum_{k \in \mathbb{Z}} \sum_{m=\pm2} {{\cal{A}}}^{\GW}_{n,k,m}(f) \; e^{i \; {\Psi}_{nkm}(f)}\,.
\label{full-SPA-family}
\ee
The decomposed Fourier amplitude is
\begin{align}
{{\cal{A}}}^{\GW}_{n,k,m}(f) &= \frac{\sqrt{{2\pi}} \; {\cal{A}}^*_{n,k,m}}{\left\lvert n\ddot{\Phi}^\orb_{n}+n\delta \ddot{\phi}_{n}+n\ddot{\Phi}^{\mbox{\tiny log}}_{n}+k\ddot{\iota}_{n}+m\ddot{\psi}_{n}\right\rvert^{1/2}}\,,
\label{Fourier-Amplitude}
\end{align}
where ${\cal{A}}^*_{n,k,m}(\xi_n)$, $\ddot{\Phi}^{\orb}_{n} = \ddot{\Phi}^{\orb}(\xi_{n})$, $\delta\ddot{\phi}_{n} =\delta\ddot{\phi}(\xi_{n})$, $\ddot{\Phi}^{\mbox{\tiny log}}_{n} = \ddot{\Phi}^{\mbox{\tiny log}}(\xi_{n})$, $\ddot{\iota}_{n} = \ddot{\iota}(\xi_n)$ and $\ddot{\psi}_{n} = \ddot{\psi}(\xi_n)$ are to be evaluated at the stationary point of Eq.~\eqref{SP-xi}. These second time derivatives are presented explicitly in Appendix~\ref{app-ddotphases}. 

The decomposed Fourier phase is 
\be
{\Psi}_{nkm}(f) ={\Psi}^{\nonprec}_n-n\delta\phi_n-k\iota_n-m\psi_n\,,
\label{Fourier-phase}
\ee
where the nonprecessing Fourier phase is given by
\begin{align}
{\Psi}^{\nonprec}_n(f) &\equiv 2 \pi f t_n - n \Phi^{\orb}_n - n \Phi^{{\mbox{\tiny log}}}_n+ \delta \Psi_n - \frac{\pi}{4}\,.
\end{align}
The time-frequency function $t_n$ is given in Appendix \ref{app-tofxi}, the orbital phase $\Phi^{\orb}_n$ in Eq.~\eqref{phic}, and the log-phase term $\Phi^{{\mbox{\tiny log}}}_n$ in Eq.~\eqref{Philog-def} all as a function of $\xi$. The Thomas phase $\delta \phi$ is given in Eqs.~\eqref{deltaphi-atan},~\eqref{sec-thomas} and~\eqref{phii-eq}, while the inclination and the polarization angles are given by Eqs.~\eqref{iota-def} and~\eqref{pol-def} respectively, all as a function of $\bm{\hat{L}}$. The angular momentum is given in Eqs.~\eqref{Lxfinal}-\eqref{Lzfinal} in terms of $\phi_{1,2}$, which in turn is given in Eq.~\eqref{phii-eq} as a function of $\xi$. All of these expressions must be evaluated at the stationary point $\xi = v_{n}$ of Eq.~\eqref{SP-xi}, which is why we included an $n$ subindex.

The nonprecessing Fourier phase can be simplified to
\begin{align}
{\Psi}^{\nonprec}_{n}(f) &= 2 \pi f t_{c} - n\phi_{c} - n \Phi^{{\mbox{\tiny log}}}_{n} + \delta \Psi_n-\! \frac{\pi}{4}\! 
+ \!\frac{3n}{256 \eta} v_n^{-5} 
\nn \\
&\times \sum_{i=0}^{16} 
\left[{\Psi}_{i} + {\Psi}_{i}^{\ell}  \ln{v_n} + {\Psi}_{i}^{\ell^{2}} (\ln{v_n})^{2} \right] v_n^{i}\nn\\
&+{\cal{O}}(c^{-17})\,,
\label{sum-SPA-Phase} 
\end{align}
where $t_{c}$ is the time of coalescence, $\phi_{c}$ is the phase of coalescence, and the coefficients $({\Psi}_{i},{\Psi}_{i}^{\ell},{\Psi}_{i}^{\ell^{2}})$ are given in Appendix~\ref{app-PhiofF} for convenience. As before, recall that formally, the sum in Eq.~\eqref{sum-SPA-Phase} is consistent only up to $3.5$PN order, but we here artificially keep terms up to $8$PN order, so that any dephasings found when comparing to numerical waveforms are induced purely by spin-precession effects.

The SPA phase correction $\delta \Psi_{n}$ is the first subleading modification to the SPA condition, i.e.~to Eq.~\eqref{SPA-condition}. This term arises by retaining the third time derivative in the Taylor expansion of the argument of the exponential in Eq.~\eqref{hnkmf}. Reference~\cite{Droz:1999qx} calculated this correction to leading PN order for the $n=2$ harmonic in the nonspinning case. We have here extended this to 3PN order beyond leading for arbitrary harmonic number:
\begin{align}
\delta \Psi_{n} &= \frac{184}{45 \; n}  \; \eta \; v_{n}^{5} \left[1 + \frac{61}{46} a_{2} \; v_{n}^2  
+ \frac{89}{46} a_{3} \; v_{n}^3 \right.\nn 
\\
&\left. + \left( \frac{123}{46} a_{4} - \frac{131}{184} a_{2}^{2} \right) v_{n}^{4} 
+ \!\left(\! -\frac{175}{92} a_{3} a_{2} + \frac{163}{46} a_{5}\! \right) v_{n}^{5}\right.\nn 
\\
&\left.+ \left( \frac{147}{46} b_{6} + \frac{627}{46} b_{6} \ln{v_{n}} + \frac{233}{368} a_{2}^{3} - \frac{225}{92} a_{4} a_{2} \right. \right. \nn 
\\
&\left. \left.-\frac{227}{184} a_{3}^{2} + \frac{209}{46} a_{6} \right) v_{n}^{6} \right]\,.  
 \label{SPA-Cor}
\end{align}
This phase correction improves the agreement between our SPA waveforms and numerical ones for nonspinning systems by as much as an order of magnitude at low frequencies. One could include spin corrections to this equation that, for example, arise from the secular correction to the Thomas phase in Eq.~\eqref{sec-thomas}. We have empirically found, however, that these effects are much smaller than ${\cal{O}}(\epsilon_{s}^{2})$ errors already contained in the time-domain waveform approximant.  

The nonspinning ingredients that go into the SPA phase in Eq.~\eqref{Fourier-phase} are artificially of higher PN order than what one is allowed to formally keep. For example, $t_{n} = t(\xi_{n})$, $\Phi^{\orb}_{n} = \Phi^{\orb}(\xi_{n})$ and ${\Psi}^{\nonprec}_n = {\Psi}^{\nonprec}(\xi_{n})$ are given to 8PN order in the appendices. As already discussed, whenever possible we keep all such nonspinning ingredients to 8PN order so as to minimize the dephasing for nonspinning inspirals between our SPA waveform and the DFT of the time-domain numerical waveform described of Sec.~\ref{GW-time-numerical-comp}. Indeed, we find that keeping terms up to this order reduces the dephasing to ${\cal{O}}(10^{-5})$ radians in the nonspinning case. Therefore, any dephasings we show next for spinning systems are exclusively due to spin-precession effects and not due to any disagreement between the SPA and numerical waveforms for the backgroung nonspinning motion. 

\subsubsection{Restricted PN SPA}

Now that we have an analytical understanding of the full SPA waveform, we can apply some further approximations to simplify the waveform family. A typical approximation is the {\emph{restricted PN}} one, where one retains only the leading-order PN terms in the Fourier amplitude, but includes all known PN order terms in the Fourier phase. We calculate this waveform here. 

The restricted PN approximation amounts to retaining only the $n=2$ term in the sum of Eq.~\eqref{full-SPA-family}, namely
\begin{align}
\tilde{h}_{\RSP}(f) &= \sqrt{\frac{\pi}{\left\vert\ddot{\Phi}^{\orb}\right\rvert}} 
\frac{\eta M \xi^2}{D_L} 
e^{i (2\pi f t - 2 \Phi^\orb - 2 \Phi^{{\mbox{\tiny log}}} + \delta \Psi^{\SP} - \frac{\pi}{4})}
\nn \\
&\sum_{k = 0}^{2} \sum_{m=\pm2}  {\cal{A}}^*_{2,k,m} 
\left\lvert 1 + \frac{ \delta\ddot{\phi}}{\ddot{\Phi}^\orb} + \frac{k}{2}\frac{\ddot{\iota}}{\ddot{\Phi}^\orb}+\frac{m}{2}\frac{\ddot{\psi}}{\ddot{\Phi}^\orb}\right\rvert^{-\frac{1}{2}}
\nn \\
& e^{- i(2\delta\phi + k\iota + m\psi)}\,.
\label{RPN1}
\end{align}
where  $\ddot{\Phi}^{\orb}$ is to be substituded with its leading-order PN value $\ddot{\Phi}^{\orb} = 96 \eta \xi^{11}/(5 M^{2})$, $\xi$ is to be evaluated at $\xi_{2}$ as given in Eq.~\eqref{SP-xi}, and we have eliminated the subindex $n$, since all quantities here are to be evaluated at $n=2$. To leading PN order the amplitudes ${\cal{A}}^*_{2,k,m}$ are given by Eqs. (E14)-(E16) of~\cite{Klein:2013qda} with $\xi=0$. Notice that, even though one can safely ignore $\ddot{\Phi}^{\mbox{\tiny log}}\sim c^{-19}$, one {\emph{cannot}} expand the square root in the second line of Eq.~\eqref{RPN1} because $\ddot{\Phi}^{\orb}$ is of the same PN order as $\delta\ddot{\phi}$, $\ddot{\iota}$ and $\ddot{\psi}$.  

\begin{figure*}[ht]
\begin{center}
\includegraphics[width=\columnwidth,clip=true]{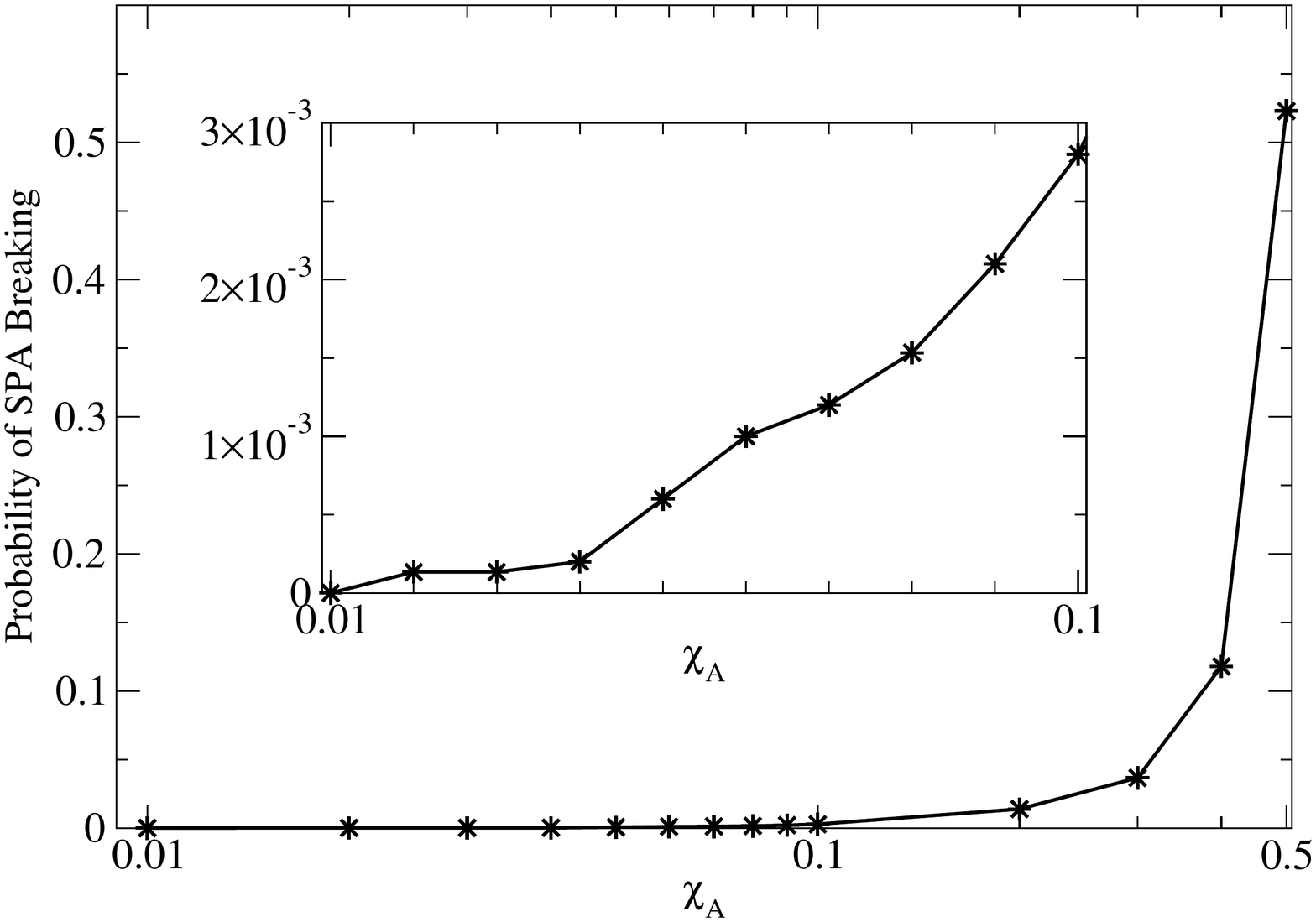}
\includegraphics[width=\columnwidth,clip=true]{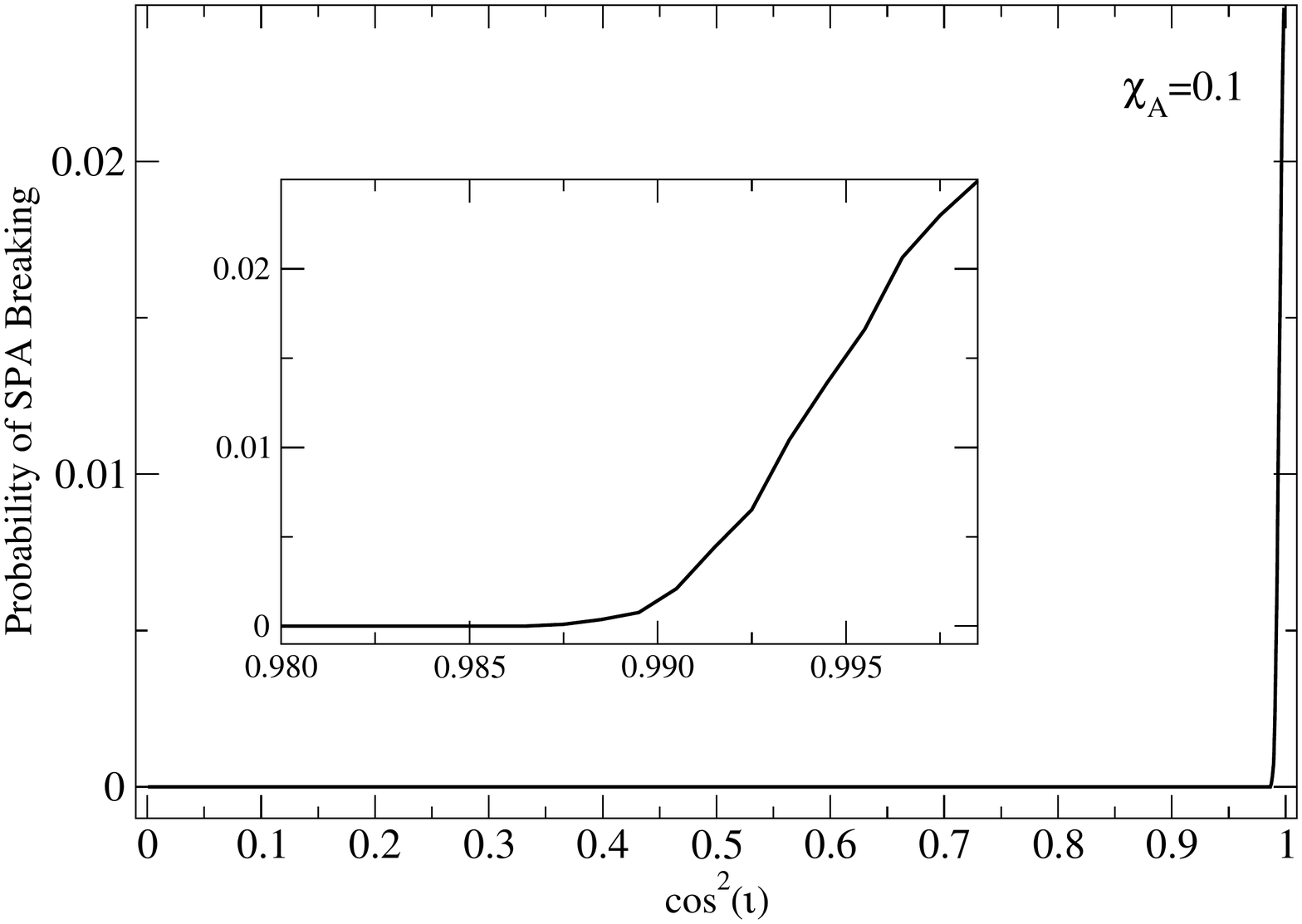}
\caption{\label{fig:Phiddot} Probability that the SPA fails as a function of the dimensionless spin parameter of the system (left) and as a function of the squared of the cosine of the inclination angle (right). Observe that the probability of the SPA breaking down is extremely small for the systems considered here. Observe also that this probability increases for systems with $\bm{\hat{N}}$ and $\bm{\hat{L}}$ aligned or antialigned.}
\end{center}
\end{figure*}

We can rewrite the above in the more suggestive form
\begin{align}
\tilde{h}_{\RSP}(f) &= {h}_{\nonprec}(f) \; {h}_{\precc}(f) \,,
\label{restricted-SPA-family}
\end{align}
where we have defined the nonprecession contribution
\be
{h}_{\nonprec} =  \sqrt{\frac{5}{96}}  \pi^{-2/3}  \frac{\eta^{1/2} M^{5/6}}{D_L}  f^{-7/6}
e^{i {\Psi}^{\nonprec}_2(f)}\label{non-prec-waveform}\,,
\ee
and the precession correction
\begin{align}
{h}_{\precc}(f) &= e^{-2 i \delta\phi}
\sum_{k = 0}^{2} \sum_{m=\pm2}  {\cal{A}}^*_{2,k,m}  e^{- i(k\iota + m\psi)}
\nn \\
&\left\lvert 1 + \frac{5 M^{2}}{96 \eta} \xi^{-11} \left(\delta\ddot{ \phi} + \frac{k}{2} \ddot{\iota} +\frac{m}{2} \ddot{\psi} \right)\right\rvert^{-1/2}\label{prec-waveform},
\end{align}
where the second time derivatives of the phases are given in Eqs.~\eqref{phiddot-lin-S}-\eqref{psiddot-lin-S}, we have used that $\ddot{\Phi}^{\orb} = 96 \eta \xi^{11}/(5 M^{2})$ to leading PN order and ignored $\ddot{\Phi}^{\mbox{\tiny log}}$ again, since it is subdominant. The Thomas phase, the inclination angle and the polarization angle can be found in Eqs.~\eqref{deltaphi-atan},~\eqref{sec-thomas},~\eqref{phii-eq},~\eqref{iota-def} and~\eqref{pol-def}. The amplitudes ${\cal{A}}_{n,k,m}$ can be found in Eqs. (E14)-(E16) of~\cite{Klein:2013qda} with $\xi=0$. The nonprecessing phase is given in Eq.~\eqref{sum-SPA-Phase} with $n=2$ evaluated at $v_2=(\pi M f)^{1/3}$ with the coefficients $({\Psi}_{i},{\Psi}_{i}^{\ell},{\Psi}_{i}^{\ell^{2}})$ given in Appendix~\ref{app-PhiofF}.

\subsection{Applicability of the SPA}

Both the restricted and full SPA waveform families defined above rely on the assumptions of the SPA being valid. In particular, these solutions require that the first nonvanishing time derivative of the argument of the exponential in Eq.~\eqref{hnkmf} be the second time derivative. If this is not the case, then the denominator in Eqs.~\eqref{Fourier-Amplitude} or~\eqref{RPN1} would vanish and the SPA amplitude would diverge. When this is the case, a more sophisticated approximation to the generalized Fourier integral is required, such as the method of steepest descent and uniform asymptotics~\cite{Bender,Klein:2013qda}. 

Whether the second time derivative of the phase vanishes or not depends sensitively on the system considered. The parameters that affect this the most for small spin systems are the angles associated with $\bm{\hat{N}}$ and those associated with $\bm{\hat{L}}$. Indeed, as we can see in Appendix~\ref{app-ddotphases}, $\ddot{\Phi}^{\orb}$, $\delta\ddot{\phi}$, $\ddot{\iota}$, and $\ddot{\psi}$ are all of the same PN order. However, $\ddot{\Phi}^{\orb}$ is of $\mathcal{O} (\epsilon_s^0)$, whereas the others are of $\mathcal{O} (\epsilon_s)$. Furthermore, $\ddot{\Phi}^{\orb}$ is always positive whereas the others oscillate around zero. Thus, for the second time derivative of the phase to vanish, a factor of $\mathcal{O} (\epsilon_s^{-1})$ has to multiply $\delta\ddot{\phi}$, $\ddot{\iota}$, or $\ddot{\psi}$ in order for these phases to be comparable to $\ddot{\Phi}^{\orb}$. By looking at their expressions, we can see that this is the case when the angle between $\bm{\hat{N}}$ and $\bm{\hat{L}}$ is of $\mathcal{O} (\epsilon_s^{-1})$.

To verify this scaling, we carried out a set of Monte Carlo studies by randomizing over all parameters, except for the dimensionless spin parameters that we kept equal for both NSs and fixed for each run. The left panel of Fig.~\ref{fig:Phiddot} shows the probability that the SPA will break down as a function of the spin parameter. Observe that for systems with $\chi_{A} < 0.1$, this probability is smaller than $0.2 \%$. This is because the SPA breaks only when large amounts of precession are present. In particular, the SPA breaks when $\bm{\hat{N}}$ and $\bm{\hat{L}}$ are coaligned, so that the system wobbles the most as seen from the detector. This can be seen on the right panel of Fig.~\ref{fig:Phiddot}, which shows the probability the SPA will break as a function of the square of the cosine of the inclination angle, where we have fixed both $\chi_{A}$ and $\iota$. Observe that for systems with the angle between $\bm{\hat{N}}$ and $\bm{\hat{L}}$ larger than $4^{\circ}$, this probability is less than $2\%$. 

One should note that the systems for which the SPA would break down are precisely those that could lead to a coincident short gamma-ray burst (short GRB) and GW observation. One of the possible progenitors of short GRBs are NS mergers. The electromagnetic observation of such a GRB would require $\bm{\hat{N}}$ and $\bm{\hat{L}}$ to be almost exactly aligned or antialigned. In both cases, the SPA waveforms constructed here and elsewhere in the literature may be ineffective at extracting a GW signal. However, we know in advance which systems will have a failing SPA, and thus, an analysis using our waveforms could switch to the full numerical solution in this (rather small) corner of parameter space. The degree to which this is feasible and effective will be studied in a data analysis framework elsewhere.

\subsection{Numerical comparison}
\label{Tukey-disc}

Having constructed a full and restricted analytical SPA model for the  Fourier transform of the analytical waveform response we now compare it to the purely numerical waveform. The two former ones are given by Eqs.~\eqref{full-SPA-family} and~\eqref{restricted-SPA-family} respectively. The latter is computed by first applying a window function to the numerical, time-domain waveform of Sec.~\ref{GW-time-numerical-comp}, then discretizing this waveform and finally computing the Fourier transform using the FFTW routine. We ensure that the number of points used in the discretized waveform time series is large enough that the Nyquist frequency is at least 5 times larger than the highest frequency of interest in our analysis.

We use a Tukey function, with parameters such that the window varies from zero to unity between $f_{\GW,1}$ and $f_{\GW,2}$ remains unity until $f_{\GW} = 500$~Hz and then falls off to zero between $f_{\GW} = 500$~Hz and $f_{\GW,3}$, where $f_{\GW}$ is the frequency of the dominant harmonic, at twice the orbital frequency. We choose the frequencies $f_{\GW,1}$ and $f_{\GW,2}$ to be $8.5$~Hz and $9.5$~Hz respectively when studying waveforms that only include the $n=2$ harmonic, or $19/7$~Hz and $20/7$~Hz, respectively, when studying ones that include all harmonics\footnote{When including all harmonics, a smaller starting frequency is required, so that the highest harmonic is included at the beginning of the frequency band, i.e.~at $10$~Hz.}. The window goes to zero at $f_{\GW,3}$ which corresponds roughly to twice the orbital frequency of a test particle in the innermost stable circular orbit of a Schwarzschild BH with mass equal to the total mass of the test system. We could have chosen a different maximum frequency, but this would not have changed the results we will show in this section. We investigated a variety of filters, including the Planck-taper function of~\cite{McKechan:2010kp}, but found that a Tukey window with the above parameters is optimal for minimizing spectral leakage inside the frequency regime of interest. 

For the  comparisons that follow, we choose the same test system as in Sec.~\ref{comparison}, which is representative of all systems considered, and concentrate on the frequency region $(10,400)$~Hz. We stop all comparisons at $400$~Hz because (i) finite size effects cannot be neglected beyond that frequency and (ii) most of the SNR is contained in this frequency region for binary NS coalescences. The alignment of the time-domain waveforms at $10$~Hz does not guarantee alignment in the frequency domain. We have freedom to choose the phase and time of coalescence to perform such an alignment. However, in this section, we will present the waveforms as they are, i.e.~without aligning them in frequency space. This implies that the errors shown here are an overestimate of the inaccuracies in our waveforms, i.e.~they are conservative.
\subsubsection{Leading-order PN amplitude}

\begin{figure*}[th]
\begin{center}
\includegraphics[width=\columnwidth,clip=true]{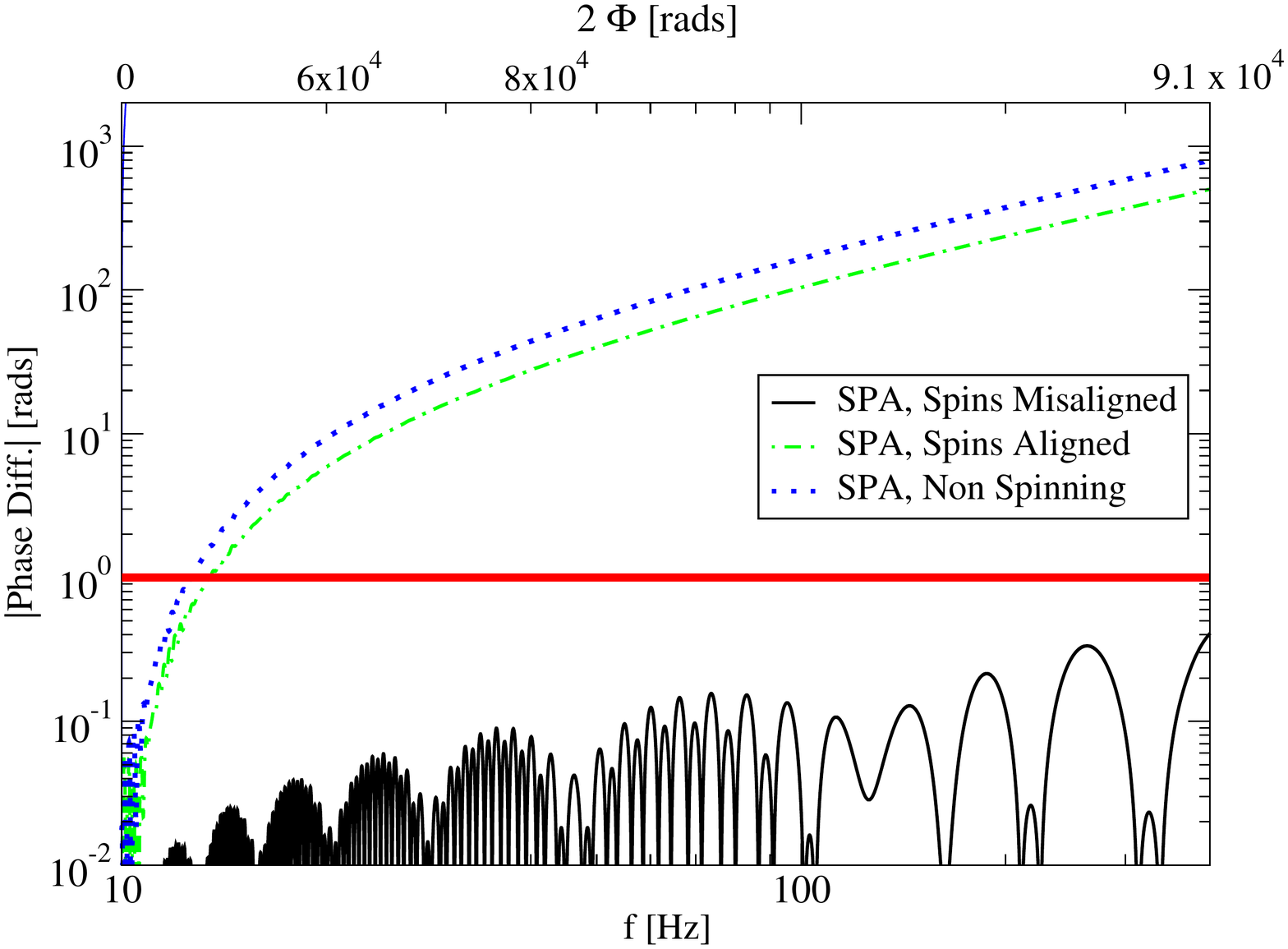} \quad
\includegraphics[width=\columnwidth,clip=true]{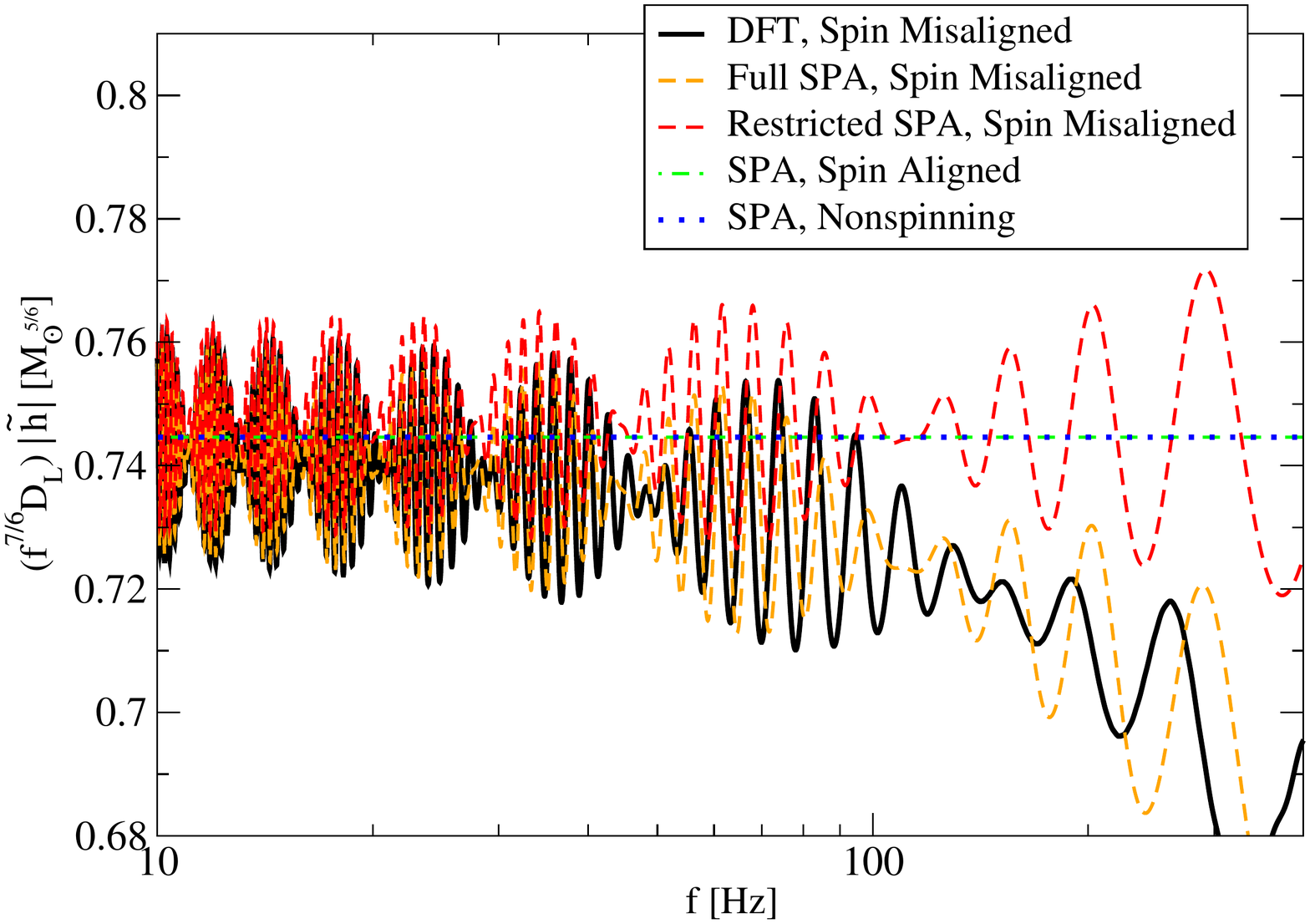} \\
\caption{\label{fig:Four-Phase-Amp}
Left: Fourier dephasings between the DFT and either the SPA waveform (solid black curve), a spin-aligned SPA waveform (dot-dashed green curve), and a nonspinning SPA waveform (dotted blue curve), together with a reference line (thick red curve) at 1 rad, as a function of frequency on the bottom x axis, and accumulated time-domain GW phase on the top x axis. Observe that the SPA phase approximates the DFT one to high accuracy over the entire inspiral evolution, which spans nearly $10^5$ radians of GW phase. 
Right: DFT (solid black curve), full SPA (dashed orange curve), restricted PN SPA (dashed red curve), spin-aligned SPA (dot-dashed green curve), and nonspinning SPA (dotted blue curve) Fourier amplitudes. Observe that the full and restricted SPA amplitudes can faithfully model the precession amplitude oscillations, while the spin-aligned and nonspinning models cannot. Observe also that the overall frequency dependence of the DFT amplitude is not quite $f^{-7/6}$ due to PN corrections to the orbital frequency evolution, and that it is correctly modeled in the full SPA.}
\end{center}
\end{figure*}

Let us begin by focusing on waveforms that contain only the leading-order PN amplitude. The numerical waveform will then be the DFT of Eq.~\eqref{full-h-of-t}, keeping only the $n=2$ (dominant) harmonic. The full SPA waveform is similarly given by Eq.~\eqref{full-SPA-family}, keeping only the leading-order $n=2$ harmonic, while the restricted SPA is given by Eq.~\eqref{restricted-SPA-family}. Notice that the restricted SPA waveform is less accurate than the full SPA waveform because the former keeps only the leading-order PN terms in the amplitude of the SPA Fourier transform, which amounts to setting $\ddot{\Phi}^{\orb}(\xi)$ to its leading-order PN expression. 

The left panel of Fig.~\ref{fig:Four-Phase-Amp} shows the difference in Fourier phase between the DFT and either the SPA (solid black curve), an SPA Fourier phase that assumes aligned spins (dot-dashed green curve), or an SPA that sets all spins to zero (dotted blue curve)\footnote{The latter two waveforms are well-known in the literature~\cite{Poisson:1995ef,Arun:2008kb}, and they can be obtained by setting either the spin angular momentum to be aligned with the orbital angular momentum or the spins to zero in Eq.~\eqref{restricted-SPA-family}. In either case there is no precession.}. The bottom horizontal axis shows the GW frequency, while the top axis shows the accumulated GW phase in the time domain. A horizontal line (thick, solid red curve) at 1 rad is also shown for reference.  Observe that while the aligned-spin and the nonspinning SPA phases build dephasings of ${\cal{O}}(10^{3})$ rads very quickly, the new SPA waveform of this paper remains in phase to better than $0.4$ rads in $\sim 10^{5}$ rads of GW evolution. Notice that the dephasing is insensitive to whether we use a restricted PN or a full PN SPA waveform, as the phase is exactly the same for both. Notice also that this dephasing is dominated by the error in the $t(\xi)$ function, i.e.~the error in the time inversion is roughly $0.2$ milliseconds at $400$~Hz (see Fig.~\ref{fig:t-of-f}), and thus, the error in $2 \pi f t(f) \approx 0.5$ rads at $400$~Hz. 

The right panel of Fig.~\ref{fig:Four-Phase-Amp} shows the Fourier amplitude multiplied by $(D_{L} f^{7/6})$ as a function of GW frequency in Hz for the DFT waveform (solid black curve), the full SPA waveform (dashed orange curve), the restricted PN SPA waveform (dashed red cube), the spin-aligned SPA waveform (dot-dashed green curve), and the nonspinning SPA waveform (dotted blue curve). Observe that the latter two are flat since these waveforms neglect precession altogether. Observe also that both the restricted and the full SPA waveforms can capture the precession amplitude oscillations present in the DFT amplitude. The full SPA amplitude, however, does better than the restricted one. Both the restricted and the full SPA amplitudes, however, dephase with respect to the DFT amplitude after roughly $35$ cycles of precession oscillations. This dephasing could be eliminated if one extended the results of this paper to include all ${\cal{O}}(\epsilon_{s}^{2})$ effects. In practice, this dephasing will induce a systematic error in the determination of spin parameters, but we expect this systematic error to be small. 

\begin{figure}[ht]
\begin{center}
\includegraphics[width=\columnwidth,clip=true]{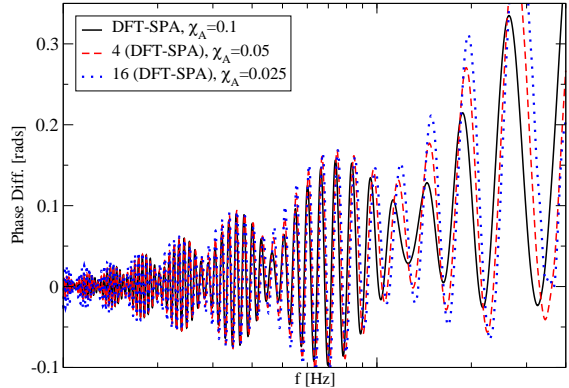}
\caption{\label{fig:Convergence} Phase difference as a function of GW frequency in Hz. The solid black curve corresponds to the test system with equal spins of $\chi^{(1)} = 0.1$, while the dashed red and dotted blue ones to the same system but with equal spins of $\chi^{(2)} = \chi^{(1)}/2 = 0.05$ and $\chi^{(3)} = \chi^{(1)}/4$, respectively. The latter are multiplied by a factor of $(\chi^{(1)}/\chi^{(2)})^{2} = 4$ and $(\chi^{(1)}/\chi^{(3)})^{2} = 16$. Observe that the curves lie roughly on top of each other, indicating that uncontrolled remainders in the SPA waveform phase are of ${\cal{O}}(\epsilon_{s}^{2})$ as expected.}
\end{center}
\end{figure}
Figure~\ref{fig:Convergence} verifies that the error contained in the SPA waveform indeed scales with spin squared. This figure shows the Fourier phase difference as a function of GW frequency in Hz between the DFT and either the restricted SPA waveforms with equal spins of $\chi^{(1)} = 0.1$ (solid black curve), $\chi^{(2)} = \chi^{(1)}/2 = 0.05$ (red dashed curve) or $\chi^{(3)} = \chi^{(1)}/4 = 0.025$ (blue dotted curve). The differences computed for the $\chi^{(2)}$ and $\chi^{(3)}$ systems were multiplied by a factor of $(\chi^{(1)}/\chi^{(2)})^{2} = 4$ and $(\chi^{(1)}/\chi^{(3)})^{2} = 16$ respectively. If the uncontrolled remainder in the SPA is of ${\cal{O}}(\epsilon_{s}^{2})$, we would expect these phase differences to roughly lay on top of each other; Fig.~\ref{fig:Convergence} verifies this expectation. Notice that the restricted SPA waveform contains errors of ${\cal{O}}(\epsilon_{s})$ in the waveform amplitude because of the truncation of $\ddot{\Phi}$ at leading PN order. Therefore, we do not expect the same scaling to be true for the Fourier amplitude of the restricted SPA.

\begin{figure}[th]
\begin{center}
\includegraphics[width=\columnwidth,clip=true]{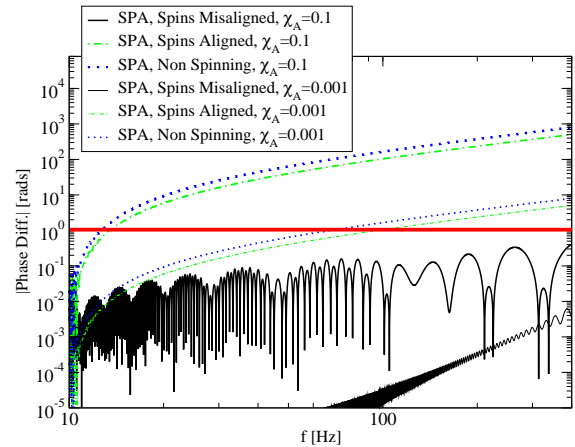}
\caption{\label{fig:Fourier-Phase-Spin-Study} Dephasing between the DFT Fourier phase and the SPA (solid black curves), the spins-aligned SPA (dot-dashed green curves) and the nonspinning SPA (dotted blue curves), for systems with dimensionless spins of $10^{-1}$ (thicker) and $10^{-3}$ (lighter). Observe that the dephasing between the DFT and the spin-aligned or nonspinning SPA waveforms reaches roughly $10$ rads even for spins of $10^{-3}$. }
\end{center}
\end{figure}
Figure~\ref{fig:Four-Phase-Amp} suggests that nonspinning and spin-aligned SPA waveforms are inadequate for parameter estimation of precessing and spinning inspirals, but is there a sufficiently small value of $\chi_{1,2}$ for which this would not be the case? Figure~\ref{fig:Fourier-Phase-Spin-Study} investigates this question by plotting the dephasing between the DFT and either the SPA (solid black curves), the spin-aligned SPA (dot-dashed green curves) or the nonspinning SPA (dotted blue curves) waveforms, for systems with equal spins of $0.1$ (thicker curves) and spins of $10^{-3}$ (thinner curves). Observe that only when the spins become smaller than ${\cal{O}}(10^{-3})$ does the dephasing of the spin-aligned or nonspinning SPA waveforms become comparable to 1 rad. This suggests that parameter estimation systematics would be introduced if one used spin-aligned waveforms to analyze precessing signals when the dimensionless spin magnitude of the latter exceeds $10^{-3}$. 

\begin{figure}[th]
\begin{center}
\includegraphics[width=\columnwidth,clip=true]{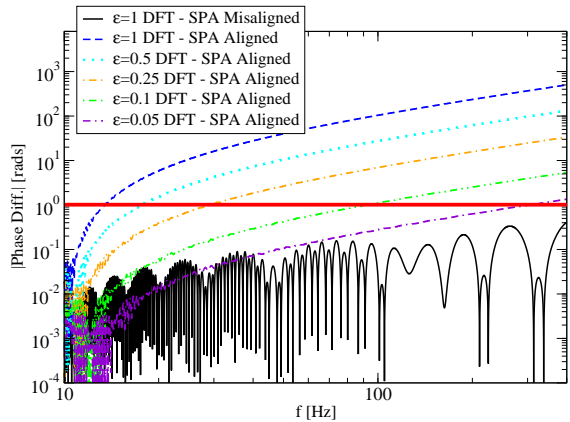}
\caption{\label{fig:Fourier-Phase-Spin-Misalignment-Study} Dephasing between a spin-aligned SPA phase and the DFT Fourier phase of a system with misalignment angles $(\theta_{S_{A}},\phi_{S_{A}})$ equal to the test system's but multiplied by $\epsilon$. In all cases, the spin magnitude is $\chi_{A} = 0.1$ and all other parameters are identical to the test system's. Different curves correspond to different values of $\epsilon$. We include, for reference, a (thick, red) horizontal line at a dephasing of $1$ rad, and the dephasing between the DFT and the SPA misaligned waveform of this paper for $\epsilon = 1$. Observe that only for values of $\epsilon < 0.05$, which corresponds to an angle of only $4^{\circ}$ between the spin and the orbital angular momenta, is the dephasing of ${\cal{O}}(1 \; {\rm{rad}})$ when using a spin-aligned SPA template.}
\end{center}
\end{figure}
One may also wonder whether there are small enough misalignment angles $(\theta_{S_{A}},\phi_{S_{A}})$ that the DFT Fourier phase can be well modeled by a spin-aligned SPA waveform.  Figure~\ref{fig:Fourier-Phase-Spin-Misalignment-Study} addresses this question by plotting the dephasing between spin-aligned SPA waveform and the DFT of a waveform corresponding to a system with the same parameters as those of the test system, but with misalignment angles between $\mb{\hat{S}}_{A}$ and $\mb{\hat{L}}$ multiplied by a small factor $\epsilon$. Different curves in this figure correspond to different values of $\epsilon$. For reference, we include a (thick, red) line at $1$ rad of dephasing. We also include the dephasing between the DFT of a waveform for a system with $\epsilon = 1$ and the new SPA waveforms computed in this paper (black solid curve). Observe that for the spin-aligned SPA waveform to dephase by less than $1$ rad, $\epsilon < 0.05$, implying misalignment angles of less than $4^{\circ}$. Even for such small misalignment angles, the spin-aligned SPA waveform will miss all the precession-induced amplitude modulations, inducing systematic errors in parameter estimation.   

\subsubsection{Full waveform}

Let us now focus on waveforms that include all PN amplitude corrections, including higher harmonics. The numerical waveform will then be the DFT of Eq.~\eqref{full-h-of-t}, keeping all known modes, i.e.~up to $n=7$. This then implies that numerical solutions to the time evolution of the angular momenta must be started at sufficiently low orbital frequency, such that the highest harmonic $(n=7)$ contributes at $10$~Hz (see discussion in Sec.~\ref{Tukey-disc}). The full SPA waveform is similarly given by Eq.~\eqref{full-SPA-family}, keeping all known harmonics, while the restricted SPA is again given by Eq.~\eqref{restricted-SPA-family}. 

\begin{figure*}[th]
\begin{center}
\includegraphics[width=\columnwidth,clip=true]{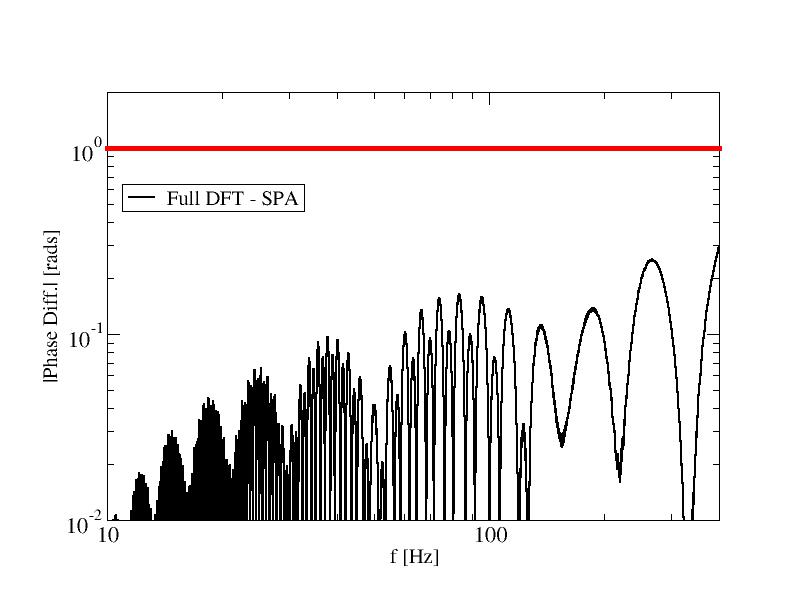} \quad
\includegraphics[width=\columnwidth,clip=true]{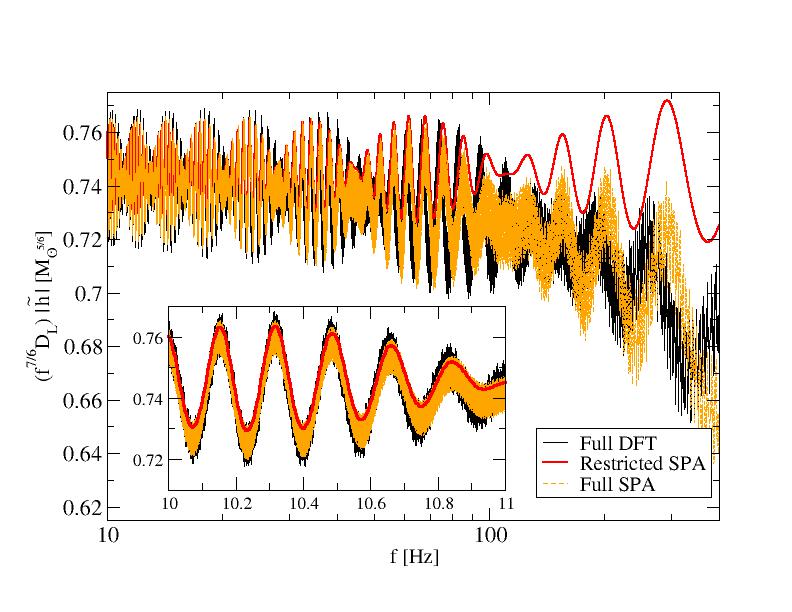} \\
\caption{\label{fig:Four-Phase-Amp-Full} 
Left: difference between the full DFT and SPA Fourier phases (solid black curve) as a function of GW frequency in Hz, together with a reference line (thick red curve) at 1 rad. Observe that the SPA phase approximates the Full DFT one to high accuracy over the entire inspiral evolution. 
Right: full DFT (solid black curve), full SPA (dashed orange curve), and restricted PN SPA (solid red curve) Fourier amplitudes. Observe that the full and restricted SPA amplitudes can faithfully model the precession amplitude oscillations, with the latter losing accuracy as the frequency increases.}
\end{center}
\end{figure*}
The left panel of Fig.~\ref{fig:Four-Phase-Amp-Full} shows the difference between the full DFT and SPA (solid black curve) Fourier phases as a function of GW frequency in Hz, for the test system described in Sec.~\ref{comparison}. For reference, we include a horizontal (thick red) curve at $1$ rad. We only show a single curve for the SPA waveform, since the Fourier phase is the same for the restricted and full SPA templates. Observe that the dephasing is still roughly $10^{-1}$ rads, as we found earlier when looking at a single harmonic.

The right panel of Fig.~\ref{fig:Four-Phase-Amp-Full} shows the full DFT (black solid curve), the full SPA (dashed orange curve) and the restricted SPA (solid red curve) Fourier amplitudes, normalized to $f^{7/6} D_{L}$, as a function of GW frequency in Hz. The inset shows a zoom to the region close to $10$ Hz. This figure presents a number of interesting features. 
First, notice that the full DFT and SPA amplitude curves are thick. This thickness is not due to numerical noise, but rather due to the beating of higher harmonics that induce high frequency oscillations. This feature is not present in the restricted SPA amplitude, as this neglects higher amplitude harmonics all together.  
Second, notice that the average of the full DFT Fourier amplitude tends to decrease with frequency. This feature is captured by the full SPA amplitude, because it includes high PN order effects that induce this trend. The restricted SPA amplitude, however, cannot recover this trend, since it retains the lowest PN order terms only. 
Third, in spite of not perfectly matching the full DFT waveform, the restricted SPA amplitude does a superb job at catching the initial oscillations of the full DFT up to roughly $100$ Hz. This is important because most of the power accumulates between $10$ and $100$ Hz for NS binary inspirals. 

\section{Data Analysis Comparisons}
\label{gw-comps}

In this section we perform a more detailed data analysis comparison for a variety of different systems. This comparison will be based on the {\emph{faithfulness}} measure, namely
\begin{align}
F_{h_1,h_2} &\equiv
\frac{\left(h_{1}\left|\right.h_{2}\right)}{\sqrt{\left(h_{1}\left|\right.h_{1}\right)
\left(h_{2}\left|\right.h_{2}\right)}}\,,
\label{match}
\end{align}
where $h_{1,2}$ are different waveforms with the 
{\emph{same}} physical parameters. The inner-product is defined in the usual way:
\begin{align}
 \left(h_{1}\left|\right.h_{2}\right) &\equiv 
4 \Re \int_{f_{\min}}^{f_{\max}} \frac{\tilde{h}_1 \tilde{h}_2^*}{S_{n}} \; 
df\,,
\end{align}
where $\Re[\cdot]$ is the real part operator, $(f_{\min},f_{\max})$ are the
boundaries of the detector's sensitivity band, $S_{n}$ is the detector's 
spectral noise density, and $\tilde{h}$ denotes the Fourier transform of $h$. 
Given this definition, the {\emph{fitting factor}} $FF_{\tilde{h}_{1}\tilde{h}_{2}}$ 
is nothing but the faithfulness maximized over all template parameters, 
which then implies that $FF_{\tilde{h}_{1}\tilde{h}_{2}} \geq F_{\tilde{h}_{1}\tilde{h}_{2}}$.

We concentrate here on NS binary inspirals, and thus, on sources suitable for detection with
ground-based instruments, such as LIGO. When calculating overlaps through Eq.~\eqref{match}, we choose $f_{\min}=10$ Hz and 
$f_{\max}=10$ kHz, with observation times of about $3 \times 10^{5}$ seconds since the lowest harmonic evolves from $f_{\orb} = 10/7$~Hz to coalescence. We here employ an Advanced~LIGO noise curve given by~\cite{Cornish:2011ys}
\begin{align}
 S_n(f) &=  10^{-49} \Bigg\{ \bar{f}^{-4.14} - \frac{5}{\bar{f}} 
 \nonumber\\
 &+ 111 \left[ 1 - \bar{f}^2 \left( 1 - \frac{1}{2} \bar{f}^2 \right) \right] 
 \left( 1 - \frac{1}{2} \bar{f}^2 \right)^{-1} \Bigg\}\,,
\end{align}
where we have defined the dimensionless frequency $\bar{f} = f/(215 \; {\rm{Hz}})$. We use a Tukey window in all our waveforms with the parameters described in Sec.~\ref{Tukey-disc}. The Tukey window is sufficiently slowly varying that we can include it in the amplitudes of our SPA waveforms. 

The image of the faithfulness measure is in the interval $[-1,1]$; it quantifies how 
well waveforms agree with each other, with unity representing perfect agreement. 
All integrations are done numerically, with errors of ${\cal{O}}(10^{-5})$; thus, we consider that a match of 
$F_{\tilde{h}_{1}\tilde{h}_{2}} =  0.9999$ is consistent with unity. Conventionally, a fitting factor about $97\%$ is generally considered to be sufficient for detection. Therefore, a faithfulness of $98\%$ certainly implies a fitting factor of at least $98\%$, which is also good enough for detection.

The faithfulness measure will be evaluated using a full DFT waveform as the signal and either the
full SPA or the restricted SPA as the template. That is, the full DFT waveform will be our reference
waveform, to which the other two template families will be compared. 

Before proceeding, let us add one last word of caution. Faithfulness comparisons are {\emph{conservative}} because the match is not maximized over physical parameters, 
such as the total mass, mass ratio, spin magnitudes, or angles. Higher matches would indeed 
be obtained if we allowed the templates to have different physical parameters from the signal, 
as one does in parameter estimation. Such higher matches, of course, will come at the cost
of a systematic bias in the recovered parameters. We leave such a study for future work. 

Instead of working with a particular system (like the test one of Sec.~\ref{comparison}), we perform a Monte-Carlo simulation with 1000 points in parameter space, with all system parameters randomized, except for the dimensionless spin magnitudes, which will be set to be equal to each other and constant along each run. We consider systems with individual masses in the range $(1.2,2)$~$M_\odot$, appropriate for NSs, with a flat distribution in log space. The distribution of unit vectors is chosen to be uniform on the sphere.

Figure~\ref{fig:F} already showed the median match and $68\%$ (1$\sigma$) interval regions between the full DFT-full SPA waveform (red solid curve), the full DFT-restricted SPA waveform (green dot-dashed curve), the full DFT-full fpin-aligned SPA waveform (blue dashed curve), and the full DFT-full nonspinning SPA waveform (magenta dotted curve) as a function of spin magnitude $\chi_{A}$. We observed that the faithfulness of the new precessing waveform families is above $99\%$ for all systems considered, which is dramatically better than the faithfulness when using spin-aligned or nonspinning templates. In fact, this figure suggests that the latter two could introduce serious systematic errors in parameter estimation. The performance of the full and restricted SPA families is very similar because their difference is dominantly due to odd amplitude harmonics, which are proportional to the dimensionless mass difference, a small number for binary NSs. This results in higher harmonic corrections being unimportant when the faithfulness reaches $F \gtrsim 0.999$.

\begin{figure*}[htb]
\begin{center}
\includegraphics[width=\columnwidth,clip=true]{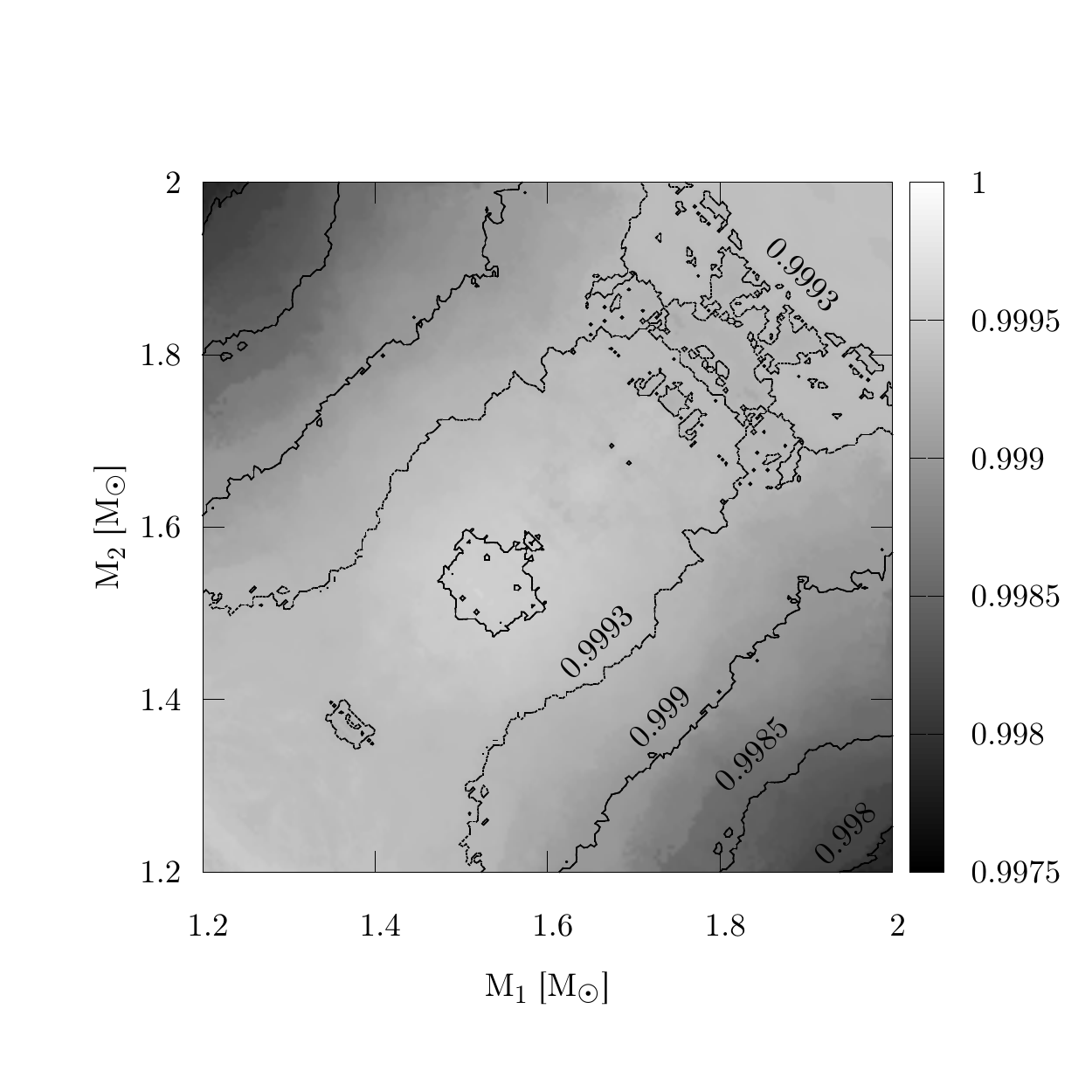} \quad
\includegraphics[width=\columnwidth,clip=true]{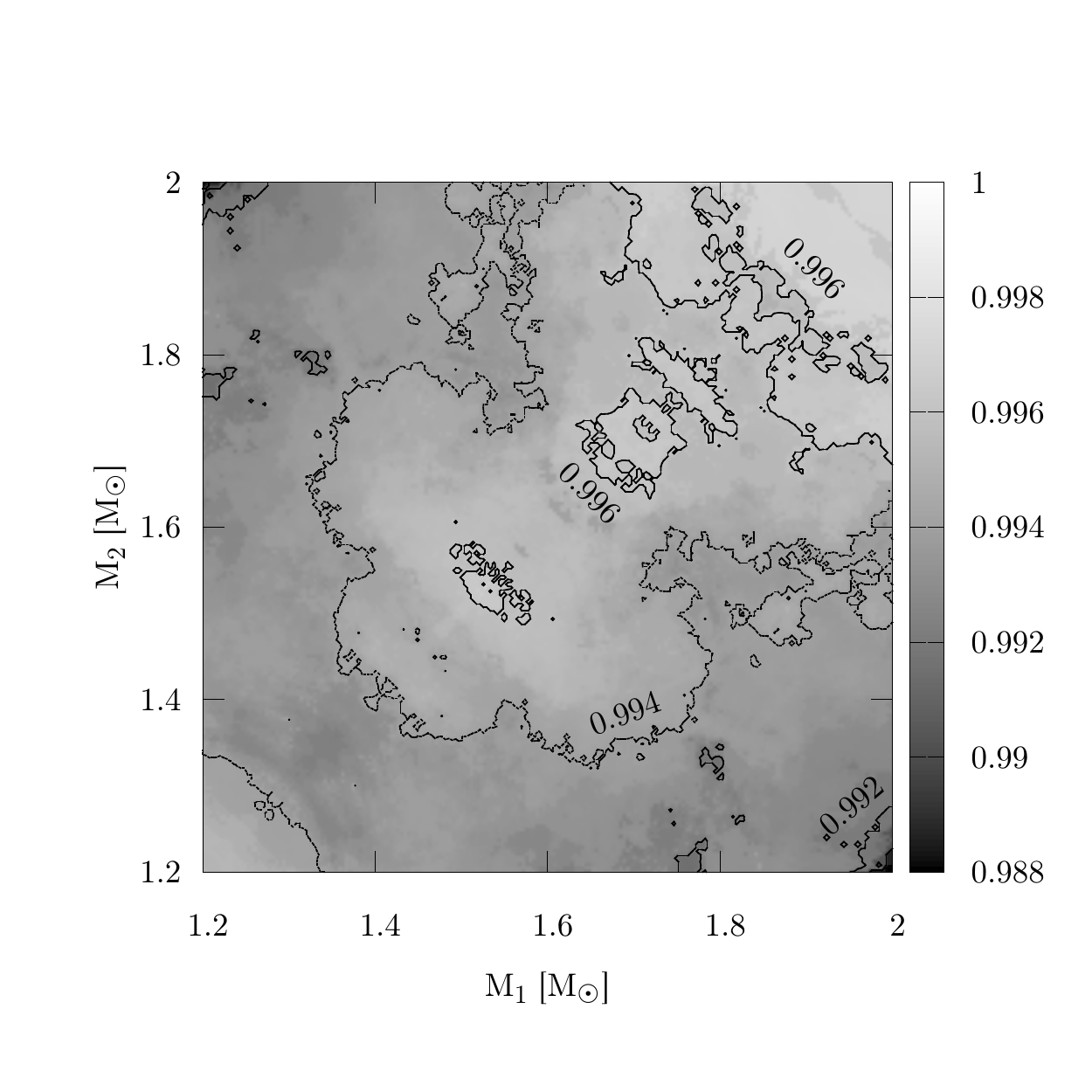} 
\caption{\label{fig:Contour} Faithfulness contour plot in the $(M_{1},M_{2})$ plane for systems with equal spin magnitudes of $0.04$ (left panel) and $0.1$ (right panel). Each point in the contour plot corresponds to the median faithfulness inside a circle centered at the given point with a radius of $0.1 M_{\odot}$. We considered a total of $1000$ system with all parameters (except for the spin a magnitudes) randomized over. Observe that the faithfulness is highest along the equal mass line because precession is suppressed there by symmetry.}
\end{center}
\end{figure*}

One may wonder how the faithfulness behaves as a function of intrinsic parameters. The two that are perhaps most important are the component masses; in this case, $M_{1}$ and $M_{2}$ have a rather limited range because we are considering NS binaries.  Figure~\ref{fig:Contour} presents contour plots of the faithfulness in the $(M_{1},M_{2})$ plane for systems with equal dimensionless spin magnitudes of $0.04$ (left panel) and $0.1$ (right panel). We considered 1000 systems with all parametrized randomized, except for the spin magnitudes. The value of the faithfulness at any given point is the median value inside a circle centered at that point with radius $0.1 M_{\odot}$. Observe that the faithfulness is largest along the equal-mass symmetry line. Perhaps this is because in the equal-mass limit, odd harmonics are suppressed and precession becomes simple harmonic, when neglecting spin-spin interactions. 

\section{Beyond the early inspiral}
\label{finite-size}

The starting frequency of our numerical comparisons corresponds to the beginning of the LIGO sensitivity band, but the ending frequency ($400$ Hz) does not correspond to the end of this band or to the merger frequency. Rather, it corresponds to the frequency at which finite-size effects can no longer be neglected~\cite{Read:2009yp,Hinderer:2009ca,Markakis:2010mp}. In this section we summarize how such effects could be taken into account, extending the waveforms obtained in this paper beyond 400Hz.

The finite-size effects give rise to two types of deformations: (i) multipole and, (ii) tidal. The former are described by the NS's multipole moments, which measure how much the object is deformed away from sphericity. The latter are caused by the companion's external field and they are quantified by the tidal Love number~\cite{Hinderer:2007mb,PhysRevD.80.084018,PhysRevD.80.084035} that characterizes the deformability of the NS. 

The leading-order multipole effect is the quadrupole one which is an ${\cal{O}}(\epsilon_s^2)$ effect entering the GW as a 2PN phase correction given by~\cite{PhysRevD.57.5287,PhysRevD.71.124043}
\begin{align}
\delta{\Psi}^{Q}_n&=\frac{75 n}{256 v_{n}}\left(\frac{Q_1}{M_1^2M_2}  + \frac{M_1}{M_2} \chi_1^2\right)
\left[3 \left({\mb{\hat{S}}}_{1} \cdot {\mb{\hat{L}}}\right)^{2} - 1 \right]
\nn \\
&+1\leftrightarrow 2 
\label{quad-def}\,,
\end{align}
where recall that $v_{n}=(2 \pi M f/n)^{1/3}$ and $n$ is the harmonic number. In the above equation, $Q_A$ is the quadrupole moment of each binary component\footnote{As in Ref.~\cite{Yagi:2013awa}, we normalize $Q_{A}$ such that Eq.~\eqref{quad-def} gives the contribution to the phase from deviations of the quadrupole moment from the BH value.}. In principle there are higher-order multipole deformations that affect the waveform phase, but these are proportional to higher powers of spin, and thus, negligible.

On the other hand, the tidal deformations caused by the gravitational field of the companion, will result in a 5PN phase correction~\cite{Flanagan:2007ix}
\begin{align}
\delta{\Psi}^{\lambda}_n&=-\frac{9 n}{32 \eta}\left(1+12\frac{M_2}{M_1}\right)\frac{\lambda_1}{M^5}v_{n}^5
+ 1\leftrightarrow 2\label{tidal-def}\,,
\end{align}
where $\lambda_A$ is the tidal Love number of the $A$th binary component. Corrections to the above equation up to 2.5PN can be found in~\cite{Vines:2011ud,PhysRevD.85.123007}.

Even though we can use Eqs.~\eqref{quad-def} and~\eqref{tidal-def} to minimize the error induced by the point-particle approximation, we still cannot extend the region of validity of inspiral PN waveforms beyond a certain frequency. For sure one cannot use simple inspiral waveform beyond the frequency at which the two NSs touch. This contact frequency can be approximated by that at which the NS's separation is equal to the sum of their radii, neglecting tidal deformations. In Table~\ref{table:f-touch} we give an estimate of that frequency for our test system [see Sec.~\ref{comparison}] for various equations of state (EoS). We see that for all EoS considered here, the regime we study can be considered to be the early inspiral, where finite-size effects have not yet started to have an observable signature in the waveforms.

Beyond this frequency the system enters the highly nonlinear and dynamical merger phase, the outcome of which is still rather uncertain~\cite{PhysRevD.71.084021,Anderson:2007kz,Liu:2008xy,Hotokezaka:2011dh,Kiuchi:2012mk,Faber:2012rw,Giacomazzo:2013uua}. Depending on the EoS and the mass ratio, the merger remnant might be a hypermassive NS, which then collapses to a BH, or it could simply collapse directly to a BH. In either case, nonlinear dynamics play an important role and would have to be mimicked in some phenomenological way if one wishes to construct an effective analytical template.
\begin{table}
\begin{centering}
\begin{tabular}{ccccc}
\hline
\hline
\noalign{\smallskip}
EoS &&  \multicolumn{1}{c}{$R_1({\rm{km}})$} &  \multicolumn{1}{c}{$R_2({\rm{km}})$}
&  \multicolumn{1}{c}{$f({\rm{Hz}})$}  \\
\hline
\noalign{\smallskip}
APR~\cite{PhysRevC.58.1804} && 12.2 & 12.2 & 1425\\
SLy~\cite{refId0} && 11.4 & 11.6  & 1553\\
LS220~\cite{Lattimer1991331} && 12.7 & 13.4  & 1292\\
Shen~\cite{Shen1998435} && 14.5 & 14.9  & 1089\\
\noalign{\smallskip}
\hline
\hline
\end{tabular}
\end{centering}
\caption{Frequency at which the separation of the two NSs is equal to the sum of their radii for various EoS.}
\label{table:f-touch}
\end{table}
%

\section{Discussion}
\label{conclusion}

We provided computationally inexpensive, analytical waveforms for spinning binaries with small spins. We do so by first obtaining an analytic perturbative solution to the spin precession equations and, then, by analytically Fourier transforming the resulting time domain waveform. 

Each step of the calculation has been aided by a series of approximations which we have tried to make explicit throughout the main body of this paper. In what follows, we discuss how to improve on these approximations and comment on the implications:
\begin{enumerate}
\item {\bf{PN Order}}:
Equations~\eqref{Ldot}-\eqref{S2dot} describe the conservative evolution of the spin angular momenta, which are here modeled to next-to-next-to-leading order in the spin-orbit interaction and to leading order in the spin-spin interaction. The spin-orbit terms lead to corrections in the waveform phase up to $3.5$PN relative order, while the spin-spin terms lead to correction at $2$PN order. When higher PN order corrections to these equations are computed, one could include these in the formalism presented here in a straightforward manner.
\item {\bf{Adiabatic Approximation}}:
Equations~\eqref{Ldot}-\eqref{S2dot} are valid in an {\emph{adiabatic approximation}}, because they are derived after orbit averaging the full evolution equations~\cite{Thorne:1984mz}, thus ignoring oscillations that could appear on an orbital time scale. These oscillations could lead to nonsecular effects that we expect would be greatly subdominant relative to all other uncontrolled remainders already present in our waveforms. But if the adiabatic approximation is not valid or if eccentricity is present, these oscillations could lead to secularly growing effects that do not average out and lead to corrections larger than other uncontrolled remainders. One could investigate whether the adiabatic approximation is valid or not by comparing our solution to a numerical one obtained from the full equations.
\item {\bf{Spin Magnitude}}:
Perhaps the most important approximation made in this paper is that of the magnitude of the spins being small. Figure~\ref{fig:F} shows that even for values of $\chi \sim 0.2$, the faithfulness is still above $99\%$. Extending our calculation to second order in spin is straightforward. One would first include the spin-spin interactions in Eqs.~\eqref{Ldot}-\eqref{S2dot}, and then extend their solution to ${\cal{O}}(\epsilon_s^{2})$.
\item {\bf{Separation of Timescales}}:
Another important approximation used to derive our waveforms is that all time scales in the problem separate, so that multiple scale analysis can be employed when solving the ${\cal{O}}(\epsilon_{s})$ precession equations. One could improve on this approximation by keeping the next order term in $\epsilon_{p}$ terms. In practice, however, we find that this improvement is truly negligible, because it would be of ${\cal{O}}(\epsilon_{s} \times \epsilon_{p}^{2})$. 
\item {\bf{SPA}}: The validity and suitability of the SPA may be compromised for two reasons: (i) the leading-order correction to the SPA may be necessary to accurately model the Fourier transform; (ii) precession may lead to a complete break-down of the SPA. As for (i), we extended the results of~\cite{Droz:1999qx} to 3PN beyond leading, which were found to be important; further extensions are straightforward, but seem unnecessary. As for (ii), Fig.~\ref{fig:Phiddot} explicitly shows that the probability the SPA will break for binary neutron star systems is remarkably tiny.
\item {\bf{Tidal Interactions}}:
All the equations used in this paper, and therefore, the waveforms derived here, neglect finite-size effects. In particular, we have neglected all multipole deformations as well as tidal effects that NSs can experience during coalescence. One could easily include these corrections by adding the terms given in Eqs.~\eqref{quad-def} and~\eqref{tidal-def}.
\end{enumerate}

The waveform family presented here is a definite step towards modeling spinning binaries and constructing waveforms that can be used for parameter estimation. Figure~\ref{fig:F} has explored how efficient this waveform family is by calculating the faithfulness between full numerical waveforms and analytical ones, but a full data analysis parameter estimation study is still missing. In the future, one could carry out such a study by injecting a generically spinning, precessing signal, and measuring how well it could be recovered by the templates calculated here through Bayesian tools~\cite{Cornish:2007ifz,Littenberg:2009bm}.

The waveform family computed in this paper fails to capture the strong precession effects induced by rapidly spinning compact objects with arbitrary spin orientations, like BHs. Astrophysical BHs can easily have dimensionless spin magnitudes larger than $0.1$, thus violating our small spins approximation. One could extend the calculation of this paper to next order in ${\cal{O}}(\epsilon_s)$ and study whether the extended solution is now accurate enough to model moderately spinning BHs. 

Another approach would be to use a slightly different background to perturb about, instead of nonspinning binaries. Reference~\cite{Apostolatos:1994mx} showed that when only a single component of a spinning binary is actually spinning, then the precession equations can be solved exactly and the resulting motion is simple precession. Introducing a small spin on the companion and perturbing about simple precession, one may obtain waveforms that could model BH-NS binaries. We leave the study of this system for future work.

\acknowledgments 

We thank Luc Blanchet for useful discussions on PN theory. K.C. and N. Y. acknowledge support from NSF Grant No. PHY-1114374 and NASA Grant No. NNX11AI49G, under subaward 00001944. A. K. and N. C. acknowledge support from the NSF Award No. PHY-1205993 and NASA Grant No. NNX10AH15G. 

\appendix

\section{Radiation-Reaction Coefficients}
\label{app-coeffomegadot}

The PN equation for the evolution of the orbital frequency is given by
\begin{align}
\dot{\omega}&=\frac{a_0}{M^2}(M\omega)^{11/3}\left\{1+\sum_{i=2}^{11} [a_i+b_i \ln{(M\omega)}] (M\omega)^{i/3}\right\}\nn\\
&+{\cal{O}}(c^{-12})\,.\label{omegadot}
\end{align}
The evolution equation including spins is known up to 3.5PN order~\cite{PhysRevD.80.044010,PhysRevD.80.024002,PhysRevD.84.064041,Bohe:2013cla}, where the nonzero coefficients of the log-independent terms are
\allowdisplaybreaks 
\begin{align}
\label{a0}
a_0&=\frac{96}{5} \eta\,,\\
a_2&=-\frac{743}{336}-\frac{11}{4}\eta\,,\\
a_3&=4\pi-\beta_{3}\,,\\
a_4&=\frac{34103}{18144}+\frac{13661}{2016}\eta+\frac{59}{18}\eta^2-\sigma_4\,,\\
a_5&=-\frac{4159}{672}\pi-\frac{189}{8}\pi \eta -\beta_{5}\,,\\
a_6&=\frac{16447322263}{139708800}+\frac{16}{3}\pi^2-\frac{856}{105}\ln{16}-\frac{1712}{105}\gamma_E-\beta_6\nn\\
&+\eta\left(\frac{451}{48}\pi^2-\frac{56198689}{217728}\right)+\eta^2\frac{541}{896}-\eta^3\frac{5605}{2592}\,,\\
a_7&=-\frac{4415}{4032}\pi+\frac{358675}{6048}\pi\eta+\frac{91495}{1512}\pi\eta^2-\beta_{7}\,.
\end{align}
with
\begin{align}
\beta_{3}&=\frac{1}{M^2}\sum_{A\neq B}\left(\frac{113}{12}+\frac{25}{4}\frac{M_B}{M_A}\right)\bm{S}_A\cdot\bm{\hat{L}}\,,
\\
\beta_{5}&=\frac{1}{M^2}\sum_{A\neq B}\left[\left(\frac{31319}{1008}-\frac{1159}{24}\eta\right)\right.\nn\\
&\left.+\frac{M_B}{M_A}\left(\frac{809}{84}-\frac{281}{8}\eta\right)\right]\bm{S}_A\cdot \bm{\hat{L}}\,,
\\
\beta_6&=\frac{\pi}{M^2}\sum_{A\neq B}\left(\frac{75}{2}+\frac{151}{6}\frac{M_B}{M_A}\right)\bm{S}_A\cdot\bm{\hat{L}}\,,\\
\beta_{7}&=\frac{1}{M^2}\sum_{A\neq B}\left[\left(\frac{130325}{756}-\frac{796069}{2016}\eta+\frac{100019}{864}\eta^2\right)\right.\nn\\
&\left.+\frac{M_B}{M_A}\left(\frac{1195759}{18144}-\frac{257023}{1008}\eta+\frac{2903}{32}\eta^2\right)\right]\bm{S}_A\cdot \bm{\hat{L}}\,,
\\
\sigma_4&=\frac{1}{\mu M^3}\left[\frac{247}{48}\bm{S}_1\cdot\bm{S}_2-\frac{721}{48}\left(\bm{S}_1\cdot\bm{\hat{L}}\right)\left(\bm{S}_2\cdot\bm{\hat{L}}\right)\right]\nn\\
&+\sum_{A}\frac{1}{M^2M_A^2}\left[\frac{233}{96}S_A^2-\frac{719}{96}\left(\bm{S}_A\cdot\bm{\hat{L}}\right)^2\right]\,.
\end{align}
where $\gamma_E$ is the Euler constant. To this same order, the nonzero coefficients of the log-dependent terms are~\cite{PhysRevD.80.044010,PhysRevD.80.024002,PhysRevD.84.064041,Bohe:2013cla}
\begin{align}
b_6&=-\frac{1712}{315}\,.
\end{align}
In reality, the evolution of the orbital frequency in Eq.~\eqref{omegadot} is consistent only up to 2.5PN order, since it does not include 3PN and 3.5PN spin-spin terms. The latter, to our knowledge, have not been calculated to date. 

One can extend the evolution equation of the orbital frequency [Eq.~\eqref{omegadot}] beyond $3.5$PN order by including higher-order terms in the binding energy and GW luminosity known in the nonspinning, point-particle limit, namely~\cite{Mino:1997bx,Varma:2013kna}, and then keeping higher orders in the Taylor expansion of $\dot{\omega} = {\cal{L}}_{\GW} (dE/d\omega)^{-1}$. Note that we do not include here BH absorption terms, as we are interested in binary NS systems. Also, the $4$PN spin-orbit term has been recently calculated~\cite{Marsat:2013caa} . Then, the log-independent coefficients are 
\begin{widetext}
\begin{align}
a_{8} &= \frac{3971984677513}{25427001600}+\frac{127751}{1470} \ln{2} - \frac{47385}{1568} \ln{3} + \frac{124741}{4410} \gamma_{E} -\frac{361}{126} \pi^{2}  + \frac{82651980013}{838252800} \eta - \frac{1712}{315} \eta \ln{2}
\nn \\
& - \frac{856}{315} \gamma_{E}\eta  - \frac{31495}{8064} \pi^{2} \eta + \frac{54732199}{93312} \eta^{2}- \frac{3157}{144} \pi^{2} \eta^2  - \frac{18927373}{435456} \eta^{3} -\frac{95}{3888} \eta^{4} -\beta_{8} 
\,,
\\
a_{9} &= \frac{343801320119}{745113600} \pi- \frac{13696}{105} \pi \ln{2} -  \frac{6848}{105} \pi \gamma_{E}  - \frac{51438847}{48384} \pi \eta + \frac{205}{6}  \pi^{3} \eta + \frac{42680611}{145152} \pi \eta^{2}  +  \frac{9731}{1344} \pi \eta^{3}\,,
\\
a_{10} &= \frac{29619150939541789}{36248733480960}  -\frac{107638990}{392931} \ln{2}+ \frac{616005}{3136} \ln{3}- \frac{11821184}{1964655} \gamma_{E} - \frac{21512}{1701} \pi^{2} - \frac{884576519037433}{228843014400} \eta \nn\\
&+ \frac{2105111}{8820}\eta \ln{2} - \frac{15795}{3136}\eta\ln{3} + \frac{3090781}{26460} \gamma_{E}  \eta+ \frac{14555455}{217728}\pi^{2} \eta  + \frac{1175999369413}{914457600} \eta^{2}- \frac{4708}{945}\eta^{2}\ln{2}\nn\\
& - \frac{126809}{3024} \pi^{2} \eta^{2}- \frac{2354}{945} \gamma_{E}\eta^{2} - \frac{9007327699}{11757312} \eta^{3}  + \frac{9799}{384} \pi^{2} \eta^{3} + \frac{51439207}{1741824} \eta^{4} - \frac{34613}{186624} \eta^{5}\,,
\\
a_{11} &=  \frac{91347297344213}{81366405120} \pi+ \frac{5069891}{17640} \pi \ln{2}- \frac{142155}{784} \pi \ln{3}  + \frac{311233}{5880} \pi\gamma_{E}   - \frac{1903651780081}{4470681600} \pi \eta- \frac{6848}{315} \pi \eta\ln{2}\nn\\
&  - \frac{3424}{315} \pi \gamma_{E}\eta- \frac{26035}{16128}  \pi^{3}\eta + \frac{1760705531}{290304} \pi \eta^{2} - \frac{112955}{576} \pi^{3} \eta^{2}- \frac{7030123}{13608} \pi \eta^{3} + \frac{49187}{6048} \pi \eta^{4}  \,,
\end{align}
\end{widetext}
where
\begin{align}
\beta_{8}&=\frac{\pi}{M^2}\sum_{A\neq B}\left[\left(\frac{76927}{504}-\frac{220055}{672}\eta\right)\right.\nn\\
&\left.+\frac{M_B}{M_A}\left(\frac{1665}{28}-\frac{50483}{224}\eta\right)\right]\bm{S}_A\cdot \bm{\hat{L}}\,,
\end{align}
is the $4$PN spin-orbit term and the log-dependent coefficients are
\begin{align}
b_{8} &= - \frac{856}{315} \eta + \frac{124741}{4410}\,,
\\
b_{9} &= - \frac{6848}{105} \pi\,,
\\
b_{10} &= \frac{3090781}{26460} \eta -  \frac{2354}{945} \eta^{2} - \frac{11821184}{1964655}\,,
\\
b_{11} &= \frac{311233}{5880} \pi - \frac{3424}{315} \pi \eta\,.
\end{align}
We stop the expansion at $5.5$PN order for convenience only; if deemed necessary, one could include higher-order terms~\cite{Fujita:2012cm,Varma:2013kna}. 

\section{Coefficients in the Time-Domain Expansion of the PN Parameter $\xi$}
\label{app-coeffxi}

The PN parameter $\xi\equiv(M\omega)^{1/3}$ with $5.5$PN radiation reaction, is given as a function of time to $5.5$PN by 
\begin{align}
\xi&=\zeta\!\left\{1+\!\sum_{i=2}^{11}\xi_i\zeta^i+\!\sum_{i=6}^{11}\xi^\ell_i\zeta^i\ln{(\zeta)}+\!\sum_{i=8}^{11}\xi^{\ell^2}_i\zeta^i[\ln{(\zeta)}]^2\right\}\nn\\
&+{\cal{O}}(c^{-12})\,,
\label{full-xi-exp}
\end{align}
where the nonzero PN coefficients are 
\allowdisplaybreaks
\begin{align}
\xi_{2}&=-\frac{a_2}{6}\,,
\\
\xi_{3}&=-\frac{a_3}{5}\,,
\\
\xi_{4}&=-\frac{1}{4}a_4+\frac{5}{24}a_2^2\,,
\\
\xi_{5}&=-\frac{1}{3}a_5+\frac{3}{5}a_3a_2\,,
\\
\xi_{6}&=-\frac{1}{2}a_6-\frac{3}{4}b_6+\frac{23}{24}a_4a_2+\frac{12}{25}a_3^2-\frac{67}{144}a_2^3\,,
\\
\xi_{7}&=-a_7+2a_5a_2+2a_4a_3-3a_3a_2^2\,,
\\
\xi_{8}&=\frac{1}{12}a_6a_2+\frac{3}{8}b_6a_2+\frac{1}{15}a_5a_3+\frac{1}{32}a_4^2-\frac{59}{288}a_4a_2^2\nn\\
&-\frac{29}{150}a_3^2a_2+\frac{971}{10368}a_2^4\,,
\\
\xi_{9}&=a_9-3b_9-\frac{5}{3}a_7a_2-\frac{9}{5}a_6a_3+\frac{33}{5}b_6a_3-\frac{11}{6}a_5a_4\nn\\
&+\frac{20}{9}a_5a_2^2+\frac{47}{10}a_4a_3a_2+\frac{102}{125}a_3^3-\frac{358}{135}a_3a_2^3\,,
\end{align}
\allowdisplaybreaks
\begin{widetext}
\begin{align}
\xi_{10}&=\frac{1}{2}a_{10}-\frac{3}{4}b_{10}-\frac{7}{6}a_8a_2+\frac{3}{2}b_8a_2-\frac{2}{5}a_7a_3-\frac{5}{8}a_6a_4+\frac{25}{16}a_6a_2^2+\frac{39}{16}b_6a_4-\frac{287}{96}b_6a_2^2-\frac{1}{3}a_5^2+\frac{49}{30}a_5a_3a_2\nn\\
&+\frac{191}{192}a_4^2a_2-\frac{2941}{1728}a_4a_2^3-\frac{67}{60}a_3^2a_2^2+\frac{9139}{20736}a_2^5\,,
\\
\xi_{11}&=-\frac{2}{3}a_9a_2+\frac{3}{2}b_9a_2-\frac{1}{5}a_8a_3+a_7a_4+\frac{5}{9}a_7a_2^2+\frac{2}{3}a_6a_5+\frac{2}{3}a_6a_3a_2+\frac{3}{2}b_6a_5-\frac{28}{5}b_6a_3a_2-\frac{16}{9}a_5a_4a_2-\frac{4}{25}a_5a_3^2\nn\\
&-\frac{104}{81}a_5a_2^3-\frac{17}{10}a_4^2a_3+\frac{71}{45}a_4a_3a_2^2-\frac{139}{375}a_3^3a_2-\frac{53}{270}a_3a_2^4\,,
\end{align}
\end{widetext}
\allowdisplaybreaks
\begin{align}
\xi^\ell_{6}&=-\frac{3}{2}b_6\,,
\\
\xi^\ell_{8}&=a_8-2a_6a_2+\frac{1}{4}b_6a_2-2a_5a_3-a_4^2+3a_4a_2^2\nn\\
&+3a_3^2a_2-a_2^4\,,
\\
\xi^\ell_{9}&=3b_9-\frac{27}{5}b_6a_3\,,
\\
\xi^\ell_{10}&=\frac{3}{2}b_{10}-\frac{1}{2}a_8a_2\!-\!\frac{7}{2}b_8a_2+a_6a_2^2-\frac{15}{8}b_6a_4\!+\!\frac{75}{16}b_6a_2^2\nn\\
&+a_5a_3a_2+\frac{1}{2}a_4^2a_2-\frac{3}{2}a_4a_2^3-\frac{3}{2}a_3^2a_2^2+\frac{1}{2}a_2^5\,,
\\
\xi^\ell_{11}&=-2b_9a_2-\frac{4}{5}a_8a_3-\frac{3}{5}b_8a_3+\frac{8}{5}a_6a_3a_2+2b_6a_5\nn\\
&+8b_6a_3a_2+\frac{8}{5}a_5a_3^2+\frac{4}{5}a_4^2a_3-\frac{12}{5}a_4a_3a_2^2\nn\\
&-\frac{12}{5}a_3^3a_2+\frac{4}{5}a_3a_2^4\,,
\\
\xi^{\ell^2}_8&=\frac{3}{2}b_8-3b_6a_2\,,
\\
\xi^{\ell^2}_{10}&=-\frac{3}{4}b_8a_2+\frac{3}{2}b_6a_2^2\,,
\\
\xi^{\ell^2}_{11}&=-\frac{6}{5}b_8a_3+\frac{12}{5}b_6a_3a_2\,,
\end{align}

and
\be
\zeta=\left[\frac{3M}{8a_0(t_c-t)}\right]^{1/8}\,,\label{zetadef}
\ee
where $t_c$ is an integration constant at which $\zeta$ diverges. In our analysis, $t_c$ is picked such that at $t=0$ the system enters the LIGO band with $f=10$~Hz, where $f$ is the gravitational wave frequency, which is twice the orbital frequency. 

The coefficients $(a_i,b_i)$ are given in Appendix~\ref{app-coeffomegadot}.

\section{Precession Phase Coefficients}
\label{app-coeffphi}

The precession phase is given by Eq.~\eqref{phii-eq} where we have defined the nonzero coefficients
\begin{align}
\phi_{2,A} &= -3a_2+3\eta\frac{C^{\one}_A}{C^{\zero}_A}\,,
\\
\phi_{4,A} &= 3a_4-3a_2^2+3a_2\eta\frac{C^{\one}_A}{C^{\zero}_A}-3\eta^2\frac{C^{\two}_A}{C^{\zero}_A}\,,
\\
\phi_{5,A} &= \frac{3}{2}a_5-3a_3a_2+\frac{3}{2}a_3\eta \frac{C^{\one}_A}{C^{\zero}_A}\,,
\\
\phi_{6,A} &= a_6-b_6-2a_4 a_2-a^{2}_3+a_2^3+a_4\eta\frac{C^{\one}_A}{C^{\zero}_A}-a^2_2\eta\frac{C^{\one}_A}{C^{\zero}_A}\nn
\\
&+a_2\eta^2\frac{C^{\two}_A}{C^{\zero}_A}\,,
\\
\phi_{7,A} &= \frac{3}{4}a^{}_7-\frac{3}{2}a^{}_5a_2-\frac{3}{2}a^{}_4a^{}_3+\frac{9}{4}a^{}_3a_2^2+\frac{3}{4}a_5\eta\frac{C^{\one}_A}{C^{\zero}_A}\nn
\\
&-\frac{3}{2}a_3a_2\eta\frac{C^{\one}_A}{C^{\zero}_A}+\frac{3}{4}a_3\eta^2\frac{C^{\two}_A}{C^{\zero}_A}\,,
\\
\phi_{8,A} &= \frac{3}{5}a_8 - \frac{9}{25}b_8-\frac{6}{5}a_6a_2+\frac{18}{25}b_6a_2-\frac{6}{5}a_5a_3-\frac{3}{5}a_4^2\nn\\
&+\frac{9}{5}a_4a^2_2+\frac{9}{5}a_3^2a_2-\frac{3}{5}a_2^4+\frac{3}{5}a_6\eta\frac{C^{\one}_A}{C^{\zero}_A}-\frac{9}{25}b_6\eta\frac{C^{\one}_A}{C^{\zero}_A}\nn
\\
&-\frac{6}{5}a_4a_2\eta\frac{C^{\one}_A}{C^{\zero}_A}+\frac{3}{5}a_4\eta^2\frac{C^{\two}_A}{C^{\zero}_A}-\frac{3}{5}a^2_3\eta\frac{C^{\one}_A}{C^{\zero}_A}\nn
\\&+\frac{3}{5}a_2^3\eta\frac{C^{\one}_A}{C^{\zero}_A}-\frac{3}{5}a_2^2\eta^2\frac{C^{\two}_A}{C^{\zero}_A}\,,
\\
\phi^{\ell}_{3,A} &= 3a_3\,,
\\
\phi^{\ell}_{6,A} &= 3b_6\,,
\\
\phi^{\ell}_{8,A} &= \frac{9}{5}b_8-\frac{18}{5}b_6a_2+\frac{9}{5}b_6\eta\frac{C^{\one}_A}{C^{\zero}_A}\,,
\end{align}
where $A\in\{1,2\}$ and the constants $C_A^{\n}$ are given in Eqs.~\eqref{C0def}-\eqref{C2def}.

\section{Orbital Phase Coefficients}
\label{app-coeffPhic}

The orbital phase to 8PN order as a function of $\xi$ is given by Eq.~\eqref{phic}, where the nonzero PN coefficients are
\allowdisplaybreaks
\begin{align}
\Phi^{\orb}_{2}&=-\frac{5}{3}a_2\,,
\\
\Phi^{\orb}_{3}&=-\frac{5}{2}a_3\,,
\\
\Phi^{\orb}_{4}&=-5a_4+5a_2^2\,,
\\
\Phi^{\orb}_{6}&=5a_6-15b_6-10a_4a_2-5a_3^2+5a_2^3\,,
\\
\Phi^{\orb}_{7}&=\frac{5}{2}a_7-5a_5a_2-5a_4a_3+\frac{15}{2}a_3a_2^2\,,
\\
\Phi^{\orb}_{8}&=\frac{5}{3}a_8-\frac{5}{3}b_8-\frac{10}{3}a_6a_2+\frac{10}{3}b_6a_2-\frac{10}{3}a_5a_3\nn\\
&-\frac{5}{3}a_4^2+5a_4a_2^2+5a_3^2a_2-\frac{5}{3}a_2^4\,,
\end{align}
\allowdisplaybreaks
\begin{widetext}
\begin{align}
\Phi^{\orb}_{9}&=\frac{5}{4}a_9-\frac{15}{16}b_9-\frac{5}{2}a_7a_2-\frac{5}{2}a_6a_3+\frac{15}{8}b_6a_3-\frac{5}{2}a_5a_4+\frac{15}{4}a_5a_2^2+\frac{15}{2}a_4a_3a_2+\frac{5}{4}a_3^3-5a_3a_2^3\,,
\\
\Phi^{\orb}_{10}&=a_{10}-\frac{3}{5}b_{10}-2a_8a_2+\frac{6}{5}b_8a_2-2a_7a_3-2a_6a_4+3a_6a_2^2+\frac{6}{5}b_6a_4-\frac{9}{5}b_6a_2^2-a_5^2+6a_5a_3a_2+3a_4^2a_2+3a_4a_3^2\nn\\
&-4a_4a_2^3-6a_3^2a_2^2+a_2^5\,,
\\
\Phi^{\orb}_{11}&=+\frac{5}{6}a_{11}-\frac{5}{12}b_{11}-\frac{5}{3}a_9a_2+\frac{5}{6}b_9a_2-\frac{5}{3}a_8a_3+\frac{5}{6}b_8a_3-\frac{5}{3}a_7a_4+\frac{5}{2}a_7a_2^2-\frac{5}{3}a_6a_5+5a_6a_3a_2+\frac{5}{6}b_6a_5-\frac{5}{2}b_6a_3a_2\nn\\
&+5a_5a_4a_2+\frac{5}{2}a_5a_3^2-\frac{10}{3}a_5a_2^3+\frac{5}{2}a_4^2a_3-10a_4a_3a_2^2-\frac{10}{3}a_3^3a_2+\frac{25}{6}a_3a_2^4\,,
\\
\Phi^{\orb}_{12}&=-\frac{10}{7}a_{10}a_2+\frac{30}{49}b_{10}a_2-\frac{10}{7}a_9a_3+\frac{30}{49}b_9a_3-\frac{10}{7}a_8a_4+\frac{15}{7}a_8a_2^2+\frac{30}{49}b_8a_4-\frac{45}{49}b_8a_2^2-\frac{10}{7}a_7a_5+\frac{30}{7}a_7a_3a_2\nn\\
&-\frac{5}{7}a_6^2+\frac{30}{49}a_6b_6+\frac{30}{7}a_6a_4a_2+\frac{15}{7}a_6a_3^2-\frac{20}{7}a_6a_2^3-\frac{90}{343}b_6^2-\frac{90}{49}b_6a_4a_2-\frac{45}{49}b_6a_3^2+\frac{60}{49}b_6a_2^3+\frac{15}{7}a_5^2a_2\nn\\
&+\frac{30}{7}a_5a_4a_3-\frac{60}{7}a_5a_3a_2^2-\frac{30}{7}a_4^2a_2^2-\frac{60}{7}a_4a_3^2a_2+\frac{5}{7}a_4^3+\frac{25}{7}a_4a_2^4-\frac{5}{7}a_3^4+\frac{50}{7}a_3^2a_2^3-\frac{5}{7}a_2^6\,,
\\
\Phi^{\orb}_{13}&=-\frac{5}{4}a_{11}a_2+\frac{15}{32}b_{11}a_2-\frac{5}{4}a_{10}a_3+\frac{15}{32}b_{10}a_3-\frac{5}{4}a_9a_4+\frac{15}{8}a_9a_2^2+\frac{15}{32}b_9a_4-\frac{45}{64}b_9a_2^2-\frac{5}{4}a_8a_5+\frac{15}{4}a_8a_3a_2\nn\\
&+\frac{15}{32}b_8a_5-\frac{45}{32}b_8a_3a_2-\frac{5}{4}a_7a_6+\frac{15}{32}a_7b_6+\frac{15}{4}a_7a_4a_2+\frac{15}{8}a_7a_3^2-\frac{5}{2}a_7a_2^3+\frac{15}{4}a_6a_5a_2+\frac{15}{4}a_6a_4a_3\nn\\
&-\frac{15}{2}a_6a_3a_2^2-\frac{45}{32}b_6a_5a_2-\frac{45}{32}b_6a_4a_3+\frac{45}{16}b_6a_3a_2^2+\frac{15}{8}a_5a_4^2-\frac{15}{2}a_5a_4a_2^2-\frac{15}{2}a_5a_3^2a_2+\frac{15}{8}a_5^2a_3+\frac{25}{8}a_5a_2^4\nn\\
&-\frac{15}{2}a_4^2a_3a_2-\frac{5}{2}a_4a_3^3+\frac{25}{2}a_4a_3a_2^3+\frac{25}{4}a_3^3a_2^2-\frac{15}{4}a_3a_2^5\,,
\\
\Phi^{{\orb}}_{14}&=-\frac{10}{9}a_{11}a_3+\frac{10}{27}b_{11}a_3-\frac{10}{9}a_{10}a_4+\frac{5}{3}a_{10}a_2^2+\frac{10}{27}b_{10}a_4-\frac{5}{9}b_{10}a^2_2-\frac{10}{9}a_9a_5+\frac{10}{3}a_9a_3a_2+\frac{10}{27}b_9a_5\nn\\
&-\frac{10}{9}b_9a_3a_2-\frac{10}{9}a_8a_6+\frac{10}{27}a_8b_6+\frac{10}{3}a_8a_4a_2+\frac{5}{3}a_8a^2_3-\frac{20}{9}a_8a^3_2+\frac{10}{27}b_8a_6-\frac{10}{9}b_8a_4a_2-\frac{20}{81}b_8b_6-\frac{5}{9}b_8a^2_3\nn\\
&+\frac{20}{27}b_8a^3_2-\frac{5}{9}a_7^2+\frac{10}{3}a_7a_5a_2+\frac{10}{3}a_7a_4a_3-\frac{20}{3}a_7a_3a_2^2+\frac{5}{3}a_6^2a_2-\frac{10}{9}a_6b_6a_2+\frac{10}{3}a_6a_5a_3+\frac{5}{3}a_6a_4^2-\frac{20}{3}a_6a_4a_2^2\nn\\
&-\frac{20}{3}a_6a_3^2a_2+\frac{25}{9}a_6a_2^4+\frac{10}{27}b_6^2a_2-\frac{10}{9}b_6a_5a_3-\frac{5}{9}b_6a_4^2+\frac{20}{9}b_6a_4a_2^2+\frac{20}{9}b_6a_3^2a_2-\frac{25}{27}b_6a_2^4+\frac{5}{3}a_5^2a_4-\frac{10}{3}a_5^2a_2^2\nn\\
&-\frac{40}{3}a_5a_4a_3a_2-\frac{20}{9}a_5a_3^3+\frac{100}{9}a_5a_3a_2^3-\frac{20}{9}a_4^3a_2-\frac{10}{3}a_4^2a_3^2+\frac{50}{9}a_4^2a_2^3+\frac{50}{3}a_4a_3^2a_2^2-\frac{10}{3}a_4a_2^5+\frac{25}{9}a_3^4a_2\nn\\
&-\frac{25}{3}a_3^2a_2^4+\frac{5}{9}a_2^7\,,
\\
\Phi^{{\orb}}_{15}&=-a_{11}a_4+\frac{3}{2}a_{11}a^2_2+\frac{3}{10}b_{11}a_4-\frac{9}{20}b_{11}a^2_2-a_{10}a_5+3a_{10}a_3a_2+\frac{3}{10}b_{10}a_5-\frac{9}{10}b_{10}a_3a_2-a_9a_6\!+\!\frac{3}{10}a_9b_6\!+\!3a_9a_4a_2\nn\\
&+\frac{3}{2}a_9a^2_3-2a_9a^3_2-\frac{9}{10}b_9a_4a_2+\frac{6}{10}b_9a^3_2+\frac{3}{10}b_9a_6-\frac{9}{50}b_9b_6-\frac{9}{20}b_9a^2_3-a_8a_7+3a_8a_5a_2+3a_8a_4a_3-6a_8a_3a_2^2\nn\\
&+\frac{3}{10}b_8a_7-\frac{9}{10}b_8a_5a_2-\frac{9}{10}b_8a_4a_3+\frac{9}{5}b_8a_3a^2_2+3a_7a_6a_2-\frac{9}{10}a_7b_6a_2+3a_7a_5a_3+\frac{3}{2}a_7a_4^2-6a_7a_4a_2^2-6a_7a_3^2a_2\nn\\
&+\frac{5}{2}a_7a_2^4+\frac{3}{2}a_6^2a_3-\frac{9}{10}a_6b_6a_3+3a_6a_5a_4-6a_6a_5a_2^2-12a_6a_4a_3a_2-2a_6a_3^3+10a_6a_3a_2^3+\frac{27}{100}b_6^2a_3-\frac{9}{10}b_6a_5a_4\nn\\
&+\frac{9}{5}b_6a_5a_2^2+\frac{18}{5}b_6a_4a_3a_2+\frac{3}{5}b_6a_3^3-3b_6a_3a_2^3+\frac{1}{2}a_5^3-6a_5^2a_3a_2-6a_5a_4^2a_2-6a_5a_4a_3^2+10a_5a_4a_2^3+15a_5a_3^2a_2^2\nn\\
&-3a_5a_2^5-2a_4^3a_3+15a_4^2a_3a_2^2+10a_4a_3^3a_2-15a_4a_3a_2^4+\frac{1}{2}a_3^5-10a_3^3a_2^3+\frac{7}{2}a_3a_2^6\,,
\\
\Phi^{\orb}_{16}&=-\frac{10}{11}a_{11}a_5+\frac{30}{11}a_{11}a_3a_2+\frac{30}{121}b_{11}a_5-\frac{90}{121}b_{11}a_3a_2-\frac{10}{11}a_{10}a_6+\frac{30}{121}a_{10}b_6+\frac{30}{11}a_{10}a_4a_2+\frac{15}{11}a_{10}a^2_3-\frac{20}{11}a_{10}a^3_2\nn\\
&+\frac{30}{121}b_{10}a_6-\frac{180}{1331}b_{10}b_6-\frac{90}{121}b_{10}a_4a_2-\frac{45}{121}b_{10}a^2_3+\frac{60}{121}b_{10}a^3_2-\frac{10}{11}a_9a_7+\frac{30}{11}a_9a_5a_2+\frac{30}{11}a_9a_4a_3-\frac{60}{11}a_9a_3a_2^2\nn\\
&+\frac{30}{121}b_9a_7-\frac{90}{121}b_9a_5a_2-\frac{90}{121}b_9a_4a_3+\frac{180}{121}b_9a_3a_2^2-\frac{5}{11}a^2_8+\frac{30}{11}a_8b_8+\frac{30}{11}a_8a_6a_2-\frac{90}{121}a_8b_6a_2+\frac{30}{11}a_8a_5a_3\nn\\
&+\frac{15}{11}a_8a^2_4-\frac{60}{11}a_8a_4a^2_2-\frac{60}{11}a_8a^2_3a_2+\frac{25}{11}a_8a^4_2-\frac{90}{1331}b^2_8-\frac{90}{121}b_8a_6a_2+\frac{540}{1331}b_8b_6a_2-\frac{90}{121}b_8a_5a_3-\frac{45}{121}b_8a^2_4\nn\\
&+\frac{180}{121}b_8a_4a^2_2+\frac{180}{121}b_8a_3^2a_2-\frac{75}{121}b_8a^4_2+\frac{15}{11}a_7^2a_2+\frac{30}{11}a_7a_6a_3-\frac{90}{121}a_7b_6a_3+\frac{30}{11}a_7a_5a_4-\frac{60}{11}a_7a_5a_2^2\nn\\
&-\frac{120}{11}a_7a_4a_3a_2-\frac{20}{11}a_7a_3^3+\frac{100}{11}a_7a_3a_2^3+\frac{15}{11}a_6^2a_4-\frac{30}{11}a_6^2a_2^2-\frac{90}{121}a_6b_6a_4+\frac{180}{121}a_6b_6a_2^2+\frac{15}{11}a_6a_5^2\nn\\
&-\frac{120}{11}a_6a_5a_3a_2-\frac{60}{11}a_6a_4^2a_2-\frac{60}{11}a_6a_4a_3^2+\frac{100}{11}a_6a_4a_2^3+\frac{150}{11}a_6a_3^2a_2^2-\frac{30}{11}a_6a_2^5+\frac{270}{1331}b_6^2a_4-\frac{540}{1331}b_6^2a_2^2\nn\\
&-\frac{45}{121}b_6a_5^2+\frac{360}{11}b_6a_5a_3a_2+\frac{180}{121}b_6a_4^2a_2+\frac{180}{121}b_6a_4a_3^2-\frac{300}{121}b_6a_4a_2^3-\frac{450}{121}b_6a_3^2a_2^2+\frac{90}{121}b_6a_2^5-\frac{60}{11}a_5^2a_4a_2\nn\\
&-\frac{30}{11}a_5^2a_3^2+\frac{50}{11}a_5^2a_2^3-\frac{60}{11}a_5a_4^2a_3+\frac{300}{11}a_5a_4a_3a_2^2+\frac{100}{11}a_5a_3^3a_2-\frac{150}{11}a_5a_3a_2^4-\frac{5}{11}a_4^4+\frac{50}{11}a_4^3a_2^2+\frac{150}{11}a_4^2a_3^2a_2\nn\\
&-\frac{75}{11}a_4^2a_2^4+\frac{25}{11}a_4a_3^4-\frac{300}{11}a_4a_3^2a_2^3+\frac{35}{11}a_4a_2^6-\frac{75}{11}a_3^4a_2^2+\frac{105}{11}a_3^2a_2^5-\frac{5}{11}a_2^8\,,
\end{align}
\end{widetext}
\allowdisplaybreaks
\begin{align}
\Phi^{\orb,\ell}_5&=5a_5-10a_3a_2\,,
\\
\Phi^{\orb,\ell}_6&=15b_6\,,
\\
\Phi^{\orb,\ell}_8&=5b_8-10b_6a_2\,,
\\
\Phi^{\orb,\ell}_9&=\frac{15}{4}b_9-\frac{15}{2}b_6a_3\,,
\\
\Phi^{\orb,\ell}_{10}&=3b_{10}-6b_8a_2-6b_6a_4+9b_6a_2^2\,,
\\
\Phi^{\orb,\ell}_{11}&=\frac{5}{2}b_{11}-5b_9a_2-5b_8a_3-5b_6a_5+15b_6a_3a_2\,,
\\
&-\frac{30}{7}a_6b_6+\frac{90}{49}b_6^2+\frac{90}{7}b_6a_4a_2+\frac{45}{7}b_6a_3^2-\frac{60}{7}b_6a_2^3\,,
\\
\Phi^{\orb,\ell}_{12}&=-\frac{30}{7}b_{10}a_2-\frac{30}{7}b_9a_3-\frac{30}{7}b_8a_4+\frac{45}{7}b_8a_2^2\nn
\\
\Phi^{\orb,\ell}_{13}&=-\frac{15}{4}b_{11}a_2-\frac{15}{4}b_{10}a_3-\frac{15}{4}b_9a_4+\frac{45}{8}b_9a_2^2\nn
\\&-\frac{15}{4}b_8a_5+\frac{45}{4}b_8a_3a_2-\frac{15}{4}a_7b_6+\frac{45}{4}b_6a_4a_3\nn
\\&+\frac{45}{4}b_6a_5a_2-\frac{45}{2}b_6a_3a_2^2\,,
\end{align}
\begin{widetext}
\begin{align}
\Phi^{\orb,\ell}_{14}&=-\frac{10}{3}b_{10}a_45b_{10}a_2a_2-\frac{10}{3}b_9a_5+10b_9a_3a_2-\frac{10}{3}a_8b_6-\frac{10}{3}b_8a_6+\frac{20}{9}b_8b_6+10b_8a_4a_2+5b_8a^2_3-\frac{20}{3}b_8a_2^3\nn\\
&-\frac{10}{3}b_{11}a_3+10a_6b_6a_2-\frac{10}{3}b_6^2a_2+10b_6a_5a_3+5b_6a_4^2-20b_6a_4a_2^2-20b_6a_3^2a_2+\frac{25}{3}b_6a_2^4\,,
\\
\Phi^{\orb,\ell}_{15}&=-3b_{11}a_4+\frac{9}{2}b_{11}a^2_2-3b_{10}a_5+9b_{10}a_3a_2-3a_9b_6-3b_9a_6+\frac{9}{5}b_9b_6+9b_9a_4a_2+\frac{9}{2}b_9a^2_3-6b_9a^3_2\!-\!3b_8a_7\!+\!9b_8a_5a_2\nn\\
&+9b_8a_4a_3-18b_8a_3a^2_2+9a_7b_6a_2+9a_6b_6a_3-\frac{27}{10}b_6^2a_3+9b_6a_5a_4-18b_6a_5a_2^2-36b_6a_4a_3a_2-6b_6a_3^3+30b_6a_3a_2^3\,,
\\
\Phi^{\orb,\ell}_{16}&=-\frac{30}{11}b_{11}a_5+\frac{90}{11}b_{11}a_3a_2+\frac{45}{11}b_{10}a^2_3-\frac{60}{11}b_{10}a_2^3+\frac{180}{121}b_{10}b_6-\frac{30}{11}a_{10}b_6-\frac{30}{11}b_{10}a_6+\frac{90}{11}b_{10}a_4a_+\frac{90}{11}a_9b_6a_22\nn\\
&-\frac{30}{11}b_9a_7+\frac{90}{11}b_9a_5a_2+\frac{90}{11}b_9a_4a_3-\frac{180}{11}b_9a_3a^2_2-\frac{30}{11}a_8b_8+\frac{90}{121}b^2_8+\frac{90}{11}b_8a_6a_2-\frac{540}{121}b_8b_6a_2+\frac{90}{11}b_8a_5a_3\nn\\
&+\frac{45}{11}b_8a^2_4-\frac{180}{11}b_8a_4a^2_2-\frac{180}{11}b_8a^2_3a_2+\frac{75}{11}b_8a^4_2+\frac{90}{11}a_7b_6a_3+\frac{90}{11}a_6b_6a_4-\frac{180}{11}a_6b_6a_2^2-\frac{270}{121}b_6^2a_4+\frac{540}{121}b_6^2a_2^2\nn\\
&+\frac{45}{11}b_6a_5^2-\frac{360}{11}b_6a_5a_3a_2-\frac{180}{11}b_6a_4^2a_2-\frac{180}{11}b_6a_4a_3^2+\frac{300}{11}b_6a_4a_2^3+\frac{450}{11}b_6a_3^2a_2^2-\frac{90}{11}b_6a_2^5\,,
\end{align}
\end{widetext}
\allowdisplaybreaks
\begin{align}
\Phi^{\orb,\ell^{2}}_{12}&=-\frac{45}{7}b_6^2\,,
\\
\Phi^{\orb,\ell^{2}}_{14}&=15b_6^2a_2-10b_8b_6\,,
\\
\Phi^{\orb,\ell^{2}}_{15}&=\frac{27}{2}b_6^2a_3-9b_9b_6\,,
\\
\Phi^{\orb,\ell^{2}}_{16}&=-\frac{90}{11}b_{10}b_6-\frac{45}{11}b_8^2+\frac{270}{11}b_8b_6a_2+\frac{135}{11}b_6^2a_4\nn\\
&-\frac{270}{11}b_6^2a_2^2\,.
\end{align}
The constant of integration $\Phi^\orb_0$ is calculated by solving $\Phi^\orb(t=0)=0$ and the radiation reaction coefficients $(a_n,b_n)$ are given in Appendix~\ref{app-coeffomegadot}.

\section{Second Time Derivative of Phases}
\label{app-ddotphases}

In this appendix, we provide the second time derivatives of different phases that go into the Fourier amplitude of Eq.~\eqref{Fourier-Amplitude}. The second time derivative of the orbital phase as a function of $\xi$ is 
\be
\ddot{\Phi}^\orb =3\frac{\xi^2\dot{\xi}}{M}=\frac{a_0}{M^2}\xi^{11}\left[1+\sum_{i=2}^{11} (a_i+3b_i \ln{\xi}) \xi^i\right]\label{ddot-Phi}\,,
\ee
where the $(a_{i},b_{i})$ coefficients are given in Appendix~\ref{app-coeffomegadot}. The second time derivative of the Thomas phase is, at leading order in the spin parameters,
\begin{align}
\delta\ddot{\phi}&=\frac{\bm{\hat{L}}\cdot\bm{\hat{N}}}{1-\left(\bm{\hat{L}}\cdot\bm{\hat{N}}\right)^2}
\left(\bm{\hat{L}}\times\bm{\hat{N}}\right) \cdot \ddot{\bm{\hat{L}}}
\,,
\label{ddot-phi}
\end{align}
while the second time derivative of the inclination and polarization angle is, at leading order in the spin parameters,
\be
\ddot{\iota}=-\frac{\ddot{\bm{\hat{L}}}\cdot\bm{\hat{N}}}{\left[1-\left(\bm{\hat{L}}\cdot\bm{\hat{N}}\right)^2\right]^{1/2}}\,,
\label{ddot-i}
\ee
and the polarization phase is, at leading order in the spin parameters,
\begin{widetext}
\begin{align}
\ddot{\psi}&=\frac{\bm{\hat{N}}\cdot(\bm{\hat{L}}\times\bm{\hat{z}})\left[\ddot{\bm{\hat{L}}}\cdot\bm{\hat{z}}-\left(\ddot{\bm{\hat{L}}}\cdot\bm{\hat{N}}\right)(\bm{\hat{z}}\cdot\bm{\hat{N}})\right]-\bm{\hat{N}}\cdot\left(\ddot{\bm{\hat{L}}}\times\bm{\hat{z}}\right)\left[\bm{\hat{L}}\cdot\bm{\hat{z}}-(\bm{\hat{L}}\cdot\bm{\hat{N}})(\bm{\hat{z}}\cdot\bm{\hat{N}})\right]}{\left[\bm{\hat{N}}\cdot(\bm{\hat{L}}\times\bm{\hat{z}})\right]^2+\left[\bm{\hat{L}}\cdot\bm{\hat{z}}-(\bm{\hat{L}}\cdot\bm{\hat{N}})(\bm{\hat{z}}\cdot\bm{\hat{N}})\right]^2}\,.
\label{ddot-psi}
\end{align}
\end{widetext}
Equations~\eqref{ddot-phi}-\eqref{ddot-psi} depend on derivatives of the unit vector angular momentum. The second time derivative $\ddot{\bm{\hat{L}}}$ is computed from the analytical solution to the angular momentum in Eqs.~\eqref{Lxfinal}-\eqref{Lzfinal}. 

To first order in the spins, and ignoring radiation reaction, since these terms are subdominant, the above equations simplify to
\begin{widetext}
\begin{align}
\delta\ddot{\phi}&=\frac{N_{z,s}}{1-N_{z,s}^2} \frac{C_{1}^{2} \xi^{11} \eta}{M^{2}} 
\left\{\left[S_{1,y}(0)N_{x,s}-S_{1,x}(0)N_{y,s}\right]\cos{\phi_1}
+\left[S_{1,x}(0)N_{x,s} + S_{1,y}(0)N_{y,s} \right] \sin{\phi_1} \right\} + 1 \leftrightarrow 2\,,
\label{phiddot-lin-S}
\\
\ddot{\iota}&=-\frac{1}{\sqrt{1-N_{z,s}^2}} \frac{C_{1}^{2} \xi^{11} \eta}{M^{2}} 
\left\{\left[S_{1,x}(0)N_{x,s}+S_{1,y}(0)N_{y,s}\right] \cos{\phi_1}
-\left[S_{1,y}(0)N_{x,s}-S_{1,x}(0)N_{y,s}\right] \sin{\phi_1}\right\} +1 \leftrightarrow 2 \label{iddot-lin-S}\,,
\\
\ddot{\psi}&=\frac{\left(N_{x,s}\sin{\theta_0}-N_{z,s}\cos{\theta_0}\right)^2-1}{N_{y,s}^2\sin^2{\theta_0}+\left[N_{x,s}N_{z,s}\sin{\theta_0}+(1-N_{z,s}^2)\cos{\theta_0}\right]^2}
 \frac{C_{1}^{2} \xi^{11} \eta}{M^{2}} 
\left\{\left[S_{1,y}(0)N_{x,s}-S_{1,x}(0)N_{y,s}\right] \cos{\phi_1} 
\right. 
\nn \\
&+ \left.
\left[S_{1,x}(0)N_{x,s}\!+\!S_{1,y}(0)N_{y,s}\right] \sin{\phi_1}\right\} +1 \leftrightarrow 2 \,,
\label{psiddot-lin-S}
\end{align}
\end{widetext}
where $\bm{\hat{N}_s}=(N_{x,s},N_{y,s},N_{z,s})$ is the line-of-sight vector in the source frame and $\theta_0$ is the polar angle of the total angular momentum in the detector frame.

Finally, we can ignore $\ddot{\Phi}^{\mbox{\tiny log}}$ since it is proportional to $\xi^{19}$ and truly negligible.

\section{Time-Domain Waveform Amplitudes}
\label{app-gwamp}

In this appendix, we provide explicit expressions for the 2.5PN term of the amplitude coefficients ${\cal{A}}_{n,k,m}$ that parametrize the response function in Eqs.~\eqref{full-h-of-t}-\eqref{hnkm}. This term is to be added to the expressions given in Appendix E of \cite{Klein:2013qda}, that are valid to 2PN order. If we define 
\begin{align}
{\cal{A}}_F&\equiv\frac{1}{2}(1+\cos^2{\theta})\cos{2\phi}\,,\\
{\cal{B}}_F&\equiv\cos{\theta}\sin{2\phi}\,,
\end{align}
and use~\cite{Arun:2004ff} to compute the amplitudes ${\cal{A}}_{n,k,m}$, we find at 2.5PN order
\allowdisplaybreaks
\begin{widetext}
\begin{align}
{\cal{A}}_{1,1,2}&=-{\cal{A}}_{1,-1,2}=-{\cal{A}}^*_{1,1,-2}={\cal{A}}^*_{1,-1,-2}=\delta M({\cal{B}}_F-i{\cal{A}}_F)\left(\frac{262901}{3932160}+\frac{392045}{589824}\eta-\frac{188615}{2359296}\eta^2\right)\xi^5\,,
\\
{\cal{A}}_{1,2,2}&=-{\cal{A}}_{1,-2,2}={\cal{A}}^*_{1,2,-2}=-{\cal{A}}^*_{1,-2,-2}=\delta M({\cal{B}}_F-i{\cal{A}}_F)\left(\frac{1527}{327680}-\frac{52195}{147456}\eta+\frac{16105}{589824}\eta^2\right)\xi^5\,,
\\
{\cal{A}}_{1,3,2}&=-{\cal{A}}_{1,-3,2}=-{\cal{A}}^*_{1,3,-2}={\cal{A}}^*_{1,-3,-2}=-\delta M({\cal{B}}_F-i{\cal{A}}_F)\left(\frac{75691}{3932160}+\frac{209}{196608}\eta-\frac{10739}{786432}\eta^2\right)\xi^5\,,
\\
{\cal{A}}_{1,4,2}&=-{\cal{A}}_{1,-4,2}={\cal{A}}^*_{1,4,-2}=-{\cal{A}}^*_{1,-4,-2}=-\delta M({\cal{B}}_F-i{\cal{A}}_F)\left(\frac{1249}{245760}-\frac{463}{36864}\eta+\frac{581}{147456}\eta^2\right)\xi^5\,,
\\
{\cal{A}}_{1,5,2}&=-{\cal{A}}_{1,-5,2}=-{\cal{A}}^*_{1,5,-2}={\cal{A}}^*_{1,-5,-2}=\delta M({\cal{B}}_F-i{\cal{A}}_F)\left(\frac{863}{2359296}-\frac{415}{589824}\eta-\frac{35}{2359296}\eta^2\right)\xi^5\,,
\\
{\cal{A}}_{1,6,2}&=-{\cal{A}}_{1,-6,2}={\cal{A}}^*_{1,6,-2}=-{\cal{A}}^*_{1,-6,-2}=\delta M({\cal{B}}_F-i{\cal{A}}_F)\frac{7}{589824}(\eta-1)(3\eta-1)\xi^5\,,
\\
{\cal{A}}_{1,7,2}&=-{\cal{A}}_{1,-7,2}=-{\cal{A}}^*_{1,7,-2}={\cal{A}}^*_{1,-7,-2}=-\delta M({\cal{B}}_F-i{\cal{A}}_F)\frac{1}{2359296}(\eta-1)(3\eta-1)\xi^5\,,
\\
{\cal{A}}_{2,0,2}&=({\cal{A}}_F+i{\cal{B}}_F)\left[\frac{91}{24}\pi-\frac{9}{8}\pi\eta-\left(\frac{1}{16}+\frac{163}{20}\eta\right)i\right]\xi^5\,,
\\
{\cal{A}}_{2,1,2}&={\cal{A}}_{2,-1,2}=-({\cal{A}}_F+i{\cal{B}}_F)\left[\frac{7}{3}\pi-\frac{1}{6}\pi\eta-\left(\frac{13}{40}+\frac{167}{40}\eta\right)i\right]\xi^5\,,
\\
{\cal{A}}_{2,2,2}&={\cal{A}}_{2,-2,2}=({\cal{A}}_F+i{\cal{B}}_F)\left[\frac{7}{24}\pi+\frac{5}{6}\pi\eta-\left(\frac{21}{40}-\frac{9}{10}\eta\right)i\right]\xi^5\,,
\\
{\cal{A}}_{2,3,2}&={\cal{A}}_{2,-3,2}=({\cal{A}}_F+i{\cal{B}}_F)\left[\frac{1}{6}\pi-\frac{1}{2}\pi\eta+\left(\frac{11}{40}-\frac{47}{40}\eta\right)i\right]\xi^5\,,
\\
{\cal{A}}_{2,4,2}&={\cal{A}}_{2,-4,2}=-({\cal{A}}_F+i{\cal{B}}_F)\left[\frac{1}{48}\pi-\frac{1}{16}\pi\eta+\left(\frac{7}{160}-\frac{7}{40}\eta\right)i\right]\xi^5\,,
\\
{\cal{A}}_{2,0,-2}&=({\cal{A}}_F-i{\cal{B}}_F)\left[\frac{91}{24}\pi-\frac{9}{8}\pi\eta-\left(\frac{1}{16}+\frac{163}{20}\eta\right)i\right]\xi^5\,,
\\
{\cal{A}}_{2,1,-2}&={\cal{A}}_{2,-1,-2}=({\cal{A}}_F-i{\cal{B}}_F)\left[\frac{7}{3}\pi-\frac{1}{6}\pi\eta-\left(\frac{13}{40}+\frac{167}{40}\eta\right)i\right]\xi^5\,,
\\
{\cal{A}}_{2,2,-2}&={\cal{A}}_{2,-2,-2}=({\cal{A}}_F-i{\cal{B}}_F)\left[\frac{7}{24}\pi+\frac{5}{6}\pi\eta-\left(\frac{21}{40}-\frac{9}{10}\eta\right)i\right]\xi^5\,,
\\
{\cal{A}}_{2,3,-2}&={\cal{A}}_{2,-3,-2}=-({\cal{A}}_F-i{\cal{B}}_F)\left[\frac{1}{6}\pi-\frac{1}{2}\pi\eta+\left(\frac{11}{40}-\frac{47}{40}\eta\right)i\right]\xi^5\,,
\\
{\cal{A}}_{2,4,-2}&={\cal{A}}_{2,-4,-2}=-({\cal{A}}_F-i{\cal{B}}_F)\left[\frac{1}{48}\pi-\frac{1}{16}\pi\eta+\left(\frac{7}{160}-\frac{7}{40}\eta\right)i\right]\xi^5\,,
\\
{\cal{A}}_{3,1,2}&=-{\cal{A}}_{3,-1,2}=-{\cal{A}}^*_{3,1,-2}={\cal{A}}^*_{3,-1,-2}=\delta M({\cal{B}}_F-i{\cal{A}}_F)\left(\frac{645381}{1310720}-\frac{275649}{65536}\eta+\frac{285123}{262144}\eta^2\right)\xi^5\,,
\\
{\cal{A}}_{3,2,2}&=-{\cal{A}}_{3,-2,2}={\cal{A}}^*_{3,2,-2}=-{\cal{A}}^*_{3,-2,-2}=-\delta M({\cal{B}}_F-i{\cal{A}}_F)\left(\frac{15417}{327680}-\frac{39333}{16384}\eta+\frac{29103}{65536}\eta^2\right)\xi^5\,,
\\
{\cal{A}}_{3,3,2}&=-{\cal{A}}_{3,-3,2}=-{\cal{A}}^*_{3,3,-2}={\cal{A}}^*_{3,-3,-2}=-\delta M({\cal{B}}_F-i{\cal{A}}_F)\left(\frac{544779}{1310720}-\frac{174891}{327680}\eta+\frac{436353}{1310720}\eta^2\right)\xi^5\,,
\\
{\cal{A}}_{3,4,2}&=-{\cal{A}}_{3,-4,2}={\cal{A}}^*_{3,4,-2}=-{\cal{A}}^*_{3,-4,-2}=\delta M({\cal{B}}_F-i{\cal{A}}_F)\left(\frac{22437}{81920}-\frac{13941}{20480}\eta+\frac{17703}{81920}\eta^2\right)\xi^5\,,
\\
{\cal{A}}_{3,5,2}&=-{\cal{A}}_{3,-5,2}=-{\cal{A}}^*_{3,5,-2}={\cal{A}}^*_{3,-5,-2}=-\delta M({\cal{B}}_F-i{\cal{A}}_F)\left(\frac{57591}{1310720}-\frac{27351}{327680}\eta-\frac{4347}{1310720}\eta^2\right)\xi^5\,,
\\
{\cal{A}}_{3,6,2}&=-{\cal{A}}_{3,-6,2}={\cal{A}}^*_{3,6,-2}=-{\cal{A}}^*_{3,-6,-2}=-\delta M({\cal{B}}_F-i{\cal{A}}_F)\frac{1701}{327680}(\eta-1)(3\eta-1)\xi^5\,,
\\
{\cal{A}}_{3,7,2}&=-{\cal{A}}_{3,-7,2}=-{\cal{A}}^*_{3,7,-2}={\cal{A}}^*_{3,-7,-2}=\delta M({\cal{B}}_F-i{\cal{A}}_F)\frac{729}{1310720}(\eta-1)(3\eta-1)\xi^5\,,
\\
{\cal{A}}_{4,0,2}&=-({\cal{A}}_F+i{\cal{B}}_F)\left[\frac{5}{3}\pi-5\pi\eta+\left(\frac{7}{2}-\frac{1193}{96}\eta-\frac{10}{3}\ln{2}+10\eta \ln{2} \right) i\right]\xi^5\,,
\\
{\cal{A}}_{4,1,2}&={\cal{A}}_{4,-1,2}=({\cal{A}}_F+i{\cal{B}}_F)\left[\frac{2}{3}\pi-2\pi \eta+\left(\frac{7}{5}-\frac{1193}{240}\eta-\frac{4}{3}\ln{2}+4 \eta\ln{2}\right) i\right]\xi^5\,,
\\
{\cal{A}}_{4,2,2}&={\cal{A}}_{4,-2,2}=({\cal{A}}_F+i{\cal{B}}_F)\left[\frac{2}{3}\pi-2\pi \eta+\left(\frac{7}{5}-\frac{1193}{240}\eta-\frac{4}{3}\ln{2}+4\eta\ln{2} \right) i\right]\xi^5\,,
\\
{\cal{A}}_{4,3,2}&={\cal{A}}_{4,-3,2}=-({\cal{A}}_F+i{\cal{B}}_F)\left[\frac{2}{3}\pi-2\pi \eta+\left(\frac{7}{5}-\frac{1193}{240}\eta-\frac{4}{3}\ln{2}+4\eta\ln{2}\right) i\right]\xi^5\,,
\\
{\cal{A}}_{4,4,2}&={\cal{A}}_{4,-4,2}=({\cal{A}}_F+i{\cal{B}}_F)\left[\frac{1}{6}\pi-\frac{1}{2}\pi \eta+\left(\frac{7}{20}-\frac{1193}{960}\eta-\frac{1}{3}\ln{2}+\eta\ln{2} \right) i\right]\xi^5\,,
\\
{\cal{A}}_{4,0,-2}&=-({\cal{A}}_F-i{\cal{B}}_F)\left[\frac{5}{3}\pi-5\pi\eta+\left(\frac{7}{2}-\frac{1193}{96}\eta-\frac{10}{3}\ln{2}+10 \eta\ln{2} \right) i\right]\xi^5\,,
\\
{\cal{A}}_{4,1,-2}&={\cal{A}}_{4,-1,-2}=-({\cal{A}}_F-i{\cal{B}}_F)\left[\frac{2}{3}\pi-2\pi \eta+\left(\frac{7}{5}-\frac{1193}{240}\eta-\frac{4}{3}\ln{2}+4\eta\ln{2} \right) i\right]\xi^5\,,
\\
{\cal{A}}_{4,2,-2}&={\cal{A}}_{4,-2,-2}=({\cal{A}}_F-i{\cal{B}}_F)\left[\frac{2}{3}\pi-2\pi \eta+\left(\frac{7}{5}-\frac{1193}{240}\eta-\frac{4}{3}\ln{2}+4 \eta\ln{2}\right) i\right]\xi^5\,,
\\
{\cal{A}}_{4,3,-2}&={\cal{A}}_{4,-3,-2}=({\cal{A}}_F-i{\cal{B}}_F)\left[\frac{2}{3}\pi-2\pi \eta+\left(\frac{7}{5}-\frac{1193}{240}\eta-\frac{4}{3}\ln{2}+4 \eta\ln{2}\right) i\right]\xi^5\,,
\\
{\cal{A}}_{4,4,-2}&={\cal{A}}_{4,-4,-2}=({\cal{A}}_F-i{\cal{B}}_F)\left[\frac{1}{6}\pi-\frac{1}{2}\pi \eta+\left(\frac{7}{20}-\frac{1193}{960}\eta-\frac{1}{3}\ln{2}+\eta\ln{2} \right) i\right]\xi^5\,,
\\
{\cal{A}}_{5,1,2}&=-{\cal{A}}_{5,-1,2}=-{\cal{A}}^*_{5,1,-2}={\cal{A}}^*_{5,-1,-2}=-\delta M({\cal{B}}_F-i{\cal{A}}_F)\left(\frac{1894375}{786432}-\frac{3723125}{589824}\eta+\frac{5569375}{2359296}\eta^2\right)\xi^5\,,
\\
{\cal{A}}_{5,2,2}&=-{\cal{A}}_{5,-2,2}={\cal{A}}^*_{5,2,-2}=-{\cal{A}}^*_{5,-2,-2}=\delta M({\cal{B}}_F-i{\cal{A}}_F)\left(\frac{749375}{589824}-\frac{469375}{147456}\eta+\frac{608125}{589824}\eta^2\right)\xi^5\,,
\\
{\cal{A}}_{5,3,2}&=-{\cal{A}}_{5,-3,2}=-{\cal{A}}^*_{5,3,-2}={\cal{A}}^*_{5,-3,-2}=\delta M({\cal{B}}_F-i{\cal{A}}_F)\left(\frac{1506875}{2359296}-\frac{1086875}{589824}\eta+\frac{686875}{786432}\eta^2\right)\xi^5\,,
\\
{\cal{A}}_{5,4,2}&=-{\cal{A}}_{5,-4,2}={\cal{A}}^*_{5,4,-2}=-{\cal{A}}^*_{5,-4,-2}=-\delta M({\cal{B}}_F-i{\cal{A}}_F)\left(\frac{101875}{147456}-\frac{66875}{36864}\eta+\frac{100625}{147456}\eta^2\right)\xi^5\,,
\\
{\cal{A}}_{5,5,2}&=-{\cal{A}}_{5,-5,2}=-{\cal{A}}^*_{5,5,-2}={\cal{A}}^*_{5,-5,-2}=\delta M({\cal{B}}_F-i{\cal{A}}_F)\left(\frac{254375}{2359296}-\frac{38125}{196608}\eta-\frac{56875}{2359296}\eta^2\right)\xi^5\,,
\\
{\cal{A}}_{5,6,2}&=-{\cal{A}}_{5,-6,2}={\cal{A}}^*_{5,6,-2}=-{\cal{A}}^*_{5,-6,-2}=\delta M({\cal{B}}_F-i{\cal{A}}_F)\frac{21875}{589824}(\eta-1)(3\eta-1)\xi^5\,,
\\
{\cal{A}}_{5,7,2}&=-{\cal{A}}_{5,-7,2}=-{\cal{A}}^*_{5,7,-2}={\cal{A}}^*_{5,-7,-2}=-\delta M({\cal{B}}_F-i{\cal{A}}_F)\frac{15625}{2359296}(\eta-1)(3\eta-1)\xi^5\,,
\\
{\cal{A}}_{7,1,2}&=-{\cal{A}}_{7,-1,2}=-{\cal{A}}^*_{7,1,-2}={\cal{A}}^*_{7,-1,-2}=\delta M({\cal{B}}_F-i{\cal{A}}_F)\frac{117649}{262144}(\eta-1)(3\eta-1)\xi^5\,,
\\
{\cal{A}}_{7,2,2}&=-{\cal{A}}_{7,-2,2}={\cal{A}}^*_{7,2,-2}=-{\cal{A}}^*_{7,-2,-2}=-\delta M({\cal{B}}_F-i{\cal{A}}_F)\frac{117649}{589824}(\eta-1)(3\eta-1)\xi^5\,,
\\
{\cal{A}}_{7,3,2}&=-{\cal{A}}_{7,-3,2}=-{\cal{A}}^*_{7,3,-2}={\cal{A}}^*_{7,-3,-2}=-\delta M({\cal{B}}_F-i{\cal{A}}_F)\frac{2235331}{11796480}(\eta-1)(3\eta-1)\xi^5\,,
\\
{\cal{A}}_{7,4,2}&=-{\cal{A}}_{7,-4,2}={\cal{A}}^*_{7,4,-2}=-{\cal{A}}^*_{7,-4,-2}=\delta M({\cal{B}}_F-i{\cal{A}}_F)\frac{117649}{737280}(\eta-1)(3\eta-1)\xi^5\,,
\\
{\cal{A}}_{7,5,2}&=-{\cal{A}}_{7,-5,2}=-{\cal{A}}^*_{7,5,-2}={\cal{A}}^*_{7,-5,-2}=\delta M({\cal{B}}_F-i{\cal{A}}_F)\frac{117649}{11796480}(\eta-1)(3\eta-1)\xi^5\,,
\\
{\cal{A}}_{7,6,2}&=-{\cal{A}}_{7,-6,2}={\cal{A}}^*_{7,6,-2}=-{\cal{A}}^*_{7,-6,-2}=-\delta M({\cal{B}}_F-i{\cal{A}}_F)\frac{117649}{2949120}(\eta-1)(3\eta-1)\xi^5\,,
\\
{\cal{A}}_{7,7,2}&=-{\cal{A}}_{7,-7,2}=-{\cal{A}}^*_{7,7,-2}={\cal{A}}^*_{7,-7,-2}=\delta M({\cal{B}}_F-i{\cal{A}}_F)\frac{117649}{11796480}(\eta-1)(3\eta-1)\xi^5\,.
\end{align}
\end{widetext}

%
\section{Coefficients in the Time-Frequency Inversion}
\label{app-tofxi}

The evolution for the PN expansion parameter $\xi$ as given in Eq.~\eqref{xidot} can be solved to obtain a relation between time and $\xi$
\begin{align}
t&=-\frac{3}{8}\frac{M}{a_0\xi^8}\left[1+\!\sum_{i=2}^{16} t_i \xi^i+\!\sum_{i=6}^{16} t^{\ell}_i \xi^i \ln{\xi}+\!\sum_{i=8}^{16} t^{\ell^{2}}_i \xi^i(\ln{\xi})^2\right]\nn\\
&+{\cal{O}}(c^{-17})\,,
\end{align}
where we have defined the nonzero PN coefficients
\allowdisplaybreaks
\begin{align}
t_2&=-\frac{4}{3}a_2\,,
\\
t_3&=-\frac{8}{5}a_3\,,
\\
t_4&=-2a_4+2a_2^2\,,
\\
t_5&=-\frac{8}{3}a_5+\frac{16}{3}a_3a_2\,,
\\
t_6&=-4a_6-6b_6+8a_4a_2+4a_3^2-4a_2^3\,,
\\
t_7&=-8a_7+16a_5a_2+16a_4a_3-24a_3a_2^2\,,
\\
t_9&=8a_9-24b_9-16a_7a_2-16a_6a_3+48b_6a_3-16a_5a_4\nn\\
&+24a_5a_2^2+48a_4a_3a_2+8a_3^3-32a_3a_2^3\,,
\end{align}
\allowdisplaybreaks
\begin{widetext}
\begin{align}
t_{10}&=4a_{10}-6b_{10}-8a_8a_2+12b_8a_2-8a_7a_3-8a_6a_4+12a_6a_2^2+12b_6a_4-18b_6a_2^2-4a_5^2+24a_5a_3a_2+12a_4^2a_2\nn\\
&+12a_4a_3^2-16a_4a_2^3-24a_3^2a_2^2+4a_2^5\,,
\\
t_{11}&=\frac{8}{3}a_{11}-\frac{8}{3}b_{11}-\frac{16}{3}a_9a_2+\frac{16}{3}b_9a_2-\frac{16}{3}a_8a_3+\frac{16}{3}b_8a_3-\frac{16}{3}a_7a_4+8a_7a_2^2-\frac{16}{3}a_6a_5+16a_6a_3a_2+\frac{16}{3}b_6a_5\nn\\
&-16b_6a_3a_2-\frac{32}{3}a_5a_2^3+16a_5a_4a_2+8a_5a_3^2+8a_4^2a_3-32a_4a_3a_2^2-\frac{32}{3}a_3^3a_2+\frac{40}{3}a_3a_2^4\,,
\\
t_{12}&=-4a_{10}a_2+3b_{10}a_2-4a_9a_3+3b_9a_3-4a_8a_4+6a_8a_2^2+3b_8a_4-\frac{9}{2}b_8a_2^2-4a_7a_5+12a_7a_3a_2-2a_6^2+3a_6b_6\nn\\
&+12a_6a_4a_2+6a_6a_3^2-8a_6a_2^3-\frac{9}{4}b_6^2-9b_6a_4a_2-\frac{9}{2}b_6a_3^2+6b_6a_2^3+6a_5^2a_2+12a_5a_4a_3-24a_5a_3a_2^2+2a_4^3\nn\\
&-12a_4^2a_2^2-24a_4a_3^2a_2+10a_4a_2^4-2a_3^4+20a_3^2a_2^3-2a_2^6\,,
\\
t_{13}&=-\frac{16}{5}a_{11}a_2+\frac{48}{25}b_{11}a_2-\frac{16}{5}a_{10}a_3+\frac{48}{25}b_{10}a_3-\frac{16}{5}a_9a_4+\frac{24}{5}a_9a_2^2+\frac{48}{25}b_9a_4-\frac{72}{25}b_9a_2^2-\frac{16}{5}a_8a_5+\frac{48}{5}a_8a_3a_2\nn\\
&+\frac{48}{25}b_8a_5-\frac{144}{25}b_8a_3a_2-\frac{16}{5}a_7a_6+\frac{48}{25}a_7b_6+\frac{48}{5}a_7a_4a_2+\frac{24}{5}a_7a_3^2-\frac{32}{5}a_7a_2^3+\frac{48}{5}a_6a_5a_2+\frac{48}{5}a_6a_4a_3\nn\\
&-\frac{96}{5}a_6a_3a_2^2-\frac{144}{25}b_6a_5a_2-\frac{144}{25}b_6a_4a_3+\frac{288}{25}b_6a_3a_2^2+\frac{24}{5}a_5^2a_3+\frac{24}{5}a_5a_4^2-\frac{96}{5}a_5a_4a_2^2-\frac{96}{5}a_5a_3^2a_2+8a_5a_2^4\nn\\
&-\frac{96}{5}a_4^2a_3a_2-\frac{32}{5}a_4a_3^3+32a_4a_3a_2^3+16a_3^3a_2^2-\frac{48}{5}a_3a_2^5\,,
\\
t_{14}&=-\frac{8}{3}a_{11}a_3+\frac{4}{3}b_{11}a_3-\frac{8}{3}a_{10}a_4+4a_{10}a_2^2+\frac{4}{3}b_{10}a_4-2b_{10}a_2^2-\frac{8}{3}a_9a_5+8a_9a_3a_2+\frac{4}{3}b_9a_5-4b_9a_3a_2-\frac{8}{3}a_8a_6\nn\\
&+\frac{4}{3}a_8b_6+8a_8a_4a_2+4a_8a_3^2-\frac{16}{3}a_8a_2^3+\frac{4}{3}b_8a_6-\frac{4}{3b_8}b_6-4b_8a_4a_2-2b_8a_3^2+\frac{8}{3}b_8a_2^3-\frac{4}{3}a_7^2+8a_7a_5a_2+8a_7a_4a_3\nn\\
&-16a_7a_3a_2^2+4a_6^2a_2-4a_6b_6a_2+8a_6a_5a_3+4a_6a_4^2-16a_6a_4a_2^2-16a_6a_3^2a_2+\frac{20}{3}a_6a_2^4+2b_6^2a_2-4b_6a_5a_3-2b_6a_4^2\nn\\
&+8b_6a_4a_2^2+8b_6a_3^2a_2-\frac{10}{3}b_6a_2^4+4a_5^2a_4-8a_5^2a_2^2-32a_5a_4a_3a_2-\frac{16}{3}a_5a_3^3+\frac{80}{3}a_5a_3a_2^3-\frac{16}{3}a_4^3a_2-8a_4^2a_3^2\nn\\
&+\frac{40}{3}a_4^2a_2^3+40a_4a_3^2a_2^2-8a_4a_2^5+\frac{20}{3}a_3^4a_2-20a_3^2a_2^4+\frac{4}{3}a_2^7\,,
\\
t_{15}&=-\frac{16}{7}a_{11}a_4+\frac{24}{7}a_{11}a_2^2+\frac{48}{49}b_{11}a_4-\frac{72}{49}b_{11}a_2^2-\frac{16}{7}a_{10}a_5+\frac{48}{7}a_{10}a_3a_2+\frac{48}{49}b_{10}a_5-\frac{144}{49}b_{10}a_3a_2-\frac{16}{7}a_9a_6\nn\\
&+\frac{48}{49}a_9b_6+\frac{48}{7}a_9a_4a_2+\frac{24}{7}a_9a_3^2-\frac{32}{7}a_9a_2^3+\frac{48}{49}b_9a_6-\frac{288}{343}b_9b_6-\frac{144}{49}b_9a_4a_2-\frac{72}{49}b_9a_3^2+\frac{96}{49}b_9a_2^3-\frac{16}{7}a_8a_7\nn\\
&+\frac{48}{7}a_8a_5a_2+\frac{48}{7}a_8a_4a_3-\frac{96}{7}a_8a_3a_2^2+\frac{48}{49}b_8a_7-\frac{144}{49}b_8a_5a_2-\frac{144}{49}b_8a_4a_3+\frac{288}{49}b_8a_3a_2^2+\frac{48}{7}a_7a_6a_2\nn\\
&-\frac{144}{49}a_7b_6a_2+\frac{48}{7}a_7a_5a_3+\frac{24}{7}a_7a_4^2-\frac{96}{7}a_7a_4a_2^2-\frac{96}{7}a_7a_3^2a_2+\frac{40}{7}a_7a_2^4+\frac{24}{7}a_6^2a_3-\frac{144}{49}a_6b_6a_3+\frac{48}{7}a_6a_5a_4\nn\\
&-\frac{96}{7}a_6a_5a_2^2-\frac{192}{7}a_6a_4a_3a_2-\frac{32}{7}a_6a_3^3+\frac{160}{7}a_6a_3a_2^3+\frac{432}{343}b_6^2a_3-\frac{144}{49}b_6a_5a_4+\frac{288}{49}b_6a_5a_2^2+\frac{576}{49}b_6a_4a_3a_2\nn\\
&-\frac{480}{49}b_6a_3a_2^3+\frac{96}{49}b_6a_3^3+\frac{8}{7}a_5^3-\frac{96}{7}a_5^2a_3a_2-\frac{96}{7}a_5a_4^2a_2-\frac{96}{7}a_5a_4a_3^2+\frac{160}{7}a_5a_4a_2^3+\frac{240}{7}a_5a_3^2a_2^2-\frac{48}{7}a_5a_2^5\nn\\
&-\frac{32}{7}a_4^3a_3+\frac{240}{7}a_4^2a_3a_2^2+\frac{160}{7}a_4a_3^3a_2-\frac{240}{7}a_4a_3a_2^4+\frac{8}{7}a_3^5-\frac{160}{7}a_3^3a_2^3+8a_3a_2^6\,,
\\
t_{16}&=-2a_{11}a_5+6a_{11}a_3a_2+\frac{3}{4}b_{11}a_5-\frac{9}{4}b_{11}a_3a_2-2a_{10}a_6+\frac{3}{4}a_{10}b_6+6a_{10}a_4a_2+3a_{10}a_3^2-4a_{10}a_2^3+\frac{3}{4}b_{10}a_6\nn\\
&-\frac{9}{16}b_{10}b_6-\frac{9}{4}b_{10}a_4a_2-\frac{9}{8}b_{10}a_3^2+\frac{3}{2}b_{10}a_2^3-2a_9a_7+6a_9a_5a_2+6a_9a_4a_3-12a_9a_3a_2^2+\frac{3}{4}b_9a_7-\frac{9}{4}b_9a_5a_2\nn\\
&-\frac{9}{4}b_9a_4a_3+\frac{9}{2}b_9a_3a_2^2-a_8^2+\frac{3}{4}a_8b_8+6a_8a_6a_2-\frac{9}{4}a_8b_6a_2+6a_8a_5a_3+3a_8a_4^2-12a_8a_4a_2^2-12a_8a_3^2a_2+5a_8a_2^4\nn\\
&-\frac{9}{32}b_8^2-\frac{9}{4}b_8a_6a_2+\frac{27}{16}b_8b_6a_2-\frac{9}{4}b_8a_5a_3-\frac{9}{8}b_8a_4^2+\frac{9}{2}b_8a_4a_2^2+\frac{9}{2}b_8a_3^2a_2-\frac{15}{8}b_8a_2^4+3a_7^2a_2+6a_7a_6a_3\nn\\
&-\frac{9}{4}a_7b_6a_3+6a_7a_5a_4-12a_7a_5a_2^2-24a_7a_4a_3a_2-4a_7a_3^3+20a_7a_3a_2^3+3a_6^2a_4-6a_6^2a_2^2-\frac{9}{4}a_6b_6a_4+\frac{9}{2}a_6b_6a_2^2\nn\\
&+3a_6a_5^2-24a_6a_5a_3a_2-12a_6a_4^2a_2-12a_6a_4a_3^2+20a_6a_4a_2^3+30a_6a_3^2a_2^2-6a_6a_2^5+\frac{27}{32}b_6^2a_4-\frac{27}{16}b_6^2a_2^2-\frac{9}{8}b_6a_5^2\nn\\
&+9b_6a_5a_3a_2+\frac{9}{2}b_6a_4^2a_2+\frac{9}{2}b_6a_4a_3^2-\frac{15}{2}b_6a_4a_2^3-\frac{45}{4}b_6a_3^2a_2^2+\frac{9}{4}b_6a_2^5-12a_5^2a_4a_2-6a_5^2a_3^2+10a_5^2a_2^3-12a_5a_4^2a_3\nn\\
&+60a_5a_4a_3a_2^2+20a_5a_3^3a_2-30a_5a_3a_2^4-a_4^4+10a_4^3a_2^2+30a_4^2a_3^2a_2-15a_4^2a_2^4+5a_4a_3^4-60a_4a_3^2a_2^3+7a_4a_2^6-a_2^8\nn\\
&-15a_3^4a_2^2+21a_3^2a_2^5\,,
\end{align}
\end{widetext}
\allowdisplaybreaks
\begin{align}
t^{\ell}_6&=-12b_6\,,
\\
t^{\ell}_8&=8a_8-16a_6a_2-16a_5a_3-8a_4^2+24a_4a_2^2\nn\\
&+24a_3^2a_2-8a_2^4\,,
\\
t^{\ell}_9&=24b_9-48b_6a_3\,,
\\
t^{\ell}_{10}&=12b_{10}-24b_8a_2-24b_6a_4+36b_6a_2^2\,,
\\
t^{\ell}_{11}&=8b_{11}-16b_9a_2-16b_8a_3-16b_6a_5+48b_6a_3a_2\,,
\\
t^{\ell}_{12}&=-12b_{10}a_2-12b_9a_3-12b_8a_4+18b_8a_2^2-12a_6b_6\nn\\
&+9b_6^2+36b_6a_4a_2+18b_6a_3^2-24b_6a_2^3\,,
\end{align}
\allowdisplaybreaks
\begin{widetext}
\begin{align}
t^{\ell}_{13}&=-\frac{48}{5}b_{11}a_2-\frac{48}{5}b_{10}a_3-\frac{48}{5}b_9a_4+\frac{72}{5}b_9a_2^2-\frac{48}{5}b_8a_5+\frac{144}{5}b_8a_3a_2-\frac{48}{5}a_7b_6+\frac{144}{5}b_6a_5a_2+\frac{144}{5}b_6a_4a_3\nn\\
&-\frac{288}{5}b_6a_3a_2^2\,,
\\
t^{\ell}_{14}&=-8b_{11}a_3-8b_{10}a_4+12b_{10}a_2^2-8b_9a_5+24b_9a_3a_2-8a_8b_6-8a_6b_8+8b_8b_6+24b_8a_4a_2+12b_8a_3^2-16b_8a_2^3\nn\\
&+24a_6b_6a_2-12b_6^2a_2+24b_6a_5a_3+12b_6a_4^2-48b_6a_4a_2^2+20b_6a_2^4-48b_6a_3^2a_2\,,
\\
t^{\ell}_{15}&=-\frac{48}{7}b_{11}a_4+\frac{72}{7}b_{11}a_2^2-\frac{48}{7}b_{10}a_5+\frac{144}{7}b_{10}a_3a_2-\frac{48}{7}a_9b_6-\frac{48}{7}b_9a_6+\frac{288}{49}b_9b_6+\frac{144}{7}b_9a_4a_2+\frac{72}{7}b_9a_3^2\nn\\
&-\frac{96}{7}b_9a_2^3-\frac{48}{7}b_8a_7+\frac{144}{7}b_8a_5a_2+\frac{144}{7}b_8a_4a_3-\frac{288}{7}b_8a_3a_2^2+\frac{144}{7}a_7a_6a_2+\frac{144}{7}a_6b_6a_3-\frac{432}{49}b_6^2a_3+\frac{144}{7}b_6a_5a_4\nn\\
&-\frac{288}{7}b_6a_5a_2^2-\frac{576}{7}b_6a_4a_3a_2-\frac{96}{7}b_6a_3^3+\frac{480}{7}b_6a_3a_2^3\,,
\\
t^{\ell}_{16}&=-6b_{11}a_5+18b_{11}a_3a_2-6a_{10}b_6-6a_6b_{10}+\frac{9}{2}b_{10}b_6+18b_{10}a_4a_2+9b_{10}a_3^2-12b_{10}a_2^3-6b_9a_7+18b_9a_5a_2\nn\\
&+18b_9a_4a_3-36b_9a_3a_2^2-6a_8b_8+18a_8b_6a_2+\frac{9}{4}b_8^2+18b_8a_6a_2-\frac{27}{2}b_8b_6a_2+18b_8a_5a_3+9b_8a_4^2-36b_8a_4a_2^2\nn\\
&-36b_8a_3^2a_2+15b_8a_2^4+18a_7b_6a_3+18a_6b_6a_4-36a_6b_6a_2^2-\frac{27}{4}b_6^2a_4+\frac{27}{2}b_6^2a_2^2+9b_6a_5^2-72b_6a_5a_3a_2-36b_6a_4^2a_2\nn\\
&-36b_6a_4a_3^2+60b_6a_4a_2^3+90b_6a_3^2a_2^2-18b_6a_2^5\,,
\end{align}
\end{widetext}
\allowdisplaybreaks
\begin{align}
t^{\ell^{2}}_8&=12b_8-24b_6a_2\,,
\\
t^{\ell^{2}}_{12}&=-18b_6^2\,,
\\
t^{\ell^2}_{14}&=-24b_8b_6+36b_6^2a_2\,,
\\
t^{\ell^2}_{15}&=-\frac{144}{7}b_9b_6+\frac{216}{7}b_6^2a_3\,,
\\
t^{\ell^2}_{16}&=-18b_{10}b_6-9b_8^2+54b_8b_6a_2+27b_6^2a_4-54b_6^2a_2^2\,.
\end{align}
These coefficients depend on $(a_{i},b_{i})$, which can be found in Appendix~\ref{app-coeffomegadot}. Some of these coefficients depend on spins, and we evaluate them at their initial spin values, as explained at the end of Sec.~\ref{Si-sol}.

\section{Coefficients in the SPA Phase}
\label{app-PhiofF}

The nonprecessing part of the SPA phase is given by Eq.~\eqref{sum-SPA-Phase} where we have defined the nonzero coefficients
\allowdisplaybreaks
\begin{align}
{\Psi}_{2} &= - \frac{20}{9} a_{2}\,,
\\
{\Psi}_{3} &= - 4 a_{3}\,,
\\
{\Psi}_{4} &= -10 a_4 +10a_{2}\,,
\\
{\Psi}_{5} &=   \frac{40}{9} a_{5} - \frac{80}{9} a_{3} a_{2}\,,
\\
{\Psi}_{6} &=  20 a_{6} - 30 b_{6} -40 a_{4} a_{2}  - 20 a_{3}^{2} + 20 a_{2}^{3}\,,
\\
{\Psi}_{7} &= 20 a_{7} -40 a_{5} a_{2} - 40 a_{4} a_{3}+ 60 a_{3} a_{2}^{2}\,,
\\
{\Psi}_{8} &= \frac{40}{9} a_{8} - \frac{40}{9} b_{8}- \frac{80}9 a_{6}a_2 +\frac{80}{9} b_{6}a_2 - \frac{80}{9} a_{5} a_{3} \nn\\
& - \frac{40}{9} a_{4}^{2}  + \frac{40}{3} a_{4} a_{2}^{2} + \frac{120}{9}  a_{3}^2a_{2}- \frac{40}{9} a_{2}^4\,,
\\
{\Psi}_{9} &=-10a_9+\frac{75}{2}b_9+20a_7a_2+20a_6a_3-75b_6a_3\nn\\
&+20a_5a_4-30a_5a_2^2-60a_4a_3a_2-10a_3^3+40a_3a_2^3\,,  
\end{align}
\allowdisplaybreaks
\begin{widetext}
\begin{align}
{\Psi}_{10} &=-4a_{10}+\frac{42}{5}b_{10}+8a_8a_2-\frac{84}{5}b_8a_2+8a_7a_3+8a_6a_4-12a_6a_2^2-\frac{84}{5}b_6a_4+\frac{126}{5}b_6a_2^2+4a_5^2-24a_5a_3a_2\nn\\
&-12a_4^2a_2-12a_4a_3^2+16a_4a_2^3+24a_3^2a_2^2-4a_2^5\,,
\\
{\Psi}_{11} &=-\frac{20}{9}a_{11}+\frac{10}{3}b_{11}+\frac{40}{9}a_9a_2-\frac{20}{3}b_9a_2+\frac{40}{9}a_8a_3-\frac{20}{3}b_8a_3+\frac{40}{9}a_7a_4-\frac{20}{3}a_7a_2^2+\frac{40}{9}a_6a_5-\frac{40}{3}a_6a_3a_2\nn\\
&-\frac{20}{3}b_6a_5+20b_6a_3a_2-\frac{40}{3}a_5a_4a_2-\frac{20}{3}a_5a_3^2+\frac{80}{9}a_5a_2^3-\frac{20}{3}a_4^2a_3+\frac{80}{3}a_4a_3a_2^2+\frac{80}{9}a_3^3a_2-\frac{100}{9}a_3a_2^4\,,
\\
{\Psi}_{12} &=\frac{20}{7}a_{10}a_2-\frac{165}{49}b_{10}a_2+\frac{20}{7}a_9a_3-\frac{165}{49}b_9a_3+\frac{20}{7}a_8a_4-\frac{30}{7}a_8a_2^2-\frac{165}{49}b_8a_4+\frac{495}{98}b_8a_2^2+\frac{20}{7}a_7a_5-\frac{60}{7}a_7a_3a_2\nn\\
&+\frac{10}{7}a_6^2-\frac{165}{49}a_6b_6-\frac{60}{7}a_6a_4a_2-\frac{30}{7}a_6a_3^2+\frac{40}{7}a_6a_2^3+\frac{4185}{1372}b_6^2+\frac{495}{49}b_6a_4a_2+\frac{495}{98}b_6a_3^2-\frac{330}{49}b_6a_2^3-\frac{30}{7}a_5^2a_2\nn\\
&-\frac{60}{7}a_5a_4a_3+\frac{120}{7}a_5a_3a_2^2-\frac{10}{7}a_4^3+\frac{60}{7}a_4^2a_2^2+\frac{120}{7}a_4a_3^2a_2-\frac{50}{7}a_4a_2^4+\frac{10}{7}a_3^4-\frac{100}{7}a_3^2a_2^3+\frac{10}{7}a_2^6\,,
\\
{\Psi}_{13} &=2a_{11}a_2-\frac{39}{20}b_{11}a_2+2a_{10}a_3-\frac{39}{20}b_{10}a_3+2a_9a_4-3a_9a_2^2-\frac{39}{20}b_9a_4+\frac{117}{40}b_9a_2^2+2a_8a_5-6a_8a_3a_2-\frac{39}{20}b_8a_5\nn\\
&+\frac{117}{20}b_8a_3a_2+2a_7a_6-\frac{39}{20}a_7b_6-6a_7a_4a_2-3a_7a_3^2+4a_7a_2^3-6a_6a_5a_2-6a_6a_4a_3+12a_6a_3a_2^2+\frac{117}{20}b_6a_5a_2\nn\\
&+\frac{117}{20}b_6a_4a_3-\frac{117}{20}b_6a_3a_2^2-3a_5^2a_3-3a_5a_4^2+12a_5a_4a_2^2+12a_5a_3^2a_2-5a_5a_2^4+12a_4^2a_3a_2+4a_4a_3^3-20a_4a_3a_2^3\nn\\
&-10a_3^3a_2^2+6a_3a_2^5\,,
\\
{\Psi}_{14} &=\frac{40}{27}a_{11}a_3-\frac{100}{81}b_{11}a_3+\frac{40}{27}a_{10}a_4-\frac{20}{9}a_{10}a_2^2-\frac{100}{81}b_{10}a_4+\frac{50}{27}b_{10}a_2^2+\frac{40}{27}a_9a_5-\frac{40}{9}a_9a_3a_2-\frac{100}{81}b_9a_5\nn\\
&+\frac{100}{27}b_9a_3a_2+\frac{40}{27}a_8a_6-\frac{100}{81}a_8b_6-\frac{40}{9}a_8a_4a_2-\frac{20}{9}a_8a_3^2+\frac{80}{27}a_8a_2^3-\frac{100}{81}b_8a_6+\frac{380}{243}b_8b_6+\frac{100}{27}b_8a_4a_2\nn\\
&-\frac{200}{81}b_8a_2^3+\frac{50}{27}b_8a_3^2+\frac{20}{27}a_7^2-\frac{40}{9}a_7a_5a_2-\frac{40}{9}a_7a_4a_3+\frac{80}{9}a_7a_3a_2^2-\frac{20}{9}a_6^2a_2+\frac{100}{27}a_6b_6a_2-\frac{40}{9}a_6a_5a_3-\frac{20}{9}a_6a_4^2\nn\\
&+\frac{80}{9}a_6a_4a_2^2+\frac{80}{9}a_6a_3^2a_2-\frac{100}{27}a_6a_2^4-\frac{190}{81}b_6^2a_2-\frac{190}{81}b_6^2a_2+\frac{100}{27}b_6a_5a_3+\frac{50}{27}b_6a_4^2-\frac{200}{27}b_6a_4a_2^2-\frac{200}{27}b_6a_3^2a_2\nn\\
&+\frac{250}{81}b_6a_2^4-\frac{20}{9}a_5^2a_4+\frac{40}{9}a_5^2a_2^2+\frac{160}{9}a_5a_4a_3a_2+\frac{80}{27}a_5a_3^3-\frac{400}{27}a_5a_3a_2^3+\frac{80}{27}a_4^3a_2+\frac{40}{9}a_4^2a_3^2-\frac{200}{27}a_4^2a_2^3\nn\\
&-\frac{200}{9}a_4a_3^2a_2^2+\frac{40}{9}a_4a_2^5-\frac{100}{27}a_3^4a_2+\frac{100}{9}a_3^2a_2^4-\frac{20}{27}a_2^7\,,
\\
{\Psi}_{15} &=\frac{8}{7}a_{11}a_4-\frac{12}{7}a_{11}a_2^2-\frac{204}{245}b_{11}a_4+\frac{306}{245}b_{11}a_2^2+\frac{8}{7}a_{10}a_5-\frac{24}{7}a_10a_3a_2-\frac{204}{245}b_{10}a_5+\frac{612}{245}b_{10}a_3a_2+\frac{8}{7}a_9a_6\nn\\
&-\frac{204}{245}a_9b_6-\frac{24}{7}a_9a_4a_2-\frac{12}{7}a_9a_3^2+\frac{16}{7}a_9a_2^3-\frac{204}{245}b_9a_6+\frac{7884}{8575}b_9b_6+\frac{612}{245}b_9a_4a_2+\frac{306}{245}b_9a_3^2-\frac{408}{245}b_9a_2^3+\frac{8}{7}a_8a_7\nn\\
&-\frac{24}{7}a_8a_5a_2-\frac{24}{7}a_8a_4a_3+\frac{48}{7}a_8a_3a_2^2-\frac{204}{245}b_8a_7+\frac{612}{245}b_8a_5a_2+\frac{612}{245}b_8a_4a_3-\frac{1224}{245}b_8a_3a_2^2-\frac{24}{7}a_7a_6a_2\nn\\
&+\frac{612}{245}a_7b_6a_2-\frac{24}{7}a_7a_5a_3-\frac{12}{7}a_7a_4^2+\frac{48}{7}a_7a_4a_2^2+\frac{48}{7}a_7a_3^2a_2-\frac{20}{7}a_7a_2^4-\frac{12}{7}a_6^2a_3+\frac{612}{245}a_6b_6a_3-\frac{24}{7}a_6a_5a_4\nn\\
&+\frac{48}{7}a_6a_5a_2^2+\frac{96}{7}a_6a_4a_3a_2+\frac{16}{7}a_6a_3^3-\frac{80}{7}a_6a_3a_2^3-\frac{11826}{8575}b_6^2a_3+\frac{612}{245}b_6a_5a_4-\frac{1224}{245}b_6a_5a_2^2-\frac{2448}{245}b_6a_4a_3a_2\nn\\
&-\frac{408}{245}b_6a_3^3+\frac{408}{49}b_6a_3a_2^3-\frac{4}{7}a_5^3+\frac{48}{7}a_5^2a_3a_2+\frac{48}{7}a_5a_4^2a_2+\frac{48}{7}a_5a_4a_3^2-\frac{80}{7}a_5a_4a_2^3-\frac{120}{7}a_5a_3^2a_2^2+\frac{24}{7}a_5a_2^5\nn\\
&+\frac{16}{7}a_4^3a_3-\frac{120}{7}a_4^2a_3a_2^2-\frac{80}{7}a_4a_3^3a_2+\frac{120}{7}a_4a_3a_2^4-\frac{4}{7}a_3^5+\frac{80}{7}a_3^3a_2^3-4a_3a_2^6\,,
\\
{\Psi}_{16} &=\frac{10}{11}a_{11}a_5-\frac{30}{11}a_{11}a_3a_2-\frac{285}{484}b_{11}a_5+\frac{855}{484}b_{11}a_3a_2+\frac{10}{11}a_{10}a_6-\frac{285}{484}a_{10}b_6-\frac{30}{11}a_{10}a_4a_2-\frac{15}{11}a_{10}a_3^2+\frac{20}{11}a_{10}a_2^3\nn\\
&-\frac{285}{484}b_{10}a_6+\frac{12285}{21296}b_{10}b_6+\frac{855}{484}b_{10}a_4a_2+\frac{855}{968}b_{10}a_3^2-\frac{285}{242}b_{10}a_2^3+\frac{10}{11}a_9a_7-\frac{30}{11}a_9a_5a_2-\frac{30}{11}a_9a_4a_3+\frac{60}{11}a_9a_3a_2^2\nn\\
&-\frac{285}{484}b_9a_7+\frac{855}{484}b_9a_5a_2+\frac{855}{484}b_9a_4a_3-\frac{855}{242}b_9a_3a_2^2+\frac{5}{11}a_8^2-\frac{285}{484}a_8b_8-\frac{30}{11}a_8a_6a_2+\frac{855}{484}a_8b_6a_2-\frac{30}{11}a_8a_5a_3\nn\\
&-\frac{15}{11}a_8a_4^2+\frac{60}{11}a_8a_4a_2^2+\frac{60}{11}a_8a_3^2a_2-\frac{25}{11}a_8a_2^4+\frac{12285}{42592}b_8^2+\frac{855}{484}b_8a_6a_2-\frac{36855}{21296}b_8b_6a_2+\frac{855}{484}b_8a_5a_3+\frac{855}{968}b_8a_4^2\nn\\
&-\frac{855}{242}b_8a_4a_2^2-\frac{855}{242}b_8a_3^2a_2+\frac{1425}{968}b_8a_2^4-\frac{15}{11}a_7^2a_2-\frac{30}{11}a_7a_6a_3+\frac{855}{484}a_7b_6a_3-\frac{30}{11}a_7a_5a_4+\frac{60}{11}a_7a_5a_2^2\nn\\
&+\frac{120}{11}a_7a_4a_3a_2+\frac{20}{11}a_7a_3^3-\frac{100}{11}a_7a_3a_2^3-\frac{15}{11}a_6^2a_4+\frac{30}{11}a_6^2a_2^2+\frac{855}{484}a_6b_6a_4-\frac{855}{242}a_6b_6v-\frac{15}{11}a_6a_5^2+\frac{120}{11}a_6a_5a_3a_2\nn\\
&+\frac{60}{11}a_6a_4^2a_2+\frac{60}{11}a_6a_4a_3^2-\frac{100}{11}a_6a_4a_2^3-\frac{150}{11}a_6a_3^2a_2^2+\frac{30}{11}a_6a_2^5-\frac{36855}{42592}b_6^2a_4+\frac{36855}{21296}b_6^2a_2^2+\frac{855}{968}b_6a_5^2\nn\\
&-\frac{855}{121}b_6a_5a_3a_2-\frac{855}{242}b_6a_4^2a_2-\frac{855}{242}b_6a_4a_3^2+\frac{1425}{242}b_6a_4a_2^3+\frac{4275}{484}b_6a_3^2a_2^2-\frac{855}{484}b_6a_2^5+\frac{60}{11}a_5^2a_4a_2+\frac{30}{11}a_5^2a_3^2\nn\\
&-\frac{50}{11}a_5^2a_2^3+\frac{60}{11}a_5a_4^2a_3-\frac{300}{11}a_5a_4a_3a_2^2-\frac{100}{11}a_5a_3^3a_2+\frac{150}{11}a_5a_3a_2^4+\frac{5}{11}a_4^4-\frac{50}{11}a_4^3a_2^2-\frac{150}{11}a_4^2a_3^2a_2+\frac{75}{11}a_4^2a_2^4\nn\\
&-\frac{25}{11}a_4a_3^4+\frac{300}{11}a_4a_3^2a_2^3-\frac{35}{11}a_4a_2^6+\frac{75}{11}a_3^4a_2^2-\frac{105}{11}a_3^2a_2^5+\frac{5}{11}a_2^8\,,
\end{align}
\end{widetext}
\allowdisplaybreaks
\begin{align}
{\Psi}_{5}^{\ell} &=  \frac{40}{3} a_{5} -\frac{80}{3}  a_{3} a_{2}\,,
\\
{\Psi}_{6}^{\ell} &= 60 b_{6}\,,
\\
{\Psi}^{\ell}_{8} &= -\frac{40}{3}a_8+\frac{40}{3}b_8+\frac{80}{3}a_6a_2-\frac{80}{3}b_6a_2+\frac{80}{3}a_5a_3\nn\\
&+\frac{40}{3}a_4^2-40a_4a_2^2-40a_3^2a_2+\frac{40}{3}a_2^4\,,
\\
{\Psi}^{\ell}_{9} &= -30 b_9 +60 b_6 a_3\,,
\\
{\Psi}^{\ell}_{10} &= -12 b_{10}+ 24b_8a_2+24b_6a_4-32b_6a_2^2\,,
\\
{\Psi}^{\ell}_{11} &=-\frac{20}{3}b_{11}+\frac{40}{3}b_9a_2+\frac{40}{3}b_8a_3+\frac{40}{3}b_6a_5\nn\\
&-40b_6a_3a_2\,,
\\
{\Psi}^{\ell}_{12} &= \frac{60}{7}b_{10}a_2+\frac{60}{7}b_9a_3+\frac{60}{7}b_8a_4-\frac{90}{7}b_8a_2^2+\frac{60}{7}a_6b_6\nn\\
&-\frac{495}{49}b_6^2-\frac{180}{7}b_6a_4a_2-\frac{90}{7}b_6a_3^2+\frac{120}{7}b_6a_2^3\,,
\\
{\Psi}^{\ell}_{13} &=6b_{11}a_2+6b_{10}a_3+6b_9a_4-9b_9a_2^2+6b_8a_5\nn\\
&-18b_8a_3a_2+6a_7b_6-18b_6a_5a_2-18b_6a_4a_3\nn\\
&+36b_6a_3a_2^2\,,
\end{align}
\allowdisplaybreaks
\begin{widetext}
\begin{align}
{\Psi}^{\ell}_{14} &=\frac{40}{9}b_{11}a_3+\frac{40}{9}b_{10}a_4-\frac{20}{3}b_{10}a_2^2+\frac{40}{9}b_9a_5-\frac{40}{3}b_9a_3a_2+\frac{40}{9}a_8b_6+\frac{40}{9}b_8a_6-\frac{200}{27}b_8b_6-\frac{40}{3}b_8a_4a_2-\frac{20}{3}b_8a_3^2\nn\\
&+\frac{80}{9}b_8a_2^3-\frac{40}{3}b_6a_6a_2+\frac{100}{9}b_6^2a_2-\frac{40}{3}b_6a_5a_3-\frac{20}{3}b_6a_4^2+\frac{80}{3}b_6a_4a_2^2+\frac{80}{3}b_6a_3^2a_2-\frac{100}{9}b_6a_2^4\,,
\\
{\Psi}^{\ell}_{15} &= \frac{24}{7}b_{11}a_4-\frac{36}{7}b_{11}a_2^2+\frac{24}{7}b_{10}a_5-\frac{72}{7}b_{10}a_3a_2+\frac{24}{7}a_9b_6+\frac{24}{7}b_9a_6-\frac{1224}{245}b_9b_6-\frac{72}{7}b_9a_4a_2-\frac{36}{7}b_9a_3^2+\frac{48}{7}b_9a_2^3\nn\\
&+\frac{24}{7}b_8a_7-\frac{72}{7}b_8a_5a_2-\frac{72}{7}b_8a_4a_3+\frac{144}{7}b_8a_3a_2^2-\frac{72}{7}a_7b_6a_2-\frac{72}{7}a_6b_6a_3+\frac{1836}{245}b_6^2a_3-\frac{72}{7}b_6a_5a_4+\frac{144}{7}b_6a_5a_2^2\nn\\
&+\frac{288}{7}b_6a_4a_3a_2+\frac{48}{7}b_6a_3^3-\frac{240}{7}b_6a_3a_2^3\,,
\\
{\Psi}^{\ell}_{16} &=\frac{30}{11}b_{11}a_5-\frac{90}{11}b_{11}a_3a_2+\frac{30}{11}a_{10}b_6+\frac{30}{11}b_{10}a_6-\frac{855}{242}b_{10}b_6-\frac{90}{11}b_{10}a_4a_2-\frac{45}{11}b_{10}a_3^2+\frac{60}{11}b_{10}a_2^3+\frac{30}{11}b_9a_7\nn\\
&-\frac{90}{11}b_9a_5a_2-\frac{90}{11}b_9a_4a_3+\frac{180}{11}b_9a_3a_2^2+\frac{30}{11}a_8b_8-\frac{90}{11}a_8b_6a_2-\frac{855}{484}b_8^2-\frac{90}{11}b_8a_6a_2+\frac{2565}{242}b_8b_6a_2-\frac{90}{11}b_8a_5a_3\nn\\
&-\frac{45}{11}b_8a_4^2+\frac{180}{11}b_8a_4a_2^2-\frac{75}{11}b_8a_2^4+\frac{180}{11}b_8a_3^2a_2-\frac{90}{11}a_7b_6a_3-\frac{90}{11}a_6b_6a_4+\frac{180}{11}a_6b_6a_2^2+\frac{2565}{484}b_6^2a_4-\frac{2565}{242}b_6^2a_2^2\nn\\
&-\frac{45}{11}b_6a_5^2+\frac{360}{11}b_6a_5a_3a_2+\frac{180}{11}b_6a_4^2a_2+\frac{180}{11}b_6a_4a_3^2-\frac{300}{11}b_6a_4a_2^3-\frac{450}{11}b_6a_3^2a_2^2+\frac{90}{11}b_6a_2^5\,,
\end{align}
\end{widetext}
\allowdisplaybreaks
\begin{align}
{\Psi}_{8}^{\ell^{2}} &= -20 b_8 40 b_{6} a_{2}\,,
\\
{\Psi}_{12}^{\ell^{2}}&= \frac{90}{7}b_6^2\,,
\\
{\Psi}_{14}^{\ell^{2}}&= \frac{40}{3}b_8b_6-20b_6^2a_2\,,
\\
{\Psi}_{15}^{\ell^{2}}&=\frac{72}{7}b_9b_6-\frac{108}{7}b_6^2a_3\,,
\\
{\Psi}_{16}^{\ell^{2}}&=\frac{90}{11}b_{10}b_6+\frac{45}{11}b_8^2-\frac{270}{11}b_8b_6a_2-\frac{135}{11}b_6^2a_4\nn\\
&+\frac{270}{11}b_6^2a_2^2\,.
\end{align}

\bibliography{review}
\end{document}